\documentclass[times]{arx}

\usepackage{graphicx, graphics}
\usepackage{color}
\usepackage{epsfig}
\usepackage{amssymb}
\usepackage{amsmath}
\usepackage{dsfont}

\newcommand{\vect}[1]{\boldsymbol{#1}}

\newcommand\tr{\textrm{tr}}
\newcommand\sdet{\textrm{sdet}}
\newcommand\str{\textrm{str}}

\newcommand\tZ{\tilde{Z}}
\newcommand\Se{S_\epsilon} 

\newcommand\Sed{S_\epsilon^\dagger}
\newcommand\Sedtau{S_\epsilon^{\mathcal{T}\dagger}}

 \newcommand\tSe{\mathcal{S}_\epsilon}

\newcommand\tSed{\mathcal{S}_\epsilon^\dagger}

\newcommand\tUe{\mathcal{U}_\epsilon}
\newcommand\tUed{\mathcal{U}_\epsilon^\dagger}

\newcommand\tJp{\mathcal{J}_r} 
\newcommand\tJm{\mathcal{J}_a}
\newcommand\tJpi{\mathcal{J}_r^{-1}}
\newcommand\tJmi{\mathcal{J}_a^{-1}}

\newcommand\tEp{\mathcal{E}^{(r)}}
\newcommand\tEm{\mathcal{E}^{(a)}}

\newcommand\sqtSe{\sqrt{\mathcal{S}_{\epsilon}}}
\newcommand\sqtSed{\sqrt{\mathcal{S}_{\epsilon}}^\dagger}

\newcommand\mf{\textrm{{\scriptsize MF}}}
\newcommand\tY{\tilde{Y}}
\newcommand\Yd{Y_\textrm{D}}
\newcommand\Yc{Y_\textrm{C}}
\newcommand\Ydtau{Y_{\textrm{D}}^{\mathcal{T}}}
\newcommand\Yctau{Y_{\textrm{C}}^{\mathcal{T}}}
\newcommand\tYd{\tilde{Y}_\textrm{D}}
\newcommand\tYc{\tilde{Y}_\textrm{C}}
\newcommand\tYdtau{\tilde{Y}^{\mathcal{T}}_\textrm{D}}
\newcommand\tYctau{\tilde{Y}^{\mathcal{T}}_\textrm{C}}

\newcommand\up{\uparrow}
\newcommand\down{\downarrow}

\title{Eigenfunction Statistics on Quantum Graphs}
\date{}
\author{}

\begin{document}

\maketitle

\thispagestyle{empty}

\vspace{-2cm}
  \begin{center} 
   S. Gnutzmann\textsuperscript{1}, $\, $J.P. Keating\textsuperscript{2}, $\, $F. Piotet\textsuperscript{2,3}
   \end{center}
{ \footnotesize  \emph{\begin{center} \textsuperscript{1}School of Mathematical Sciences, University of
    Nottingham , Nottingham, NG7 2RD, United Kingdom \\
  \textsuperscript{2}School of Mathematics, University of Bristol, Bristol, 
    BS8 1TW, United Kingdom \\
\textsuperscript{3}Department of Physics of Complex
    Systems, The Weizmann Institute of Science, 76100 Rehovot ,
    Israel
    \end{center} } }
 \begin{center}   
May 6, 2010
 \end{center}

 \vspace{0.5cm} \noindent
 \textbf{Abstract} 
 
 \vspace{0.2cm}
 \noindent
    We investigate the spatial statistics of the energy eigenfunctions
    on large quantum graphs.  It has previously been conjectured that these
    should be described by a Gaussian Random Wave Model, by analogy
    with quantum chaotic systems, for which such a model was proposed
    by Berry in 1977.  The autocorrelation functions we calculate for
    an individual quantum graph exhibit a universal component, which
    completely determines a Gaussian Random Wave Model, and a
    system-dependent deviation. This deviation depends on the graph
    only through its underlying classical dynamics. Classical criteria
    for quantum universality to be met asymptotically in the large
    graph limit (i.e.~for the non-universal deviation to vanish) are then extracted. We use an exact field theoretic
    expression in terms of a variant of a supersymmetric $\sigma$
    model.  A saddle-point analysis of this expression leads to the
    estimates.  In particular, intensity correlations
    are used to discuss the possible equidistribution of the
    energy eigenfunctions in the large graph limit. When
    equidistribution is asymptotically realized, our theory predicts a
    rate of convergence that is a significant refinement of previous
    estimates. The universal and system-dependent components of 
    intensity correlation functions are recovered by means of
    an exact trace formula which we analyse in the diagonal
    approximation, drawing in this way a parallel
    between the field theory and semiclassics.  Our results provide
    the first instance where an asymptotic Gaussian Random Wave Model
    has been established microscopically for eigenfunctions in a system with no
    disorder.
 
 \vspace{0.5cm} \noindent
\emph{Keywords :} 
    Quantum ergodicity, \sep  Random Wave Model, \sep Criteria
    for universality, \sep  Rate of universality, \sep Trace formulae, \sep Nonlinear supersymmetric $\sigma$ model.

\section{Introduction}
 
Gaussian Random Wave Models are commonly used to describe the
statistical properties of the energy eigenfunctions of chaotic quantum
systems. The original idea was introduced in 1977 by Berry
\cite{Berry}, who proposed that a random function $\psi$ with Gaussian
distribution
\begin{equation} \label{normal density for RWMs} \mathcal{N}(\psi)
  \propto e^{-\frac{\beta}{2} \int \psi^\ast(\vect{r_1})
    c^{-1}(\vect{r_1},\vect{r_2} | e_n) \psi(\vect{r_2}) d\vect{r_1}
    d\vect{r_2} },
\end{equation}
could, in the semiclassical limit, reproduce all the spatial
autocorrelation functions
\begin{equation} \label{Berry autocorrelation functions} C\left(
    \{\vect{x_i} \}_{i \in \mathbb{N}_q} ; \{\vect{y_j} \}_{j \in
      \mathbb{N}_p} \right) \equiv \frac{1}{|S|} \int_S \prod_{i =
    1}^q \psi_n^\ast(\vect{x_i} + \vect{q}) \prod_{j = 1}^p
  \psi_n(\vect{y_j} + \vect{q}) \ d\vect{q},
\end{equation}
of a chaotic eigenfunction $\psi_n$ of energy $e_n$.  Here, $S$ is a
small volume that shrinks in the semiclassical limit but does so
slowly enough to contain an increasing number of oscillations of
$\psi_n$, and $\beta$ in \eqref{normal density for RWMs} is 1 if
time-reversal symmetry is conserved, in which case $\psi$ is chosen
real, and 2 if this symmetry is broken, in which case $\psi$ is
complex. From a semiclassical calculation of $C(\vect{r_1},
\vect{r_2})$, Berry deduced that the covariance
$c(\vect{r_1},\vect{r_2} )$ in \eqref{normal density for RWMs} is the
free quantum propagator from $\vect{r_2}$ to $\vect{r_1}$.  

This is
one of the central conjectures in the field of quantum chaos.
Essentially, it asserts that the local statistics of quantum chaotic
eigenfunctions correspond, in the semiclassical limit, to those of
random superpositions of plane waves, and so are universal.

Following Berry, the universal Gaussian Random Wave Model has been
refined to incorporate systems-specific features.  For example, in
quantum billiards, it does not fulfill the necessary boundary conditions. In this
case, Hortikar and Srednicki \cite{Hortikar} suggested replacing the
covariance with the semiclassical approximation \cite{Berry0,
  Gutzwiller} to the propagator of the system. This Gaussian model
satisfies the boundary conditions and has the property that the direct
path contribution to the semiclassical formula corresponds to Berry's
conjecture. Further understandings and refinements of this
system-dependent Gaussian Random Wave model are given in
\cite{JuanDiego0, JuanDiego1, JuanDiego2}, for example.

It is important to emphasize that to-date effort has mainly been
directed towards deriving the consequences of the Random Wave Model
and its refinements, {\it assuming its validity}.  
Numerical tests strongly support the predictive value
of the Random Wave Model.
However, in no system has
its validity yet been established or derived microscopically.

We tackle here the problem of the validity of the Gaussian Random Wave
Model on quantum graphs (a variant of this model was introduced in \cite{GSW}).  Quantum graphs are favorable systems to gain
some insights on the mechanisms responsible for random waves models to
hold because, depending on their topology and their boundary
conditions, their behaviors range from chaotic \cite{KottosSmil,
  KottosSmil2}, where a random model is expected to hold, to
intermediate \cite{BK, BBK, KMW, BKW, K}, where such models should fail.
Without any prior assumption on the nature of the quantum graph, one
can evaluate its autocorrelation functions
\begin{equation} \label{our autocorrelation functions} 
  C\left(
    \{\vect{x_i} \}_{i \in \mathbb{N}_q} ; \{\vect{y_j} \}_{j \in
      \mathbb{N}_p} \right) \equiv \lim_{N \to \infty} \frac{1}{N}
  \sum_{n = 1}^N \prod_{i = 1}^q \psi_n^\ast(\vect{x_i}) \prod_{j =
    1}^p \psi_n(\vect{y_j}),
\end{equation}
where $\{\vect{x_i} \}_{i \in \mathbb{N}_q}$ and $\{\vect{y_j} \}_{j
  \in \mathbb{N}_p}$ are points on the graph. In fact, we focus on the autocorrelations for $q = p = 1$, and for $q = p$ with
$\{\vect{x_i} \}_{i \in \mathbb{N}_q} = \{\vect{y_j} \}_{j \in
  \mathbb{N}_q}$. The other autocorrelations are believed to vanish
due to additional complex phases that fluctuate strongly. The result
obtained for $q = p = 1$ is exact and yields a universal covariance
$c$ which defines the unique candidate for the Gaussian model on
quantum graphs.  
It should be emphasized that this does not
contradict the construction of Gaussian Random Wave
Models with a system-dependent correction in analogy to   
Urbina and Richter's guess for billiards \cite{JuanDiego0}.  
Indeed, our autocorrelation functions are
defined by averaging over the whole energy spectrum. Such an average
on Urbina and Richter's random functions also kills the
system-dependent correction and leads to a covariance given by the
free propagator, namely, to Berry's universal model. The system
dependency found in the  autocorrelation functions of higher degree
evaluated here
is of different nature. It is not a refinement of a universal 
Random Wave Model, but it rather measures how chaotic a given quantum graph
is from the energy eigenfunctions perspective. Interestingly, this
non-universal term is found to depend on the quantum graph
only through its classical dynamics. This provides us a way to estimate the
deviation from quantum universality in terms of a classical quantity,
and so to discuss criteria for the Random Wave Model to hold in the
large-graph limit, and cases where it fails, such as Neumann star
graphs \cite{BKS2}.  Our results provide the first instance where an asymptotic Gaussian Random Wave Model 
has been established microscopically for eigenfunctions in a system with no disorder.

It would be a major achievement to show that our result for the
autocorrelation functions \eqref{our autocorrelation functions} in the
case of quantum graphs also applies to other quantum systems. If this
is the case, the deviations from universality vanish in the
semiclassical limit in chaotic billiards, which explains why such
deviations have indeed never been found, whereas they must prevail
over the universal part in non-chaotic systems. For chaotic systems,
the corrections would reveal the rate of approach to universality as
$\hbar\rightarrow 0$. Finally, if such a formula was found, its
ability to describe systems with mixed phase spaces could be studied
and compared with the empirical results
\cite{BackerSchubert1,BackerSchubert2} and alternative approaches
based on bifurcation theory and singularity-dominated strong
fluctuations \cite{KP}.

The moments and autocorrelations of second degree 
(i.e.~intensity correlations) play a particularly
important role in quantum chaos, because they suffice to measure the
spreading of the energy eigenfunctions, and they can be rigorously
controlled. According to \cite{Shnirelman}, the high energy
eigenfunctions of a classically ergodic system should become uniformly
spread over the surface of constant energy, a property known as
quantum ergodicity.  This claim has found rigorous proofs in
\cite{CdV}, \cite{Bouzouina} and \cite{ZelditchZworski} for example,
where the authors consider compact manifolds with ergodic geodesic
flows, quantized ergodic maps and ergodic billiards respectively. The
main tool used in these works is an Egorov estimate, which, in the
case of quantum maps, reads
\begin{equation} \label{Egorov map} \Big\Vert U_M^{\dagger k} Op(f)
  U_M^k - Op \big( f \circ M^k \big) \Big\Vert \leq \textrm{const}
  \cdot \hbar,
\end{equation}
for $M$ a map, $f$ any smooth function on the configuration space, and
$U$ and $Op(f)$ their quantized analogs. A version of \eqref{Egorov
  map} also holds for continuous Hamiltonian systems. However, the
Egorov method does not provide any information on the rate with which
quantum ergodicity is reached. This much harder problem is
investigated in \cite{FP, MWilkinson1, Zelditch1, Zelditch2, EFKAMM,
  BackerSchubert, Schubert}.

In fact, quantum ergodicity is significantly more difficult to tackle
on quantum graphs than on other chaotic systems. The reason is the
non-existence of a deterministic classical map, and hence, of an
Egorov estimate. In \cite{BKS}, quantum ergodicity is proved for
graphs related to quantum maps by using the Egorov property on the
underlying quantum maps. On the other hand, it is shown in \cite{BKW, BKS2, K}
that some graphs, namely star graphs, are not quantum ergodic. Here,
our result for the autocorrelation functions \eqref{our autocorrelation
  functions} with $p = 2$ enables us to expound a criterion for
graphs to become quantum ergodic. A summary of our results in this
special case has already been given in \cite{PRL}. Moreover, our
method also yields the rate of quantum ergodicity in terms of the
classical dynamics, when quantum ergodicity does occur. The result
obtained is a significant refinement of the previous estimates in
\cite{EFKAMM}.

The reader not interested in the derivation of the formulae can
directly jump to Section~\ref{Criteria and Rates of Universality}
where the final formulae are given and exploited. The rest of the text
is structured as follows. In the sections~\ref{Quantum Graphs} and
\ref{Eigenfunction Statistics}, quantum graphs are defined, and the
autocorrelation functions together with other statistical quantities
of interest are introduced. In particular, a first type of trace
formulae is developed in \ref{Green Matrices and Trace Formulae} and
\ref{Long Diagonal Orbits}. An exact field theoretic expression for
the autocorrelation functions is developed in Section~\ref{Generating
  Functions}, and a second type of trace formulae is presented in
\ref{Diagonal Approximation}.  Then, two different contributions to
the exact expression for the autocorrelation functions are extracted
and calculated in sections \ref{Mean Field Theory} and 
\ref{The Gaussian Correction}. 
Section~\ref{Criteria and Rates of Universality} compares these 
two contributions and illustrates them with a few examples.
Section~\ref{Discussions} discusses our results and gives an outlook
on possible implications.

\section{Quantum Graphs} \label{Quantum Graphs}

\subsection{Definitions}

A metric graph $G$ is a set of $V\in \mathbb{N}$ points, called the
vertices, and of $B\in\mathbb{N}$ bonds of positive lengths $L =
(L_{1},\cdots,L_{B})$ linking some pairs of vertices. The topology of
a graph is determined by its connectivity matrix $C$, namely the $V
\times V$ matrix
\begin{equation}
  C_{i,j} = C_{j,i}=\# \{ \textrm{bonds connecting the
    vertices $i$ and $j$} \}.
\end{equation} 
If $C_{i,i} = 0$ and $C_{i,j} \leq 1$ for all $i,j \in \mathbb{N}_V$,
the graph is said to be simple.  The valency $v_i$ of a vertex $i \in
\mathbb{N}_V$ is defined by $v_i = \sum_{j = 1}^V C_{i,j}$. The
valencies are always all supposed positive.  A point on a graph is
specified by a pair $(b,x_{b})$, where $b \in \mathbb{N}_B$ determines
the bond and $x_{b}\in[0,L_{b}]$ determines the position of this point
on $b$.

Each bond of a metric graph can be traversed in two possible
directions, denoted by $d \in \{ +,- \}$. A pair $\beta = (b,d)$ then
denotes a directed bond, and $\hat{\beta} = (b,-d)$ stands for its
reverse partner. The vertex from which a directed bond $\beta$ emerges
is written $o\beta$ and the vertex to which it leads is written
$t\beta$. In particular, $o\beta = t\hat{\beta}$ is always fulfilled.
We suppose the set of directed bonds to be ordered so that, by abuse
of language, any directed bond $\beta$ can also be seen as an element
of $\mathbb{N}_{2B}$.

A quantum graph is a metric graph $G$ that is turned into a quantum
system. In order to do this, the $\mathbb{C}$-linear space
\begin{equation} \label{Hilbert space} \mathcal{H} = \Big\lbrace \Psi
  = \bigoplus_{b=1}^B \psi_{b} \Big\vert \psi_{b}, \psi_{b}',
  \psi_{b}'' \in L^2 \big( [0,L_{b}] \big) \Big\rbrace
\end{equation}
is introduced, and its elements are referred to as wave functions.
This space is endowed with the scalar product defined by
\begin{equation}
  (\Psi,\Phi) \equiv \sum_{b = 1}^B \int_0^{L_b} \psi^\ast_b(x) \phi_b(x) dx
\end{equation}
for any $\Psi,\Phi \in \mathcal{H}$.  The number $\psi_{b}(x_{b})$ is
interpreted as the value of the wave function $\Psi$ at the point
$(b,x_{b})$ of $G$.  One can define an operator $H$ acting on
$\mathcal{H}$ as
\begin{equation} \label{Hamiltonian} H \bigoplus_{b=1}^B \psi_{b} =
  \bigoplus_{b=1}^B - \psi^{''}_{b}.
\end{equation}
This is the expression of the free quantum particle Hamiltonian on
each bond. The restriction of $H$ on the subset $ \mathcal{H}_0
\subset \mathcal{H}$ of wave functions vanishing at the vertices is
symmetric. A wave function $\Psi \in \mathcal{H}_0$ is called a
Dirichlet wave function. A Schr\"odinger operator on a metric 
graph (and thus a quantum graph) 
can be defined as a self-adjoint extension of $H$. 
However, we will follow a slightly 
different definition using the scattering approach \cite{KottosSmil}. We first
give a brief overview of this approach and then discuss its relation
to self-adjoint extensions of $H$.

For any real number $k>0$, the solutions of the equation $H\Psi = k^2
\Psi$ form the subspace
\begin{equation}
  \tilde{\mathcal{A}}(k)  = \left\{ \bigoplus_{b=1}^B \left. \sum_{d = +,-} a_{bd} \tilde{e}_{bd}(k) \right| a_{\beta} \in \mathbb{C}, \  \forall \beta = (b,d) \in \mathbb{N}_{2B}\right\},
\end{equation}
where, for $b \in \mathbb{N}_B$ and $d \in \{ +,- \}$,
\begin{equation}
  \tilde{e}_{bd}(k) = e^{idk\left( x - \frac{L_b}{2}\right)}.
\end{equation}
A wave function in $\tilde{\mathcal{A}}(k)$ is then characterized by
$2B$ waves of wave number $k$, each of which carries a complex
amplitude $a_\beta$ corresponding to its value at the mid-point of the
bond.

Let us introduce $2B$ formal symbols $|e_\beta \rangle$, $\beta \in
\mathbb{N}_{2B}$, and the set $\mathcal{A}$ of their possible linear
combinations over $\mathbb{C}$. The set $\mathcal{A}$ is a
$2B$-dimensional $\mathbb{C}$-linear space called amplitude space, and
it is endowed with the hermitian scalar product defined by
\begin{equation} \label{e beta def} \langle e_{\beta'} | e_{\beta}
  \rangle = \delta_{\beta,\beta'}.
\end{equation}
It can be seen as the direct product $\mathcal{A} = \mathcal{A}_b
\otimes \mathcal{A}_d$ of a $B$-dimensional bond space $\mathcal{A}_b$
and a two-dimensional direction space $\mathcal{A}_d$.  For each
$k>0$, there is a natural one-to-one mapping
\begin{equation} \label{wave amplitude mapping} \Psi =
  \bigoplus_{b=1}^B\sum_{d = +,-} a_{bd} \tilde{e}_{bd}(k) \mapsto
  |\vect{a} \rangle = \sum_{\beta = 1}^{2B} a_\beta |e_\beta \rangle
\end{equation}
between $\tilde{\mathcal{A}}(k)$ and $\mathcal{A}$. If $\Psi_1 \mapsto
| \vect{a_1} \rangle$ and $\Psi_2 \mapsto | \vect{a_2}\rangle$ by this
mapping, the scalar products in the spaces $\tilde{\mathcal{A}}(k)$
and $\mathcal{A}$ translate
\begin{equation} \label{scalar products relation} (\Psi_1,\Psi_2) =
  \left\langle \vect{a_1} \left| L + \frac{\sin (kL)}{k} \sigma_1^d
    \right| \vect{a_2} \right\rangle,
\end{equation}
where $ \sigma_1^d$ stands for the first Pauli matrix acting on
$\mathcal{A}_d$, and $L$ denotes the $2B \times 2B$ diagonal matrix
\begin{equation} \label{L matrix def} L_{\beta'\beta} =
  \delta_{\beta,\beta'} L_{\beta}.
\end{equation}
Here and henceforth, the length of a directed bond is the length of
the bond on which it is supported. In particular, $L_\beta =
L_{\hat{\beta}}$ is always fulfilled.  The identity \eqref{scalar
  products relation} shows that the mapping \eqref{wave amplitude
  mapping} does not preserve length and orthogonality in general.

In the scattering approach to quantum graphs the values at each vertex $i\in
\mathbb{N}_V$ of the $v_i$ waves emerging from this vertex and of the
$v_i$ waves incoming to this vertex are related through some fixed
matrix $\sigma^i$. If $|\vect{a}^i_{\textrm{\scriptsize{out}}}
\rangle$ and $|\vect{a}^i_{\textrm{\scriptsize{in}}} \rangle$ denote
the $v_i$-dimensional vectors containing the values at vertex $i$ of
the emerging waves and of the incoming waves respectively, this
relation reads
\begin{equation} \label{scattering at vertex i}
  |\vect{a}^i_{\textrm{\scriptsize{out}}} \rangle = \sigma^i
  |\vect{a}^i_{\textrm{\scriptsize{in}}} \rangle.
\end{equation}
A wave function $\Psi \in \tilde{\mathcal{A}}(k)$ conserves the
probability current if and only if the $V$ matrices $\sigma^i$ are all
unitary.  The components of the $V$ outgoing and incoming vectors
$|\vect{a}^i_{\textrm{\scriptsize{out}}} \rangle$ and
$|\vect{a}^i_{\textrm{\scriptsize{in}}} \rangle$ can then be grouped
together to form the $2B$-dimensional vectors $
|\vect{a}_{\textrm{\scriptsize{out}}}\rangle$ and $
|\vect{a}_{\textrm{\scriptsize{in}}}\rangle$ respectively. These
vectors are related to $| \vect{a} \rangle$ in \eqref{wave amplitude
  mapping} through
\begin{equation} \label{a_out a_in and T}
  |\vect{a}_{\textrm{\scriptsize{out}}} \rangle = T^\dagger(k) |
  \vect{a} \rangle \quad \textrm{and} \quad
  |\vect{a}_{\textrm{\scriptsize{in}}} \rangle = T(k) | \vect{a}
  \rangle
\end{equation}
where $T(k)$ is the $2B \times 2B$ diagonal matrix $T(k) =
e^{ik\frac{L}{2}}$. This matrix contains the phases gained by the $2B$
waves of wave number $k$ when they travel along half the bonds on which
they are supported. It is referred to as the propagation matrix.
Moreover, the $V$ identities \eqref{scattering at vertex i} become
\begin{equation} \label{a_out a_in and S}
  |\vect{a}_{\textrm{\scriptsize{out}}} \rangle = S
  |\vect{a}_{\textrm{\scriptsize{in}}} \rangle,
\end{equation}
where $S$ is the $2B \times 2B$ unitary matrix, called scattering
matrix, defined by
\begin{equation} \label{amplitude scattering} S_{\beta'\beta} =
  \left\lbrace \begin{array}{ll} \sigma^i_{\beta'\beta} & \textrm{if }
      o\beta' = t\beta = i \\ 0 & \textrm{otherwise} \end{array}
  \right.
\end{equation}
Putting \eqref{a_out a_in and T} and \eqref{a_out a_in and S} together
yields
\begin{equation} \label{amplitude left invariant by U} U(k) |\vect{a}
  \rangle = |\vect{a} \rangle, \quad \textrm{with} \quad U(k) = T(k) S
  T(k).
\end{equation}
The $2B \times 2B$ matrix $U(k)$ is called the quantum map or evolution map of the graph. It is
unitary since both $T(k)$ and $S$ are unitary.

Equation \eqref{amplitude left invariant by U} shows that imposing the
conservation of probability current through fixed unitary matrices
$\sigma^i$ restricts the possible amplitudes $|\vect{a} \rangle$ and
the possible wave numbers $k>0$. Indeed, the secular equation
\begin{equation} \label{secular equation} \det \big( 1 - U(k) \big) =
  0
\end{equation}
must be satisfied for \eqref{amplitude left invariant by U} to admit
non-trivial solutions. This equation is satisfied for a sequence
\begin{equation}
  0 \leq k_1 < k_2 < \ldots < k_\nu < k_{\nu + 1} < \ldots \to \infty
\end{equation}
called the spectrum of the quantum graph, and the square of these
wave numbers are the quantized energies. If the bond lengths $L_1,
\ldots, L_B$ are independent over $\mathbb{Q}$, there is typically a
normalized vector $| \vect{a}^\nu \rangle$ in $ \mathcal{A}$ for any
$\nu \in \mathbb{N}$ that satisfies $U(k_\nu) | \vect{a}^\nu \rangle =
| \vect{a}^\nu \rangle$ and so that any other vector satisfying this
equation is of the form $z | \vect{a}^\nu \rangle$ for some $z\in
\mathbb{C}$. The vector $| \vect{a}^\nu \rangle$ then provides the
amplitudes of the eigenfunction $\Psi^\nu$ satisfying $H \Psi^\nu =
k_\nu^2 \Psi^\nu$ by the mapping \eqref{wave amplitude mapping}.
Incommensurability of the bond lengths and this non-degeneracy
property will be assumed henceforth.

It is well-known \cite{GS} that the mean number of allowed wave numbers
in $[0,K]$ is $N(K) \equiv K \bar{d}$, where the mean level density
$\bar{d}$ reads
\begin{equation}
  \bar{d} \equiv \frac{\tr L}{2 \pi}.
\end{equation}

For any $k>0$, the unitarity of $U(k)$ ensures the existence of an
orthonormal basis $\{ |n,k\rangle \}_{n \in \mathbb{N}_{2B}}$ of
$\mathbb{C}^{2B}$ and of $2B$ real numbers $\{ \phi_n (k) \}_{n \in
  \mathbb{N}_{2B} }$ such that
\begin{equation} \label{evol map eigen equ} U(k) |n,k\rangle =
  e^{i\phi_n(k)} |n,k\rangle.
\end{equation}
These sets can be ordered by imposing the inequalities
\begin{equation}
  -2\pi < \phi_{2B}(0) \leq \phi_{2B - 1}(0) \leq \ldots \leq \phi_2(0) \leq \phi_1(0) \leq 0
\end{equation}
and by requiring the $2B$ eigencurves $k \mapsto \phi_n(k)$ to be
$C^\infty$. This smoothness condition can indeed be realized since the
map $U(k)$ depends on $k$ in an analytic way. Taking a derivative with
respect to $k$ on both sides of \eqref{evol map eigen equ} leads to
\begin{equation} \label{der eigenphases} \phi_n'(k) = \langle n,k | L
  | n,k \rangle \in [L_{\textrm{\scriptsize{min}}},
  L_{\textrm{\scriptsize{max}}}],
\end{equation}
where $L_{\textrm{\scriptsize{min}}}$ and
$L_{\textrm{\scriptsize{max}}}$ denote the minimal and maximal bond
lengths on the graph.

A quantum graph is time-reversal invariant if its quantum map
satisfies $\tr\ \left({U(k)^{\mathcal{T}}}\right)^n = \tr\ U(k)^n$ 
for all $k \geq 0$
and integers $n$. Here and henceforth,
the generalized transposition $A^{\mathcal{T}}$
of a linear transformation $A$ is defined by 
\begin{equation}
  A^{\mathcal{T}} = \sigma_1^d A^T \sigma_1^d,
\end{equation}
$A^T$ being the transpose of $A$. It satisfies ${A^\mathcal{T}}^\mathcal{T}=A$.
Since $T(k)^{\mathcal{T}} = T(k)$, a graph 
is time-reversal invariant if and only if
its scattering matrix satisfies $\tr\ \left({S^{\mathcal{T}}}\right)^n =\tr\ S^n$
for all integers $n$. Obviously,
${S^{\mathcal{T}}} = S$ implies time-reversal invariance. Note, however
that  replacing 
\begin{equation}
  S \mapsto S'= e^{-i \theta} S e^{i \theta}  
  \qquad |\vect{a} \rangle
  \mapsto |\vect{a}' \rangle=  e^{-i \theta}|\vect{a} \rangle
  \label{gauge}
\end{equation}
where $\theta=\mathrm{diag}(\theta_1,\dots,\theta_{2B})$ is 
a diagonal real matrix 
is equivalent to
choosing a different reference phase for the amplitudes.
We will call such a transformation a (passive) gauge transformation --
it neither affects the spectrum 
nor the condition described above for time-reversal invariance.
The latter can now be reformulated:
a quantum graph is time-reversal
invariant if and only if there is a (possibly trivial)
gauge transformation 
$S\mapsto S'=e^{-i \theta} S e^{i \theta}$ such that ${{S'}^{\mathcal{T}}} = S'$.
For time-reversal invariant graphs we will henceforth assume
that the reference phases have been chosen such that $S^{\mathcal{T}}=S$
holds. There remains a residual gauge freedom to which we will
return later when we discuss the wave function statistics
in quantum graphs.

The set of all
time-reversal invariant graphs form the orthogonal symmetry class, and
the set of all quantum graphs violating this property form the unitary
symmetry class. We will frequently use the parameter $\kappa$ which takes
the values
\begin{equation}
  \label{kappa}
  \kappa
  =
  \begin{cases}
    1 & \text{in the unitary class, and}
    \\
    2 & \text{in the orthogonal class.}
  \end{cases}
\end{equation}
Note that the parameter $\kappa$ that we use here is linked to the 
parameter $\beta$ used in random-matrix theory to distinguish symmetry 
classes by $\kappa=2/\beta$.

We have already mentioned that the scattering approach 
described above is not the only way to define
quantum graphs. The other frequently used definition is based on
self-adjoint extensions of $H$ in \eqref{Hamiltonian} defined on the
Dirichlet domain $\mathcal{H}_0$ (see \cite{JB} and references therein).
A complete description of all possible  self-adjoint extensions
was given in \cite{KosSchra}.
In general each self-adjoint extension 
is equivalent to energy-dependent
matrices $\sigma^{\textrm{\scriptsize{KS}},i}(k)$ relating the
outgoing amplitudes to the incoming amplitudes of $\Psi \in
\tilde{\mathcal{A}}(k)$ at each vertex $i$ instead of \eqref{scattering
  at vertex i}. These matrices can then be grouped together to form a
global unitary scattering matrix $S^{\textrm{\scriptsize{KS}}}(k)$ as
in \eqref{amplitude scattering}, and a global quantum map
$U^{\textrm{\scriptsize{KS}}}(k) =
T(k)S^{\textrm{\scriptsize{KS}}}(k)T(k)$ satisfying the secular
equation \eqref{secular equation}. The two definitions of quantum graphs have 
a certain overlap as there is a subset of self-adjoint extensions which leads
to energy-independent scattering matrices.
It is shown in \cite{Berk}
and \cite{JB} that any scattering matrix
$S^{\textrm{\scriptsize{KS}}}$ defining a self-adjoint operator $H$
admits a limit $S^{\textrm{\scriptsize{KS}}}_\infty$ as $k$ tends to
infinity, and moreover, it is argued in \cite{Berk} that a scattering
matrix $S^{\textrm{\scriptsize{KS}}}$ and its limit
$S^{\textrm{\scriptsize{KS}}}_\infty$ share the same spectral
statistics. The coincidence of these statistics comes from the fact
that they are properties at asymptotically large wave number $k$.
Hence, one can deduce that the eigenfunction statistics of
$S^{\textrm{\scriptsize{KS}}}$ and
$S^{\textrm{\scriptsize{KS}}}_\infty$ also coincide.
As a consequence the eigenfunction statistics of
quantum graphs defined following the self-adjoint extension approach
can be recovered from the eigenfunction statistics of quantum graphs
defined through the scattering approach by substituting
$S^{\textrm{\scriptsize{KS}}}_\infty$ for
$S^{\textrm{\scriptsize{KS}}}$.

Henceforth, the scattering matrix $S$ always refers to the matrix in
\eqref{amplitude scattering} obtained from the scattering approach. It
can be any $2B\times 2B$ unitary matrix such that $S_{\beta'\beta}$
vanishes if $t\beta \neq o\beta'$. A possible choice is the so-called
Neumann scattering matrix, which is defined at each vertex $i \in
\mathbb{N}_V$ by
\begin{equation} \label{Neumann scattering matrix}
  \sigma^i_{\beta'\beta} = \frac{2}{v_i} - \delta_{\beta,\beta'}, \quad \forall t\beta = o\beta' = i.
\end{equation}
Quantum graphs with this choice of scattering matrix at each vertex
will be called
Neumann quantum graphs.

In general, a quantum graph is then specified by a pair $(G,S)$ where $G$ is a
metric graph and $S$ is a scattering matrix on $G$. The class of
possible scattering matrices $S$ on $G$ contains all the asymptotic
matrices $S^{\textrm{\scriptsize{KS}}}_\infty$ obtained from the
self-adjoint extension approach. There are however some scattering
matrices that are acceptable from the scattering point of view but not
from the second approach. An example is given by the Direct Fourier
Transform (DFT) graphs \cite{GS}, for which the scattering processes
at vertex $i \in \mathbb{N}_V$ are described by the $v_i \times v_i$
unitary matrix
\begin{equation} \label{DFT scattering matrix} \sigma^i_{\beta'\beta}
  =  \frac{1}{\sqrt{v_i}}
  e^{2\pi i \frac{n^i(\beta)\
      n^i(\beta')}{v_i}}, \quad \forall t\beta = o\beta' = i,
\end{equation}
where $n^i$ is a surjective assignment of an integer in
$\mathbb{N}_{v_i}$ to each directed bond around $i$ such that
$n^i(\hat{\beta}) = n^i(\beta)$. With these matching conditions, the
wave functions $\{ \Psi^\nu \}_{\nu \in \mathbb{N}}$ obtained from the
amplitudes $\{ |\vect{a}^\nu \rangle \}_{\nu \in \mathbb{N}}$ and the
spectrum $\{ k_\nu \}_{\nu \in \mathbb{N}}$ by \eqref{wave amplitude
  mapping} are in general not orthogonal to each other 
in $\mathcal{H}$, which
shows that $H$ acting on the wave functions satisfying \eqref{DFT
  scattering matrix} is not self-adjoint. By contrast, the Neumann
scattering matrices \eqref{Neumann scattering matrix} do lead to a
self-adjoint Laplace operator.

Both examples of scattering matrices were defined in terms of
symmetric unitary matrices at each vertex $\sigma^i={\sigma^i}^\mathcal{T}$.
As a consequence $S^{\mathcal{T}}=S$ and the
a quantum graph obeys time-reversal symmetry. One may break time-reversal
symmetry by adding a magnetic field to the graph. In the scattering
approach adding a magnetic field which is constant on every bond
is straightforward. Let $A$ be the diagonal matrix that contains the
magnetic field strengths. It obeys $A_\beta=-A_{\hat{\beta}}$. 
The corresponding quantum map is
\begin{equation}
  U(k)= e^{i(k+A)L/2} S e^{i(k+A)L/2} \equiv T(k) S_A T(k)
\end{equation} 
and the magnetic field effectively just changes the
scattering matrix $S \mapsto S_A= e^{iAL/2} S e^{iAL/2}$.
If $S=S^\mathcal{T}$ and the graph is multiply connected
(that is, it contains cycles) then the magnetic field generally 
breaks the time reversal invariance.

Henceforth, the metric  graphs $G$ considered are assumed simple. The
reason for this assumption is to simplify some notations and
calculations. However, if a graph contains a directed bond $\beta$
such that $o\beta = t\beta$, a Neumann vertex can be added on the bond
$b$ supporting $\beta$ to destroy the loop $b$ without modifying the
quantum dynamics. Similarly, if the graph has two directed bonds
$\beta, \beta' $ such that $o\beta = o\beta'$ and $t\beta = t\beta'$,
a Neumann vertex can be added on the bond $b$ supporting $\beta$ to
destroy this parallel connection without modifying the dynamics.
Hence, any graph can be made simple by adding sufficiently many
Neumann vertices, and this process does not change the quantum
dynamics. One can thus assume the graph simple without loss of
generality.

\subsection{Classical Dynamics}

With any quantum graph, one can associate a bistochastic classical map
$M$ defined by
\begin{equation}
  \label{classical map}
  M_{\beta \beta'} \equiv |U_{\beta \beta'}(k)|^2 =  |S_{\beta \beta'}|^2,
\end{equation}
where $U(k)$ is the quantum map and $S$ is the scattering matrix.
The matrix $M$ describes a Markov process on the graph, which is the
classical counterpart of the quantum dynamics defined by $S$. The
uniform vector
\begin{equation} \label{vector 1 def} |1\rangle \equiv
  \frac{1}{\sqrt{2B}} \sum_{\beta = 1}^{2B} |e_\beta \rangle
\end{equation} 
is an eigenvector of $M$ of eigenvalue 1, and its hermitian conjugate
$\langle 1 |$ is a left eigenvectors of $M$ of eigenvalue 1. Besides,
the Perron-Frobenius theorem \cite{Matrices} ensures that the spectrum
of $M$ lies on or within the complex unit disc.

A graph is said to be ergodic if and only if, for any $\beta,\beta'
\in \mathbb{N}_{2B}$, there is a discrete time $n\in \mathbb{N}$ for
which the transition probability $\langle e_{\beta'} | M^n | e_{\beta}
\rangle$ is positive. This condition is equivalent to the
non-degeneracy of the eigenvalue 1 of $M$. Any non-ergodic graph
$(G,S)$ is the union of several ergodic components, that is $(G,S) =
\bigcup_{i = 1}^k (G_i,S_i)$ for some integer $k>1$.  The eigenvalue 1
has degeneracy $k$, and the $k$ vectors that are uniform on one
component $(G_i,S_i)$ and zero on the others form a basis of this
eigenspace.

Let us write $M_\epsilon = e^{-2\epsilon} M$ for an ergodic classical
map $M$ and for some $\epsilon >0$. The sum of all classical paths
from $\beta \in \mathbb{N}_{2B}$ to $\beta' \in \mathbb{N}_{2B}$
followed with $M_\epsilon$ can be written
\begin{equation} \label{sum over classical paths}
  \left(\frac{M_\epsilon}{1 - M_\epsilon}\right)_{\beta'\beta} =
  \left( M_\epsilon + M_{\epsilon}^2 + M_{\epsilon}^3 + \ldots
  \right)_{\beta'\beta}.
\end{equation}
It becomes singular as $\epsilon$ approaches zero due to the
eigenvalue 1 of $M$.  Let $M = D_M + N_M$ be the Jordan decomposition
of $M$ into a diagonalizable part $D_M$ and a nilpotent part $N_M$
commuting with each other. Let $ \{ \lambda_j \}_{j \in
  \mathbb{N}_{2B}}$ be the $2B$ eigenvalues of $D_M$, and let $\{ | j
\rangle \}_{j \in \mathbb{N}_{2B}}$ be corresponding normalized
eigenvectors in $\mathcal{A}$ with $| 1 \rangle$ as in \eqref{vector 1
  def}. Then, it is straight forward to check that $\langle 1|j\rangle
= \delta_{1,j}$. This fact enables one to extract the singular part of
\eqref{sum over classical paths} and write
\begin{equation} \label{classical paths decomposition}
  \frac{M_{\epsilon}}{1-M_{\epsilon}} =
  \frac{e^{-2\epsilon}}{1-e^{-2\epsilon}} |1\rangle \langle 1| +
  R_\epsilon,
\end{equation}
where the remainder $R_\epsilon$ is such that $R \equiv
\lim_{\epsilon\to 0} R_\epsilon$ exists and satisfies $\langle 1| R =
0$ and $R|1\rangle = 0$. The first and second terms in the right-hand
side of \eqref{classical paths decomposition} will respectively be
referred to as uniform and massive components. If $m_i = 1 -
\lambda_i$ for $i = 2,\ldots, 2B$ denote the $2B-1$ non-zero
eigenvalues of $1-M$, the massive component satisfies
\begin{equation} \label{masses def} \tr R = \sum_{i=2}^{2B}
  \frac{1-m_i}{m_i}.
\end{equation} 
The eigenvalues $\{m_i \}_{i \in \mathbb{N}_{2B}}$ of $1-M$ are called
masses. They all lie in the closed disc of radius 1 and centered at 1
in the complex plane, and the zero mass $m_1 = 0$ is non-degenerate.

\section{Eigenfunction Statistics} \label{Eigenfunction Statistics}

\subsection{Wave function correlation functions} \label{Wave function correlation functions}

Let $(G,S)$ be a quantum graph, $\{ k_\nu \}$ be its spectrum, and $\{ \vect{a}^\nu \} \subset \mathbb{C}^{2B}$ be a set of normalized amplitude vectors defining the eigenfunctions $\{ \Psi^\nu \}$ as in \eqref{wave amplitude mapping}.  Let us
consider $2B$ complex random variables $a_\beta$ and investigate the
existence of a joint probability density function $\varphi(\vect{a}) =
\varphi(a_1, \ldots, a_{2B})$ satisfying
\begin{eqnarray} \label{statistics} \left\langle \prod_{k=0}^{p-1}
    a^{\ast}_{\beta_k} \prod_{l=0}^{q-1} a_{\beta'_l} \right\rangle
  & \equiv & \lim_{K\to\infty}\frac{1}{N(K)} \sum_{k_\nu \leq K} \frac{\tr L}{2B \langle L  \rangle_\nu} \prod_{k=0}^{p-1} a^{\nu\ast}_{\beta_k} \prod_{l=0}^{q-1} a^\nu_{\beta'_l} \\
  \label{phi def}
  & = & \int_{\mathbb{C}^{2B}} \prod_{k=0}^{p-1} a^\ast_{\beta_k} \prod_{l=0}^{q-1} a_{\beta'_l} \ \varphi(\vect{a}) \ d\vect{a}^\ast d\vect{a},
\end{eqnarray}
for any choice of $\beta_0, \ldots, \beta_{p-1}, \beta'_0, \ldots,
\beta'_{q-1} \in \mathbb{N}_{2B}$ with $p,q \in \mathbb{N}_0$.  
Here,
the measure $d\vect{a}^\ast d\vect{a}$ denotes the product of the $2B$
flat Lebesgue measures $da_\beta^\ast da_\beta$ in the complex plane,
and the notation $\langle O \rangle_\nu$ for a $2B \times 2B$ matrix
$O$ stands for $\langle O \rangle_\nu= \langle  \vect{a}^\nu | O |  \vect{a}^\nu \rangle$.

The first line in \eqref{statistics} defines the wave function
correlation functions.
The peculiar factor $\frac{\tr L}{2B \langle L \rangle_\nu}$ in
this definition is introduced
for further calculational convenience. For large graphs
with extended wave functions 
this factor is expected to be close to unity. Indeed, 
it has generally a tiny effect on the wave 
function statistics.
It will be seen later that with the inclusion of this factor 
\eqref{statistics} does not depend on the particular values of the
incommensurate bond lengths. 
Moreover, performing an average over the spectrum of the
quantum graph in presence of this factor, such as in
\eqref{statistics}, amounts to averaging the same quantity over all the
eigenfunctions $|n,k\rangle$ of $U(k)$ and then integrating over all $k
\in (0,\infty)$. Indeed, it is proven in \cite{BerkWinn} that
graphs with incommensurate bond lengths obey
\begin{equation} \label{BerkWinn Theorem} \lim_{K\to\infty}
  \frac{1}{N(K)} \sum_{k_\nu \leq K} \frac{ \tr L}{2B \langle
    L\rangle_\nu} \langle O \rangle_\nu^q = \lim_{K\to\infty}
  \frac{1}{K} \int_0^K \frac{1}{2B} \sum_{n=1}^{2B} \langle n,k | O
  |n,k \rangle^q dk.
\end{equation}
for any $2B \times 2B$ matrix $O$ and any non-negative integer $q$.

The identity \eqref{BerkWinn Theorem} shows that the joint probability
density function $\varphi(\vect{a})$ in \eqref{phi def} is normalized.
Indeed, choosing $q = 0$ in this formula leads to
\begin{equation}
  \langle 1 \rangle \equiv \lim_{K\to\infty}\frac{1}{N(K)} \sum_{k_\nu \leq K} \frac{\tr L}{2B \langle L  \rangle_\nu} = \lim_{K\to\infty} \frac{1}{K} \int_{0}^K \frac{1}{2B} \sum_{n=1}^{2B} 1 = 1.
\end{equation}
Moreover, it also provides an exact expression for the 
covariance of $\varphi(\vect{a})$.
Indeed, the equality \eqref{BerkWinn Theorem} with $q = 1$ and $O = |
e_{\beta'} \rangle \langle e_\beta | $ yields
\begin{eqnarray} \label{exact covariance}
  \langle a_{\beta'}^\ast a_{\beta} \rangle & \equiv & \lim_{K\to\infty}\frac{1}{N(K)} \sum_{k_\nu \leq K} \frac{\tr L}{2B \langle L  \rangle_\nu} \langle \nu | e_{\beta'} \rangle \langle e_\beta | \nu \rangle \nonumber\\
  & = & \lim_{K\to\infty} \frac{1}{K} \int_0^K \frac{1}{2B}
  \sum_{n=1}^{2B} \langle n,k | e_{\beta'} \rangle \langle e_\beta |
  n,k \rangle dk \ = \ \frac{\delta_{\beta,\beta'}}{2B},
\end{eqnarray}
by orthonormality of the families $\{ | n,k \rangle \}_{n \in
  \mathbb{N}_{2B}}$. This derivation of the covariance relies
on the incommensurability of bond lengths. We will show later 
in Subsection~\ref{Green Matrices
  and Trace Formulae} that
the restriction to incommensurable bond lengths can be lifted.

Further properties of the joint probability
density can be derived considering its invariance under gauge transformation of
the form described in \eqref{gauge}. This discussion has to
treat systems with and without time-reversal invariance separately and
we will start with the unitary class (broken time-reversal invariance).
In this class we are free to choose a gauge and
one expects that
all correlation functions which are not gauge invariant will vanish.
This implies that the non-trivial 
correlation functions  \eqref{statistics} 
have the same number of complex conjugated amplitudes as non-conjugated
amplitudes (that is $p=q$). It is thus sufficient to
consider the autocorrelation functions
\begin{equation} 
  \label{C alpha def} C_{[\vect{\alpha}]} \equiv
  \left\langle |a_{\alpha_0}|^2 \ldots
    |a_{\alpha_{q-1}}|^2\right\rangle .
\end{equation}
where $[\vect{\alpha}] \equiv [\alpha_0, \ldots, \alpha_{q-1}] $ is a
vector containing $q\in \mathbb{N}_0$ directed bonds $\alpha_j \in
\mathbb{N}_{2B}$. The integer $q$ is called degree of
$C_{[\vect{\alpha}]}$.\\
Of particular interest to us are the
moments 
\begin{equation}
  \label{moments}
  M_{\alpha,q}\equiv C_{[q \times \alpha]}=
  \left\langle |a_{\alpha}|^{2q} \right\rangle
\end{equation}
and the first non-trivial autocorrelation  functions  
\begin{equation}
  \label{intensitycorrelation}
  C_{\alpha \alpha'}\equiv C_{[\alpha \alpha']}=\left\langle |a_\alpha|^{2} |a_{\alpha'}|^{2} \right\rangle
\end{equation}
which form the symmetric intensity correlation matrix $C_{\alpha \alpha'}=C_{\alpha' \alpha}$.
 
For time-reversal invariant systems the property $U(k)^{\mathcal{T}}=U(k)$ 
of the quantum map 
implies that one may  always choose the phase of its eigenvectors 
$|n,k\rangle=\sum_{\beta} a_{n,\beta}(k)| e_\beta \rangle$
such that $a_{n,\beta}(k)=a_{n,\hat\beta}(k)^\ast$. An equivalent statement
is that the wave function on the graph can be chosen real. This has strong 
implications on wave function statistics -- for instance the
autocorrelation functions  $C_{[\vect{\alpha}]}$ defined in
\eqref{C alpha def}
are invariant under 
replacing any directed bond in $\vect{\alpha}=(\alpha_0\dots,\alpha_{q-1})$
by its reverse partner  $\alpha_i\mapsto \hat{\alpha}_i$. 
The joint probability density function then reduces to a product
\begin{equation}
  \varphi(\vect{a})
  = \delta^B(\vect{a}_{+}-\vect{a}_{-}^\ast) \ 
  \varphi_{\mathrm{red}}(\vect{a}_{+})
\end{equation}
where $\vect{a}_+$ ($\vect{a}_-$) is the $B$-dimensional vector 
containing the 
amplitudes for directed bond $\alpha=(b,d)$ with positive (negative) 
direction index $d$. 
For a quantum graph in the 
orthogonal class it is thus sufficient to consider only the correlation 
functions in \eqref{statistics} for which all directed bond have a positive 
direction index. As in the unitary case we also expect for the orthogonal
case that correlation functions that depend on a local gauge vanish exactly.
Note, that in the orthogonal case not all gauge transformations \eqref{gauge}
are allowed. In order to preserve the properties
$S^{\mathcal{T}}=S$ and  $a_{n,\beta}(k)=a_{n,\hat\beta}(k)^\ast$
only gauge transformations with $\theta_\beta = - \theta_{\hat{\beta}}$
are allowed. Again the only non-trivial correlation functions are
the autocorrelation functions \eqref{C alpha def}.

\begin{figure}
  \begin{center}
    \includegraphics[width=0.49\textwidth]{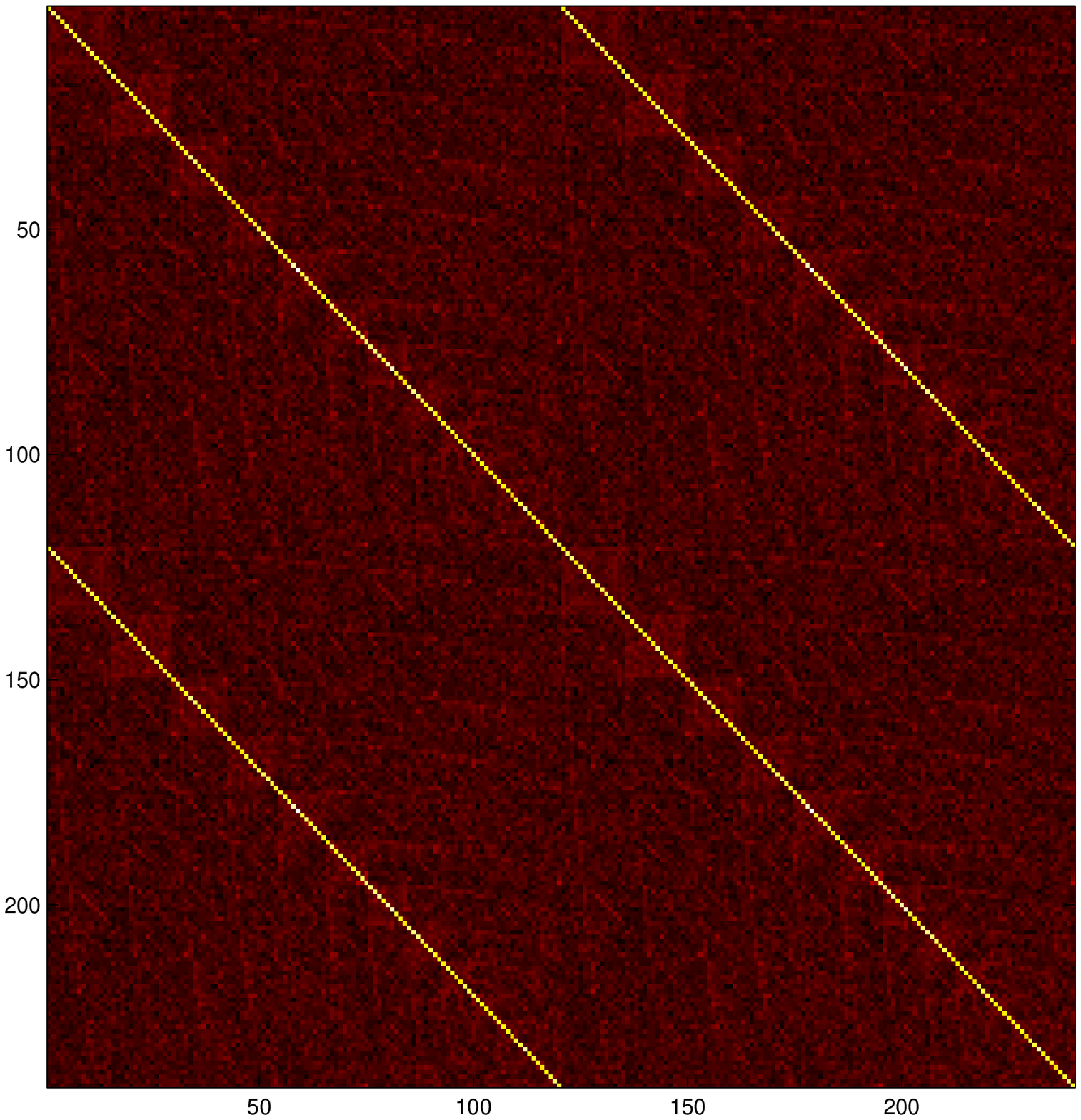}
    \includegraphics[width=0.49\textwidth]{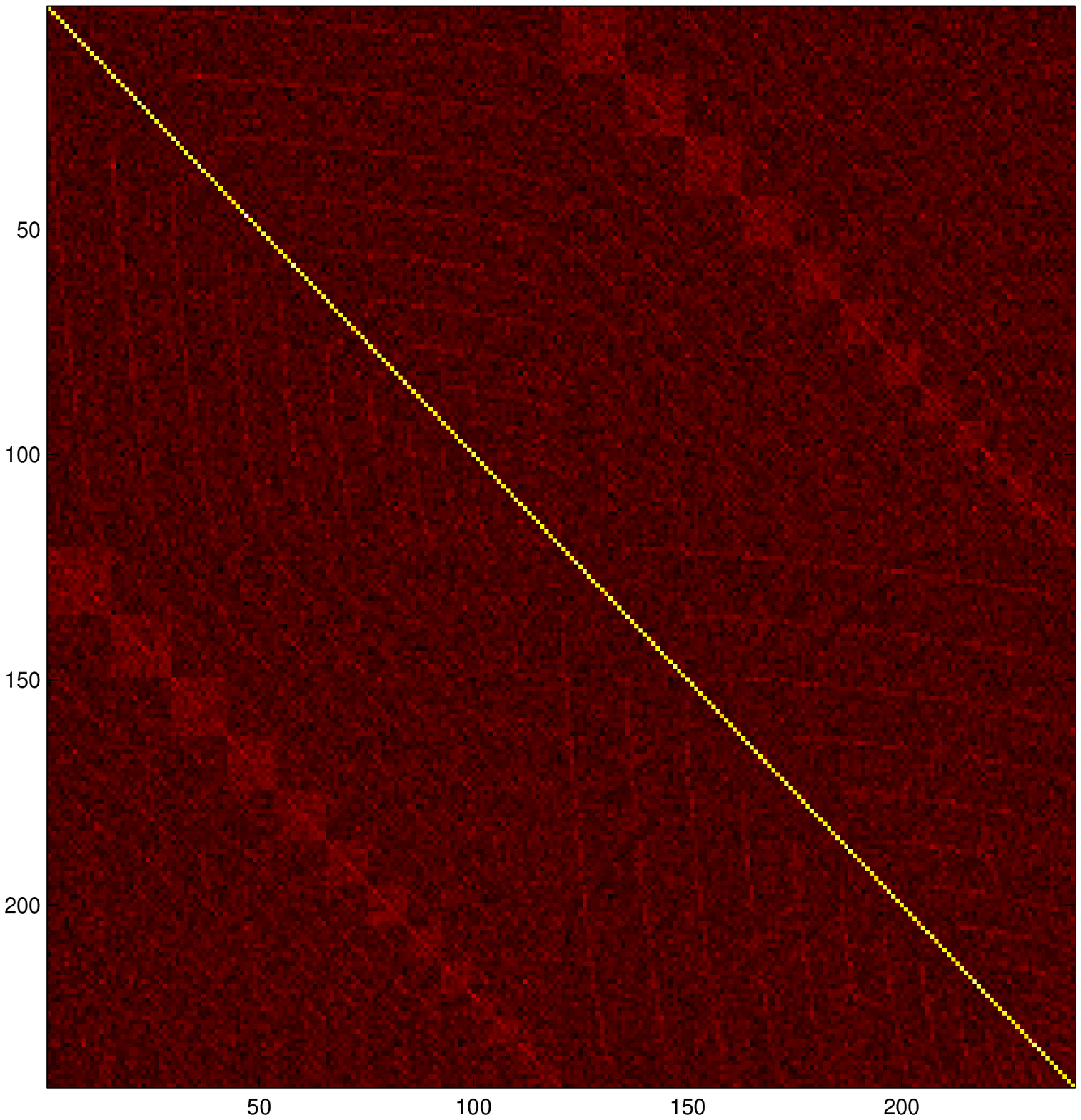}\\
    \includegraphics[width=0.49\textwidth]{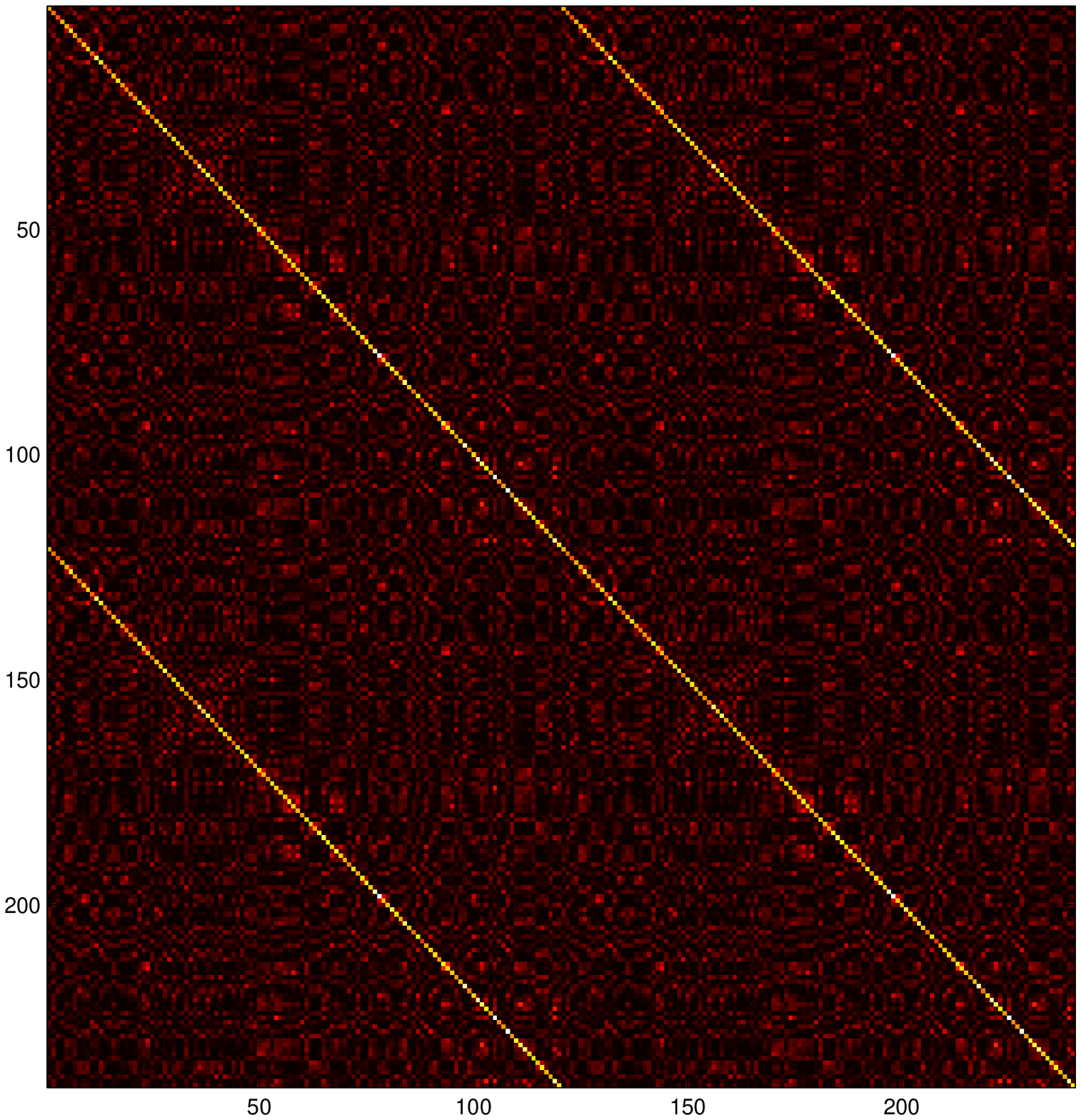}
    \includegraphics[width=0.49\textwidth]{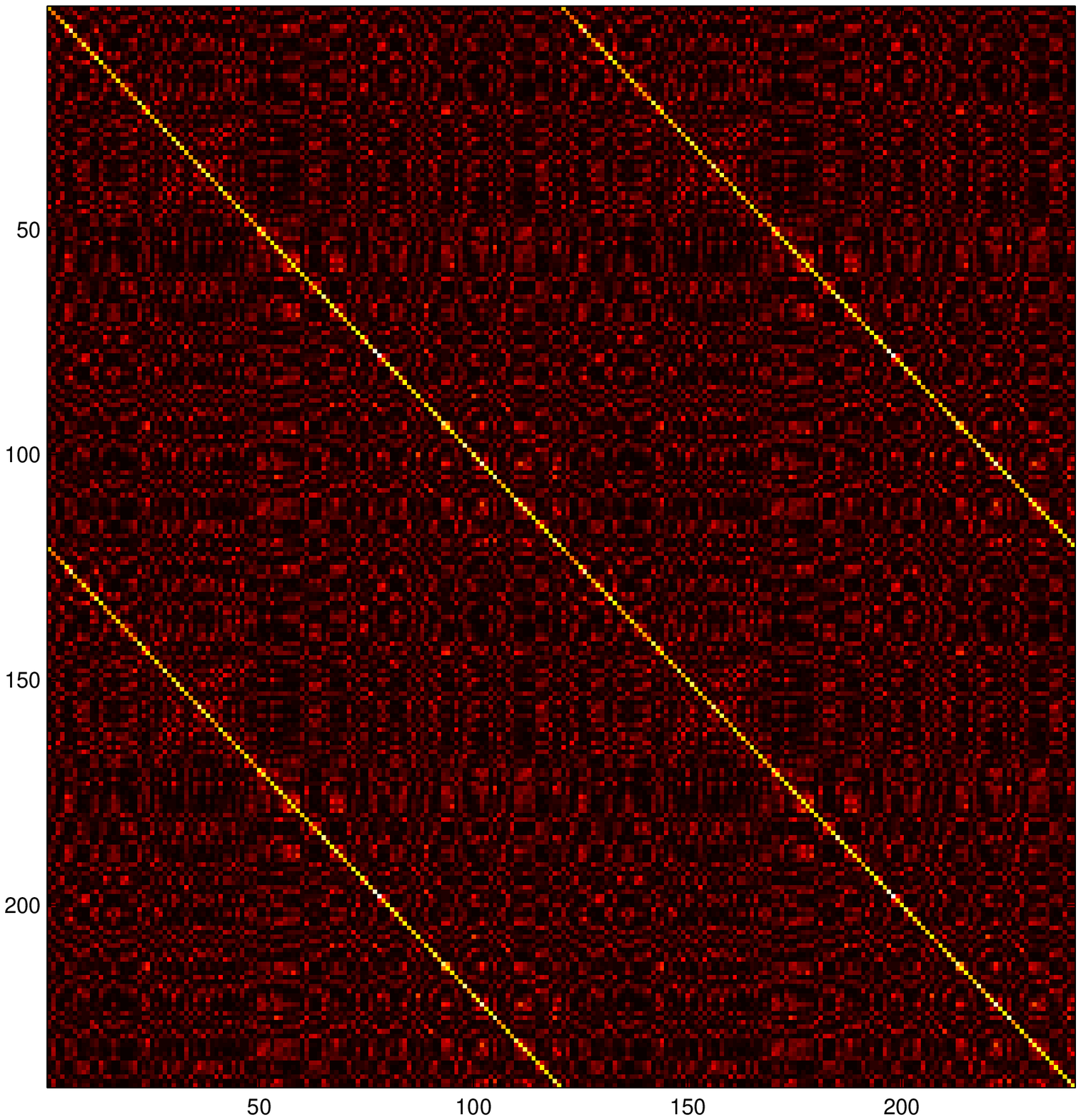}
    \caption{Numerically evaluated intensity correlation matrices 
      $C_{\alpha \alpha'}$  defined in \eqref{intensitycorrelation} for a
      complete graph with $V=16$ vertices and $B=120$ bonds. In the upper two 
      panels the DFT scattering matrices have been used at each
      vertex. The lower two panels are for Neumann scattering matrices.
      For the right two panels time-reversal symmetry has been broken by adding
      a magnetic field. The directed bonds $\alpha=(b,d)$ have been ordered
      as $\left( (1,-),(2,-),\dots,(B,-),(1,+),\dots,(B,+)\right)$. 
      For the graphs in the orthogonal class on the left side there are
      four identical blocks as  $C_{\alpha \alpha'}=
      C_{\hat{\alpha},\alpha'}=C_{\alpha \hat{\alpha}'}=C_{\hat{\alpha} 
        \hat{\alpha}'}$. 
      In the unitary case, note that the correlation matrix
      on the off-diagonal $\alpha=\hat{\alpha}'$ remains strongly
      peaked  for Neumann scattering matrices. However the four blocks are
      no longer the identical (this is not obvious from the picture).
      For DFT scattering matrices the strong off-diagonal peak 
      almost 
      disappears in the presence of a magnetic field. 
    }
    \label{intensity correlation figure}
  \end{center}
\end{figure}

\subsection{Circular and Gaussian Random Waves Models} \label{Gaussian Random Waves
  Models}

For a large well-connected quantum graph in the unitary symmetry class the
quantum map $U(k)$ does generally not  have any symmetries. Moreover,
in a complex network (e.g. a randomly  chosen connected graph) the
neighborhood of any bond looks  statistically the same. By analogy with
the circular ensembles of  random matrix theory one is inclined 
to guess that the joint probability  density function
$\varphi(\vect{a})$  for  
the eigenvectors of the $2B \times 2B$ matrix $U(k)$
defined in \eqref{phi def} is invariant
under transformations $\vect{a}\mapsto u \vect{a}$ for unitary
matrices $u$. This implies that the vectors $\vect{a}$
are uniformly distributed over the unit sphere in $\mathbb{C}^{2B}$.
We will call the guess
\begin{equation}
  \label{circular random waves unitary}
  \varphi_{\mathrm{CU}}(\vect{a}) \equiv
  \frac{(2B-1)!}{\pi^{2B}}  \delta\left(1-\Vert \vect{a} \Vert^2 \right)
\end{equation}
the Circular Random Wave Model for quantum graphs
in the unitary class. The 
moments and the intensity correlation matrix predicted by
the Circular Random Wave Model read
\begin{equation} 
  \label{unitary predictions circular} 
  \begin{array}{rll}
   \displaystyle M_{\mathrm{CU},\alpha,q}&  \displaystyle =
    \frac{q! (2B-1)!}{(2B+q-1)!}& \displaystyle =
    \frac{q!}{(2B)^q}\left(1-\frac{q(q-1)}{4B} + \mathcal{O}(B^{-2})
    \right) \\
    \displaystyle C_{\mathrm{CU}, \alpha, \alpha'}& \displaystyle =
    \frac{1+\delta_{\alpha \alpha'}}{(2B)(2B+1)}& \displaystyle =
    \frac{1+\delta_{\alpha \alpha'}}{(2B)^2}
    \left(
      1-\frac{1}{2B}+\mathcal{O}(B^{-2})
    \right)
  \end{array}
\end{equation} 
In the limit $B\rightarrow \infty$ (and constant degree $q$) one may replace the
Circular Random Wave Model by the Gaussian Random Wave Model
with the joint probability density
\begin{equation} \label{gaussian model}
  \mathcal{\varphi}_{\mathrm{GU}}(\vect{a})   \equiv 
  \frac{B^{2B}}{\pi^{2B}} \ e^{-2B\Vert \vect{a} \Vert^2} .
\end{equation}
The predictions for the moments and the intensity correlation matrix 
in the Gaussian Random Wave Model  read
\begin{equation} 
  \label{unitary predictions} 
  M_{\mathrm{GU}, \alpha,q}
  = \frac{q!}{(2B)^q} \quad \textrm{and} \quad
  C_{\mathrm{GU}, \alpha, \alpha'}=
  \frac{1+\delta_{\alpha \alpha'}}{(2B)^2}
\end{equation}
which is equivalent to the leading order of the predictions
\eqref{unitary predictions circular}  of the Circular Random Wave
Model as $B\rightarrow \infty$.

If the eigenfunction statistics \eqref{statistics} of a family of
quantum graphs in the unitary symmetry class 
are well reproduced by $\varphi_{\mathrm{CU}}(\vect{a})$ in
\eqref{circular random waves unitary} or
by $\varphi_{\mathrm{GU}}(\vect{a})$ in
\eqref{gaussian model}, these formulae provide us with a
universal Circular or Gaussian Random Wave Model, which 
gives access to all the
statistical properties of the eigenfunctions.
Notice that the exact calculation
\eqref{exact covariance} asserts that \eqref{gaussian model} is the
only possible Gaussian joint probability density function of the type
\eqref{phi def}, and hence, a non-universal Gaussian model cannot be
realized on quantum graphs.

Establishing the possible validity of the Gaussian Random Wave Model
\eqref{gaussian model} would require the calculation of
\eqref{statistics} for arbitrary products of amplitudes $a_\beta$.
Note that the Gaussian Random Wave Model is consistent
with the gauge principle, i.e. its prediction
for any correlation function
that is not explicitly gauge invariant vanishes identically.
In what follows we 
will mainly focus on the explicitly gauge invariant autocorrelation
functions \eqref{C alpha def}. However in 
subsection~\ref{Diagonal Approximation} we will show
that some low order correlation functions that are not explicitly gauge
invariant indeed vanish on the level of the diagonal approximation.

When time-reversal symmetry is conserved one has to take 
into account that the amplitudes of
counter propagating waves on the same bond are complex 
conjugates, so that the 
wave function is real. We can thus only expect 
that a universal joint probability function 
is invariant under $\vect{a}\mapsto u \vect{a}$ where $u$ is a 
unitary $2B \times  2B$ 
matrix that respects reality of the wave function or, equivalently, that
$S \mapsto u^\dagger S u$ conserves $S^\mathcal{T}=S$. Such unitary matrices
have the block structure  
\begin{equation}
  \label{orthogonaltransformation}
  u=\begin{pmatrix} 
    u_{++}^\ast & u_{+-}\\
    u_{+-}^\ast & u_{++}
  \end{pmatrix} 
\end{equation}
in terms of the direction index $d$. 
Here $u_{++}$ and $u_{+-}$ are two $B \times B$ 
matrices which are only constrained by unitarity of $u$. 
Unitary matrices with this block structure
obey $u^{\mathcal{T}}=u^\dagger=u^{-1}$ and are thus in fact orthogonal
matrices with respect to $\mathcal{T}$-transposition.\\
The Circular Random Wave Model for the orthogonal class
\begin{equation}
  \label{circular random waves orthogonal}
  \varphi_{\mathrm{CO}}(\vect{a}) \equiv  \frac{2^{B-1}(B-1)!}{\pi^{B}} 
  \delta^{B}(\vect{a}_+-\vect{a}_-^\ast)\
  \delta\left(\frac{1}{2} -\Vert\vect{a}_+\Vert^2\right)
\end{equation}
is the unique model 
which respects $\vect{a}_+=\vect{a}_-^\ast$, the normalization
$\Vert \vect{a} \Vert^2=1=2  \Vert \vect{a}_+\Vert^2$, and is invariant under
the generalized orthogonal transformations  \eqref{orthogonaltransformation}.
It gives the predictions
\begin{equation} 
  \label{orthogonal predictions circular} 
  \begin{array}{rll}
    \displaystyle M_{\mathrm{CO},\alpha,q}&  
    \displaystyle =
    \frac{q! (B-1)!}{2^q (B+q-1)!}
    & \displaystyle =
    \frac{q!}{(2B)^q}\left(1-\frac{q(q-1)}{2B} + \mathcal{O}(B^{-2})
    \right) 
    \\
    \displaystyle C_{\mathrm{CO}, \alpha, \alpha'}& \displaystyle =
    \frac{1+\delta_{\alpha \alpha'}+\delta_{\alpha \hat{\alpha}'}}{4B(B+1)}
    & \displaystyle =
    \frac{1+\delta_{\alpha \alpha'}+\delta_{\alpha \hat{\alpha}'} }{(2B)^2}
    \left(
      1-\frac{1}{B}+\mathcal{O}(B^{-2})
    \right)\ .
  \end{array}
\end{equation} 
The only difference in the leading order for large graphs is
the term $\delta_{\alpha \hat{\alpha}'}$ which ensures that the intensity 
correlation matrix is invariant under $\alpha \mapsto \hat{\alpha}$.
Note that the deviations in the next order are twice as large in
the orthogonal case.

In the limit $B\rightarrow \infty$ one may again replace the
Circular Random Wave Model by a Gaussian Random Wave Model
with the joint probability density
\begin{equation} \label{orthogonal jpdf}
  \mathcal{\varphi}_{\mathrm{GO}}(\vect{a})   \equiv  \frac{B^{B}}{\pi^{B}} \
   \delta^{B}(\vect{a}_+-\vect{a}_-^\ast)\
 e^{-2B \Vert \vect{a}_+ \Vert^2} 
\end{equation}
where only one half of the coefficients is taken from a Gaussian ensemble
while the other half remains fixed by the symmetry constraints.
The moments and the intensity correlation matrix 
in this Gaussian Random Wave Model are just the leading order terms 
from  \eqref{orthogonal predictions circular} 
\begin{equation} 
  \label{orthogonal predictions} 
  M_{\mathrm{GO}, \alpha,q}
  = \frac{q!}{(2B)^q} \quad \textrm{and} \quad
  C_{\mathrm{GO}, \alpha, \alpha'}=
  \frac{1+\delta_{\alpha \alpha'}+\delta_{\alpha \hat{\alpha}'}}{(2B)^2} .
\end{equation} 

Note that the unitary and orthogonal
universal Gaussian Random Wave Models \eqref{gaussian model}
and \eqref{orthogonal jpdf} 
do not obey the normalization condition $\Vert \vect{a} \Vert^2=1$.
In fact one has
\begin{equation}
  \left\langle \Vert \vect{a} \Vert^2 \right\rangle_{\mathrm{GU}} = 
  \left\langle \Vert \vect{a} \Vert^2 \right\rangle_{\mathrm{GO}} = 1.
\end{equation} 
only as an average property while the variances
\begin{equation}
  \left\langle \left( \Vert \vect{a} \Vert^2 -1 \right)^2 \right\rangle_{\mathrm{GU}} = 
  \frac{1}{2B} \quad \textrm{and} \quad 
  \left\langle \left( \Vert \vect{a} \Vert^2  -1 \right)^2 \right\rangle_{\mathrm{GO}} = 
  \frac{1}{B}
\end{equation} 
are positive. Similarly, $|a_\alpha^\nu|$ cannot exceed one 
while the Gaussian Random Wave Models have a finite probability
for this event. The Circular Random Wave Models take all these
constraints into account correctly. 

There is another obstruction to all the Random Waves Models 
\eqref{circular random waves unitary}, \eqref{gaussian
  model}, \eqref{circular random waves orthogonal} and \eqref{orthogonal jpdf}. The matching conditions at
vertex $i$ impose some correlation between the amplitudes supported on
the neighboring bonds. This type of local and system-dependent
correlations is ignored in the universal Random Wave Models. The most 
striking
example consists in adding a Neumann vertex on some bond $b$ of an
ergodic graph. By doing so, the bond $b$ is split into two new bonds
$b_1$ and $b_2$, which can be oriented such that $(b_1,+) \to
(b_2,+)$. Then, the Neumann condition imposes $|a_{b_1+}|^2 =
|a_{b_2+}|^2$ and $|a_{b_1-}|^2 = |a_{b_2-}|^2$. These strong
correlations contradict the predictions \eqref{unitary predictions circular} 
and \eqref{orthogonal predictions circular}. Hence, a necessary condition for
the universal Gaussian models \eqref{gaussian model} and
\eqref{orthogonal jpdf} to be fulfilled in the limit of large graphs is
that all the valencies tend to infinity.

For a finite graph one should expect that none
of these models reproduces the exact correlation functions. 
Indeed, any numerical evaluation of the wave function statistics
shows (amongst other things) an intensity fluctuation matrix that 
is far less uniform than the predictions from the Random Wave Models
(see Figure~\ref{intensity correlation figure}). The deviations
can only be expected to vanish as $B \to \infty$ and if certain other
conditions that we are going to derive are also satisfied.

\subsection{Asymptotic Quantum Ergodicity} \label{Asymptotic Quantum
  Ergodicity}

Let $G$ be a metric graph with $B$ bonds.  An observable on $G$ is a
family
\begin{equation} \label{observable wave} V = \Big\{ V_b \in C^0\big(
  [0,L_b]\big) \Big| b \in \mathbb{N}_B \Big\}
\end{equation} 
of $B$ real functions $V_b(x)$ defined on the bonds of $G$. The mean
value $\bar{V}$ of an observable $V$ is defined by
\begin{equation}
  \bar{V} \equiv \frac{2}{\tr L} \int^{\oplus}_G V \equiv \frac{2}{\tr L} \sum_{b = 1}^B \int_0^{L_b} V_b(x) dx.
\end{equation}

\noindent Notice that $\frac{\tr L}{2} = \int^{\oplus}_G 1$ is the
volume of $G$. If an observable $V$ is constant on each bond, one can
simply write $V = (V_b)_{b\in \mathbb{N}_B}$ with $V_b \in
\mathbb{R}$. The mean value of such an observable reads
\begin{equation} \label{V bar wave} \bar{V} = \frac{\sum_{b=1}^B V_b
    L_b}{\sum_{b = 1}^B L_b}
\end{equation}
and is invariant under a global scaling of the bond lengths.

Suppose now that $S \in U(2B)$ is a scattering matrix on $G$. The
quantum graph $(G,S)$ is said to be quantum ergodic if and only if
there exists a subsequence $i \mapsto \nu(i)$ of density 1 such that
\begin{equation} \label{quantum ergodicity waves} \lim_{i \to \infty}
  \frac{\big( \Psi^{\nu(i)}, V \Psi^{\nu(i)} \big)}{\big(
    \Psi^{\nu(i)},\Psi^{\nu(i)} \big)} = \bar{V}
\end{equation}
for any observable $V$. In this definition, $\Psi^\nu = \bigoplus_{b =
  1}^{B} \psi^\nu_b$ denotes an eigenfunction of $H$ of eigenvalue
$k_\nu^2$. 
By assumption, it is unique up to multiplication by complex 
numbers (or by real numbers for the orthogonal class).

The left-hand side of \eqref{quantum ergodicity waves} represents the
mean value of the observable $V$ in the eigenstate $ \Psi^{\nu(i)}$. A
straightforward calculation shows that
\begin{eqnarray} \label{mean value of V in Psi}
  (\Psi^\nu, V \Psi^\nu ) & = & \sum_{b = 1}^B \Big( |a_{b+}^\nu|^2 + |a_{b-}^\nu|^2\Big) \int_0^{L_b} V_b(x) dx \nonumber\\
  && + 2 \Re \sum_{b = 1}^B a_{b-}^{\nu \ast} a_{b+}^\nu \int_0^{L_b}
  V_b(x) e^{2ik_\nu \left( x - \frac{L_b}{2}\right)} dx
\end{eqnarray}
for the wave function $\Psi^\nu$ with wave number $k_\nu >0$ and
amplitudes $a_{b+}^\nu$ and $a_{b-}^\nu$ as in \eqref{wave amplitude
  mapping}.  Since the observable $V$ is assumed continuous on each
bond, and since $|a_{b-}^{\nu \ast} a_{b+}^\nu| \leq 1$, the second
term in the right-hand side of \eqref{mean value of V in Psi} is  $\mathcal{O}(k_\nu^{-1})$. In the high energy limit
this second term 
gives no contribution to
the left-hand side of \eqref{quantum ergodicity waves}. 
Moreover, the first term in the
right-hand side of \eqref{mean value of V in Psi} remains unchanged if
the observable $V$ is replaced with the observable $W$ defined by $W_b
\equiv L_b^{-1} \int_0^{L_b} V_b(x) dx$. These two remarks imply that,
in the definition \eqref{quantum ergodicity waves} of quantum
ergodicity, it is sufficient to consider observables that are
constant on each bond, and this will always be the case in what follows.

If the equality \eqref{quantum ergodicity waves} holds for any
observable of vanishing mean $\bar{V} = 0$, then it also holds for any
observable $W$. In order to see this, it is sufficient to observe that
$W - \bar{W}$ has vanishing mean and to apply \eqref{quantum ergodicity
  waves} to this new observable. Hence, without loss of generality,
one can also restrict attention to observables $V$ with $\bar{V} = 0$.

If the identity \eqref{quantum ergodicity waves} is satisfied for any
subsequence of eigenfunctions, the quantum graph is said to be quantum
unique ergodic. In \cite{Scars}, it is shown that 
many short closed
cycles, like the triangle $\beta_1 \to \beta_2 \to \beta_3 \to
\beta_1$ for instance, support eigenfunctions with arbitrarily high
energies. These eigenfunctions, called scars, break quantum unique
ergodicity. While these scarred eigenfunctions were obtained explicitly for
Neumann quantum graphs, quantum unique ergodicity should certainly
not be expected to hold on
general finite quantum graphs. 

Moreover, 
quantum ergodicity is generally
not realized on a finite quantum
graph as well. This notion has thus to be replaced with a weaker one
which we call asymptotic quantum ergodicity. Let us consider an infinite
sequence $\{(G_l,S_l)\}_{l\in\mathbb{N}}$ of quantum graphs with
increasing number of bonds $B_l<B_{l+1}$. We also suppose that the
bonds of any $G_l$ have bond lengths that
satisfy
\begin{equation}
  \label{length condition}
  L_b \in [L_{\textrm{\scriptsize{min}}},
  L_{\textrm{\scriptsize{max}}}]\qquad  \text{where} \qquad
  0 <L_{\textrm{\scriptsize{min}}} < L_{\textrm{\scriptsize{max}}}< \infty
\end{equation}
are independent of $l$. Such a sequence will be called increasing. We
always assume that either all the graphs $(G_l,S_l)$ are time-reversal
invariant, or they all break this symmetry. The eigenfunctions of
$(G_l,S_l)$ are denoted by $\Psi_l^\nu$, and similarly, all the
quantities introduced above are indexed by $l$. Besides, a sequence
$\{ V_l \}_{l\in\mathbb{N}}$, where $V_l$ is an observable on $G_l$,
is said to be acceptable if and only if the two conditions
\begin{equation} \label{acceptable seq of obs}
  \begin{array}{cc} \lim_{l \to \infty} \bar{V}_l \equiv \bar{V}_\infty \  \textrm{exists}, \\
    0 \leq |V_{l,b}| \leq V_{\textrm{\scriptsize{max}}} \end{array} 
\end{equation}
are fulfilled. Then, an increasing sequence
$\{(G_l,S_l)\}_{l\in\mathbb{N}}$ of quantum graphs is said to be
asymptotically quantum ergodic if and only if
\begin{equation} \label{as QE wave} \lim_{l \to \infty} \lim_{i \to
    \infty} \frac{\big( \Psi_l^{\nu(i)}, V_l \Psi_l^{\nu(i)}
    \big)}{\big( \Psi_l^{\nu(i)},\Psi_l^{\nu(i)} \big)} =
  \bar{V}_\infty
\end{equation}  
for all acceptable sequences of observables $\{ V_l
\}_{l\in\mathbb{N}}$. The limit $l \to \infty$ plays the role of the
semiclassical limit for quantum graphs.

For the sequences of graphs satisfying \eqref{as QE wave}, the rate of
convergence is also of particular interest. Therefore, we will treat a
single finite quantum graph first, and come back to convergence and
rate considerations afterwards.

A calculation similar to \eqref{mean value of V in Psi} shows that,
for an observable $V$ on $G$ constant on each bond, one has
\begin{equation} \label{scalar product psi v psi} \big( \Psi^{\nu}, V
  \Psi^{\nu} \big) = \left\langle \vect{a^\nu} \left| VL \left( 1 +
        \frac{\sin (k_\nu L)}{k_\nu L} \right) \right| \vect{a^\nu}
  \right\rangle = \langle \vect{a}^\nu | VL | \vect{a}^\nu \rangle +
  \mathcal{O}(k_\nu^{-1}),
\end{equation}
where $|\vect{a}^\nu \rangle \in \mathcal{A}$ is the vector of
amplitudes defining $\Psi^\nu$ through the construction \eqref{wave
  amplitude mapping}. There is a slight abuse of notation in this
expression. 
On
the left-hand side, $V = (V_b)_{b \in \mathbb{N}_B}$ is an observable
constant on each bond, whereas on the right-hand side $V$ stands for
the diagonal $2B \times 2B$ matrix $V_{bd,b'd'} \equiv \delta_{b,b'}
\delta_{d,d'} V_b$. Such a matrix is called observable on
$\mathcal{A}$ and has mean value
\begin{equation} \label{V bar amplitude} \bar{V} \equiv \frac{\tr
    (VL)}{\tr L} = \frac{\sum_{\beta=1}^{2B} V_\beta
    L_\beta}{\sum_{\beta=1}^{2B} L_\beta} = \frac{\sum_{b=1}^B V_b
    L_b}{\sum_{b = 1}^B L_b}.
\end{equation}
This expression coincides with the mean value \eqref{V bar wave} of
$V$ seen as an observable on $G$ constant on each bond.

From \eqref{quantum ergodicity waves}, \eqref{scalar product psi v
  psi} and \eqref{V bar amplitude}, we deduce that a quantum graph is
quantum ergodic if and only if there exists a subsequence $i \mapsto
\nu(i)$ of density 1 such that
\begin{equation} \label{quantum ergodicity amplitude} \lim_{i \to
    \infty} \frac{ \langle VL \rangle_{\nu(i)} }{ \langle L
    \rangle_{\nu(i)}} \equiv \lim_{i \to \infty} \frac{\langle
    \vect{a}^{\nu(i)} |VL | \vect{a}^{\nu(i)} \rangle}{\langle
    \vect{a}^{\nu(i)} |L | \vect{a}^{\nu(i)} \rangle} = \bar{V}
\end{equation}
for any observable $V$ on $\mathcal{A}$. As above, one can restrict
attention to observables $V$ such that $\bar{V} = 0$ without loss
of generality.

A standard theorem of ergodic theory, proven for example in
\cite{PeterWalters}, states that the quantum ergodicity property
\eqref{quantum ergodicity amplitude} is equivalent to the vanishing of
\begin{equation} \label{original fluctuations} F_V \equiv \lim_{K \to
    \infty} \frac{1}{N(K)} \sum_{k_\nu \leq K} \frac{ \langle VL
    \rangle_{\nu}^2 }{ \langle L \rangle_{\nu}^2}
\end{equation}
for all observables $V$ on $\mathcal{A}$ with $\bar{V} = 0$. Moreover,
since the bond lengths are bounded by $ L_{\textrm{\scriptsize{min}}}$
and $ L_{\textrm{\scriptsize{max}}}$ by assumption \eqref{length condition}, 
this property is also equivalent
to the vanishing of the fluctuations
\begin{equation} \label{fluctuations def} \mathcal{F}_V \equiv \left(
    \frac{2B}{\tr L}\right)^2 \sum_{\beta,\beta' = 1}^{2B} \big( VL
  \big)_\beta \big( VL \big)_{\beta'} C_{\beta \beta'}
\end{equation}
for all observables $V$ on $\mathcal{A}$ with $\bar{V} = 0$, 
where the
intensity correlation matrix $C_{\beta \beta'}$ in the right-hand side
is defined in \eqref{intensitycorrelation}.

In the case of an increasing sequence of graphs
$\{(G_l,S_l)\}_{l\in\mathbb{N}}$, asymptotic quantum ergodicity is
obeyed if and only if the sequence $\{F_{l,V_l}\}_{l\in\mathbb{N}}$
whose terms are defined as in \eqref{original fluctuations}, or
equivalently the sequence $\{\mathcal{F}_{l,V_l}\}_{l\in\mathbb{N}}$
whose terms are defined as in \eqref{fluctuations def}, converges to
zero as $l \to \infty$ for all acceptable sequences of observables $\{
V_l \}_{l\in\mathbb{N}}$. The rate of convergence is then called the
rate of quantum ergodicity.

The Gaussian Random Wave Models \eqref{gaussian model} and
\eqref{orthogonal jpdf} predict the fluctuations
\begin{equation} \label{F Gaussian prediction} 
  \mathcal{F}_V = \bar{V}^2
  + \kappa \frac{\tr (VL)^2}{(\tr\ L)^2},
\end{equation}
as can easily be shown from the Gaussian predictions for the
intensity correlation matrix
\eqref{unitary predictions} and \eqref{orthogonal predictions}. 
The parameter $\kappa$ was defined in \eqref{kappa}. The 
term proportional to $\kappa$ describes the deviation
from quantum ergodicity. For any admissible observable
and bond lengths bounded by \eqref{length condition}
the deviation predicted by the Gaussian Random Wave Models
is $\mathcal{O}(B^{-1})$.
Hence the Gaussian Random Wave Models predict that any increasing
sequence of quantum graphs is asymptotically quantum ergodic and that
the rate of convergence is larger by a factor of two if
time-reversal symmetry is conserved.

Note, that quantum ergodicity holds on average, in the sense that
\begin{equation} \label{LWL} A_V \equiv \lim_{K \to \infty}
  \frac{1}{N(K)} \sum_{k_\nu \leq K} \frac{ \langle VL \rangle_{\nu}
  }{ \langle L \rangle_{\nu}} = \bar{V}
\end{equation}
for all observables $V$. This is known as the local Weyl law.
It is easily checked to hold for any quantum
graph . Indeed, by the definition \eqref{C alpha def}, $A_V$ can also
be written
\begin{equation} \label{AV and C} 
  A_V = \frac{2B}{\tr L} \sum_{\beta =
    1}^{2B} (VL)_\beta 
  \left\langle |a_\beta |^2 \right\rangle .
\end{equation}
Then, the identity \eqref{exact covariance} shows that $
\left\langle |a_\beta |^2 \right\rangle =
(2B)^{-1}$, and the definition \eqref{V bar amplitude} of $\bar{V}$
concludes the proof of the claim. The 
restriction to incommensurable bond lengths is not necessary for
the local Weyl law, indeed we will show in the following
subsection that \eqref{exact covariance} is true for any choice of bond 
lengths.

\subsection{Green Matrices and Trace Formulae} \label{Green Matrices
  and Trace Formulae}

For $(G,S)$ a quantum graph, and for $\epsilon > 0$, one defines a
sub-unitary  quantum map $U_\epsilon(k)$ by
\begin{equation}
  U_\epsilon(k) = T(k) S_\epsilon T(k), \quad \textrm{with} \quad S_\epsilon \equiv e^{-\epsilon} S,
\end{equation}
and where $T(k)$ is the propagation matrix of $G$ given in \eqref{a_out
  a_in and T}. The retarded Green matrix (resolvent) $G(k)$ 
is the matrix-valued
function on $\mathbb{R}_+$ defined by
\begin{equation} \label{G def} G(k) \equiv \Big( 1-U_\epsilon(k)
  \Big)^{-1} = \sum_{n=1}^{2B} \frac{|n,k\rangle\langle
    n,k|}{1-e^{i(\phi_n(k) + i\epsilon)}}.
\end{equation}
It has poles in the lower complex half-plane at $\phi_n(k) = 2\pi p -
i\epsilon$ for any $p\in\mathbb{Z}$. The advanced Green matrix
$G^\dagger(k)$ is the hermitian conjugate of $G(k)$, that is
\begin{equation} \label{G dagger def} G^\dagger(k) = \Big(
  1-U_\epsilon^\dagger(k) \Big)^{-1} = \sum_{n=1}^{2B}
  \frac{|n,k\rangle\langle n,k|}{1-e^{-i(\phi_n(k) - i\epsilon)}}.
\end{equation} 
It has poles in the upper complex half-plane at $\phi_n(k) = 2\pi p +
i\epsilon$ for any $p\in\mathbb{Z}$. Making use of formula \eqref{der
  eigenphases}, it is not difficult to check that, for any integer $q
\geq 2$, and for any permutation $\sigma \in S_q$, the statistical
quantities defined in \eqref{statistics} with $p = q$ read
\begin{equation} \label{statistics and Green matrices} \langle
  a^\ast_{\beta_0} \ldots a^\ast_{\beta_{q-1}}a_{\beta'_0} \ldots
  a_{\beta'_{q-1}} \rangle = \lim_{\epsilon\to
    0}\frac{(2\epsilon)^{q-1}}{2B} \Big\langle \prod_{j=1}^{q-1}
  G(k)_{\beta_{\sigma(j)} \beta'_j} \cdot
  G^\dagger(k)_{\beta_{\sigma(0)} \beta'_0} \Big\rangle_k,
\end{equation}
where, in the right-hand side, the average over $k$ is defined by the
formula
\begin{equation}
  \big\langle f(k) \big\rangle_k \equiv \lim_{K \to \infty} \frac{1}{K} \int_0^K f(k) dk
\end{equation}
which is meaningful for any function $f$ integrable on every compact
interval $[0,K]$. The formula \eqref{statistics and Green matrices}
relies on the non-degeneracy of the spectrum, which generically
follows from the incommensurability of the bond lengths. However, it
still holds if the subsequence of levels $k_\nu$ that are degenerate
is of density zero. There are other versions of the equality
\eqref{statistics and Green matrices} where the right-hand side
involve $n_r \in \mathbb{N}$ elements of $G(k)$ and $n_a \in
\mathbb{N}$ elements of $G^\dagger(k)$ with $n_r + n_a = q$. A formula
similar to \eqref{statistics and Green matrices} is used in
\cite{FyodMirl1} to study the statistical properties of the
eigenfunctions in disordered systems. For the derivation
of exact expressions the choice of
the permutation $\sigma \in S_q$ in \eqref{statistics and Green matrices}
is mainly a matter of computational ease (and sometimes taste).
Throughout the remainder of this subsection we will show how the
different choices lead to different exact expressions.

The Green matrices $G(k)$ and $G^\dagger(k)$ can be viewed as the
results of summing geometrical series in $U_\epsilon(k)$ and
$U_\epsilon^\dagger(k)$. This gives rise to interpretations of their
components as sums of walks on the quantum graph $(G,S)$. An oriented
walk $\vec{\beta}$ is a list $(\beta_{0},\beta_{1},\ldots,\beta_{n})$
of consecutive directed bonds on the graph.  Its topological length
$|\vec{\beta}|$ is the number of vertices traversed, that is
$|\vec{\beta} |=n$.  The set of all oriented walks having topological
length $n$ is written $W_{n}$. The metric length of $\vec{\beta}$ is
\begin{equation}
  l(\vec{\beta}) \equiv \frac{ L_{\beta_{0}} }{ 2 } + \sum_{i=1}^{n-1} 
  L_{\beta_{i}} + \frac{ L_{\beta_{n}} }{ 2 }.
\end{equation}
The origin and terminus of $\vec{\beta} $ are respectively
$o\vec{\beta} \equiv \beta_{0}$ and $t\vec{\beta} \equiv \beta_{n}$.
The set of walks in $W_n$ having origin $\beta$ and terminus $\beta'$
is written $W_n(\beta,\beta')$, and $\cup_{n\in \mathbb{N}_0}
W_n(\beta,\beta') \equiv W(\beta,\beta')$. We also define the
stability amplitude
\begin{equation}
  A_{\vec{\beta}} \equiv \prod_{i=0}^{n-1} S_{ \beta_{i+1}
    \beta_{i} }.
\end{equation}
With these definitions, it is easy to see that
\begin{equation}
  G(k)_{\beta \beta'} = \sum_{ \vec{\beta} \in W(\beta',\beta)} e^{-\epsilon|\vec{\beta}|} e^{ik l(\vec{\beta}) } A_{\vec{\beta}}
\end{equation}
and
\begin{equation}
  G^\dagger(k)_{\beta \beta'} = \sum_{ \vec{\beta} \in W(\beta,\beta')} e^{-\epsilon|\vec{\beta}|} e^{-ik l(\vec{\beta}) } {A_{\vec{\beta}}}^\ast.
\end{equation}
Together with \eqref{statistics and Green matrices}, these formulae
enable one to express the autocorrelation functions
$C_{[\vect{\alpha}]}$ in \eqref{C alpha def} as sums over oriented
walks.

The different choices for the order of the left indices $\beta$ in
\eqref{statistics and Green matrices} lead to different equivalent
expressions for the autocorrelation functions $C_{[\vect{\alpha}]}$ in
terms of oriented walks. In general, showing the equivalence between
these trace formulae at the level of oriented walks turns out to be a
very difficult problem. In this subsection, these non-trivial
equivalences are illustrated by two alternative proofs of the local
Weyl law \eqref{LWL}.

In the case of the intensity correlation matrix 
$C_{\beta \beta'}$, two
permutations $\sigma \in S_2$ of the left indices in \eqref{statistics
  and Green matrices} can be chosen. The identity permutation $\sigma
= \textrm{id}$ leads to
\begin{eqnarray} \label{autocorrelation G 1 analytic}
  C_{\beta \beta'} & = & \lim_{\epsilon \to 0} \frac{\epsilon}{B}
  \left\langle G(k)_{\beta\beta}
    G^\dagger(k)_{\beta'\beta'}\right\rangle_k \\
  \label{autocorrelation G 1} & = & \lim_{\epsilon \to 0}
  \frac{\epsilon}{B} \sum_{ \vec{\beta} \in W(\beta,\beta)} \sum_{
    \vec{\beta}' \in W(\beta',\beta')} e^{-\epsilon (|\vec{\beta}| +
    |\vec{\beta}'|)} \delta_{l(\vec{\beta}), l(\vec{\beta}')}
  A_{\vec{\beta}} {A_{\vec{\beta'}}}^\ast
\end{eqnarray} 
while choosing the transposition $\sigma = (1 \ 2)$ leads to
\begin{eqnarray} \label{autocorrelation G 2 analytic}
  C_{\beta \beta'} & = & \lim_{\epsilon \to 0} \frac{\epsilon}{B}
  \left\langle G(k)_{\beta'\beta}
    G^\dagger(k)_{\beta\beta'}\right\rangle_k \\
  \label{autocorrelation G 2} & = & \lim_{\epsilon \to 0}
  \frac{\epsilon}{B} \sum_{ \vec{\beta} \in W(\beta,\beta')}
  \sum_{\vec{\beta}' \in W(\beta,\beta')} e^{-\epsilon(|\vec{\beta}| +
    |\vec{\beta}'|)} \delta_{l(\vec{\beta}), l(\vec{\beta}')}
  A_{\vec{\beta}} {A_{\vec{\beta}'}}^\ast.
\end{eqnarray}
In both cases, the Kronecker symbols originate from the average over
$k$.
\begin{figure}
  \begin{center}
    \includegraphics[width=0.8\textwidth]{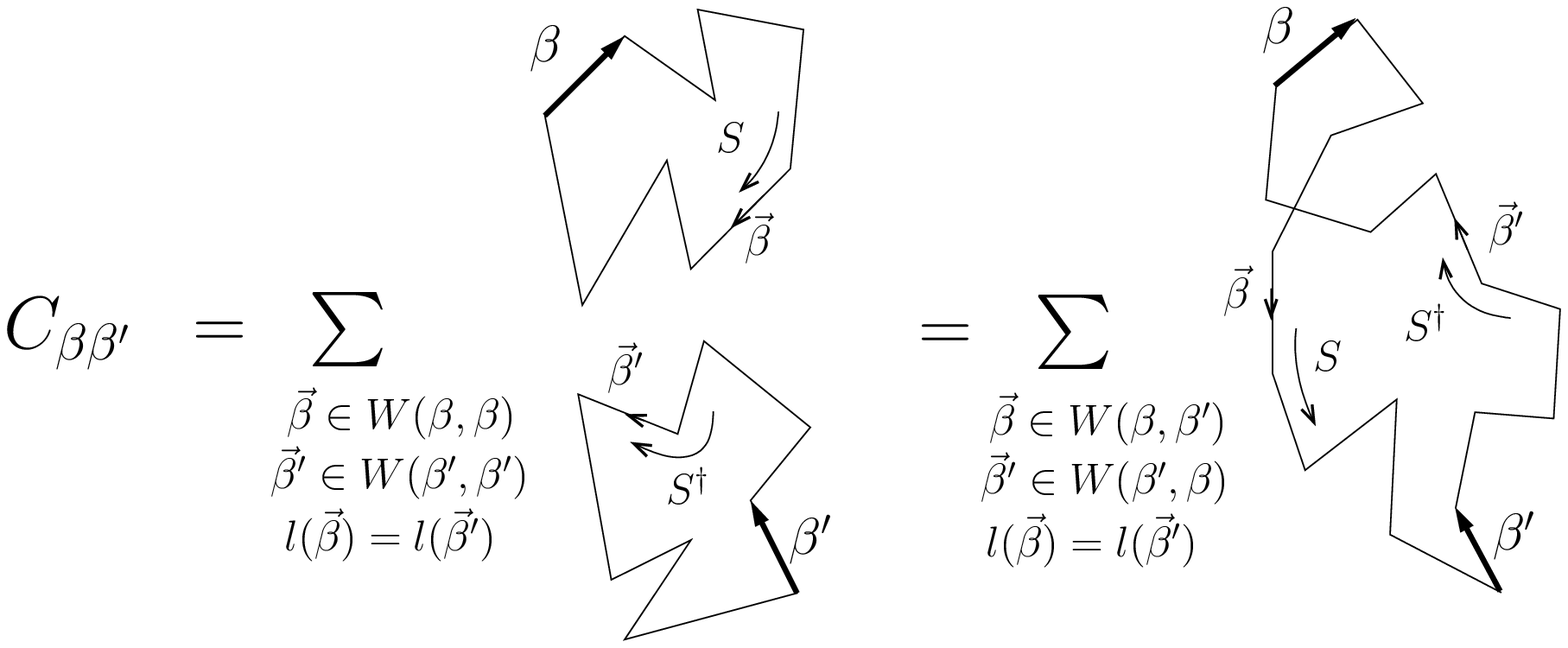}
    \caption{The two equivalent formulae \eqref{autocorrelation G 1}
      and \eqref{autocorrelation G 2} for the autocorrelation function
      $C_{\beta \beta'}$. The underlying graph has not been
      represented for sake of clarity. The trace formulae \eqref{C with
        curly G 1} and \eqref{C with curly G 2} are obtained from the
      ones represented here by adding the contributions where $S$ and
      $S^\dagger$ are swapped and by dividing by two.}
    \label{G matrices and paths}
  \end{center}
\end{figure}

These orbit expressions can also be recovered by means of the Poisson
summation formula. If $\delta_\epsilon(x)$ denotes the Lorentzian of
width $\epsilon$ centered at the origin, this formula leads to
\begin{eqnarray}
  \mathcal{G}_{\epsilon}(k) & \equiv & \sum_{n=1}^{2B}
  |n,k\rangle\langle n,k| \sum_{p = 0}^\infty  \delta_{\epsilon}\big(\phi_{n}(k) - 2\pi p
  \big) \\  \label{curly G trace formula}
  & = &  \frac{1}{2\pi}\mathds{1} +
  \frac{1}{2\pi} \sum_{q=1}^\infty \big( U(k)^q  + U^\dagger(k)^q \big) e^{-\epsilon q}
\end{eqnarray}
The trace formula \eqref{autocorrelation G 1}, or more exactly its
symmetrization obtained by replacing $A_{\vec{\beta}}
{A_{\vec{\beta}'}}^\ast$ with $\frac{1}{2} ( A_{\vec{\beta}}
{A_{\vec{\beta}'}}^\ast + {A_{\vec{\beta}}}^\ast A_{\vec{\beta}'} )$,
follows from \eqref{curly G trace formula} and the identity
\begin{equation} \label{C with curly G 1} C_{\beta \beta'} = \frac{2
    \pi^2 \epsilon}{B} \left\langle
    \mathcal{G}_{\epsilon}(k)_{\beta\beta}\
    \mathcal{G}_{\epsilon}(k)_{\beta'\beta'} \right\rangle_k
\end{equation}

This identity is a consequence of the fact that, in terms of
distributions, the product $2\pi \epsilon \delta_\epsilon(x)
\delta_\epsilon(y)$ tends to zero if $x \neq y$ and to $\delta(x)$ if
$x = y$. Similarly, the symmetric version of \eqref{autocorrelation G
  2} follows from \eqref{curly G trace formula} and
\begin{equation} \label{C with curly G 2} C_{ \beta \beta' } = \frac{2
    \pi^2 \epsilon}{B} \left\langle
    \mathcal{G}_{\epsilon}(k)_{\beta' \beta }\
    \mathcal{G}_{\epsilon}(k)_{\beta  \beta'} \right\rangle_k.
\end{equation}
The main advantage of the expressions \eqref{C with curly G 1} and
\eqref{C with curly G 2}, involving $\mathcal{G}_\epsilon(k)$, over
their analogues \eqref{autocorrelation G 1 analytic} and
\eqref{autocorrelation G 2 analytic}, which involve $G(k)$, is that
the matrix $\mathcal{G}_\epsilon(k)$ is real, whereas $G(k)$ and
$G^\dagger(k)$ have non-vanishing imaginary parts and must always
appear together in \eqref{statistics and Green matrices}. In
particular, the first moment $M_{\beta,1}=\left\langle |a_{\beta}|^2
\right\rangle$ 
can be written
\begin{equation} \label{C beta curly G}
  M_{\beta,1}
  = \frac{\pi}{B}
  \lim_{\epsilon \to 0} \left\langle
    \mathcal{G}_\epsilon(k)_{\beta\beta} \right\rangle_k,
\end{equation}
which involves a single closed oriented walk, while, in terms of
matrices $G(k)$, an additional directed bond $\beta'$ must first be
introduced in order for $M_{\beta,1}$ to be written as the sum
$\sum_{\beta' = 1}^{2B} C_{\beta\beta'}= 
\sum_{\beta' = 1}^{2B} \left\langle |a_{\beta}|^2|a_{\beta'}|^2
\right\rangle$ and the representations
\eqref{autocorrelation G 1} or \eqref{autocorrelation G 2} to be used.
From \eqref{curly G trace formula} and \eqref{C beta curly G}, one
finds directly that
\begin{equation}
  M_{\beta,1}
  = \frac{1}{2B},
\end{equation}
which, together with \eqref{AV and C}, provides a second proof of the
local Weyl law. Let us now use the trace formula \eqref{autocorrelation
  G 2} and perform the sum over the directed bond $\beta'$. It is
easy to show by induction over $n$ and $m$ that the unitarity of the
scattering matrix $S$ implies
\begin{equation}
  \sum_{\beta' = 1}^{2B} \sum_{ \vec{\beta} \in W_n(\beta,\beta')}  
  \sum_{\vec{\beta}' \in W_m(\beta,\beta')}    
  \delta_{l(\vec{\beta}), l(\vec{\beta}')} A_{\vec{\beta}} 
  {A_{\vec{\beta}'}}^\ast = \delta_{n,m}
\end{equation}
for all $n,m \in \mathbb{N}_0$. With \eqref{autocorrelation G 2}, this
gives
\begin{equation}
  M_{\beta,1}  = 
  \sum_{\beta' = 1}^{2B} C_{\beta\beta'} = \lim_{\epsilon \to 0} \frac{\epsilon}{B} \sum_{n,m = 0}^{\infty} e^{-\epsilon(n+m)} \delta_{n,m} = \frac{1}{2B},
\end{equation}
which, together with \eqref{AV and C}, yields a third proof of the
local Weyl law.

The choice for the permutation $\sigma \in S_q$ in \eqref{statistics
  and Green matrices} leads, in the case $q = 2$, to the equivalent
expressions \eqref{autocorrelation G 1} and \eqref{autocorrelation G
  2} for $C_{\beta \beta' }$ in terms of oriented walks that are
illustrated in Figure \ref{G matrices and paths}. Similar pictures
could also be drawn for $C_{[\beta_0, \ldots \beta_{q-1}]}$ when $q>2$
. Indeed, the right-hand side of \eqref{statistics and Green matrices}
with $\beta_j' = \beta_j$ for all $0 \leq j \leq q-1$ and with a fixed
permutation $\sigma \in S_q$ can be expressed as a sum over $q$
oriented walks $\vec{\beta}_1, \ldots, \vec{\beta}_q$, where each
$\vec{\beta}_j$ leads from the directed bond $\beta_j$ to the directed
bond $\beta_{\sigma(j)}$. The walk $\vec{\beta}_q$ is followed with
$S_\epsilon^\dagger$, whereas the $q-1$ other walks are followed with
$S_\epsilon$, and its metric length must equal the sum of the metric
lengths of the $q-1$ other walks. Therefore, \eqref{statistics and
  Green matrices} yields $q \cdot q!$ different ways of expressing
the autocorrelation function $C_{[\vect{\beta}]}$ of degree $q$ in
terms of oriented walks. Here, the first factor $q$ accounts for the
$q$ possible choices for the walk followed with $S_\epsilon^\dagger$.

\subsection{Long Diagonal Orbits} \label{Long Diagonal Orbits}

An expression for the fluctuations $\mathcal{F}_V$ in
\eqref{fluctuations def} can be obtained by retaining only a subset of
the whole set of pairs of oriented walks $(\vec{\beta}, \vec{\beta}')$
entering \eqref{autocorrelation G 1}. For this purpose, it is
convenient to come back to the expression \eqref{C with curly G 1} of
$C_{\beta \beta'}$ in terms of $\mathcal{G}_{\epsilon}(k)$ and write
the fluctuations of an observable $V$ with $\bar{V} = 0$ as
\begin{equation} \label{F and curly G} \mathcal{F}_V = \frac{8 B \pi^2
    \epsilon}{(\tr L)^2} \left \langle \left( \tr \big(
      \mathcal{G}_{\epsilon}(k) VL \big) \right)^2 \right\rangle_k.
\end{equation} 
The right-hand side can be written in terms of periodic orbits rather
than closed oriented walks as in \eqref{autocorrelation G 1}. A
periodic orbit is an equivalence class of closed oriented walks whose
sequences of directed bonds differ from each other by cyclic
permutations. For a periodic orbit $p$, the notions of reverse
$\hat{p}$, topological length $|p|$, metric length $l_p$ and stability
amplitude $A_p$ are inherited from the oriented walks terminology, and
the repetition number $r_p$ is the number of times $p$ retraces
itself. With this notation, one gets from \eqref{curly G trace
  formula}
\begin{equation} \label{trace gvl orbit} \tr\left(
    \mathcal{G}_{\epsilon} VL \right) = \frac{1}{\pi} \Re \sum_{p}
  e^{-\epsilon |p|} \frac{(VL)_p}{r_p} e^{ikl_p}A_p,
\end{equation}
where the sum is over all the periodic orbits on the graph and
$(VL)_p$ stands for the number obtained by accumulating the values
$(VL)_\beta$ of $VL$ along $p$. The square of the last formula admits
the spectral average
\begin{equation}
  \Big\langle \left[ \tr \big(\mathcal{G}_{\epsilon}VL \big) \right]^2 \Big\rangle_{k} =
  \frac{1}{2\pi^2} \sum_{p,q : l_p = l_q} \frac{(VL)_{p} (VL)_{q}}
  {r_{p}r_{q}} \Re(A_{p}A_{q}^\ast) e^{-\epsilon (|p|+ |q|)}.
\end{equation}
The diagonal approximation, which consists in only keeping the pairs
$q = p$ and $q = \hat{p}$ in the time-reversal invariant case, yields
\begin{equation} \label{long diag orb 1} \Big\langle \left[ \tr
    \big(\mathcal{G}_{\epsilon}VL \big) \right]^2
  \Big\rangle_{k}^{\textrm{diag}} = \frac{\kappa}{2\pi^2} \sum_{p,q :
    l_p = l_q} \frac{ \big[(VL)_{p}\big]^2 } {r_{p}^2} |A_{p}|^2
  e^{-2\epsilon |p|},
\end{equation}
where $\kappa$ is the parameter as in \eqref{F Gaussian prediction}
indicating whether time-reversal invariance is broken or conserved.
We have neglected some corrections in the diagonal approximation
which are due to repetitions and self-retracing
orbits. These can be shown not to contribute in the
present context.
The formula \eqref{long diag orb 1} is then approximated further. The
orbits for which $r_p >1$ are rare, so that we only keep the primitive
orbits, namely those with $r_p = 1$. We also take the long orbits
approximation \cite{EFKAMM}, which amounts to approximating
\begin{equation} \label{observable p 2 expectation}
  \big[(VL)_{p}\big]^2 \approx \big[ (VL)^2 \big]_p \approx |p| \frac{\tr
    (VL)^2}{2B}.
\end{equation}
Besides, the stability amplitude is known to behave like \cite{GS}
\begin{equation}
  |A_p|^2 \sim e^{-\alpha |p|},
\end{equation}
where $\alpha$ is the topological entropy. This parameter also
characterizes the number $|p|^{-1}e^{\alpha |p|}$ of periodic orbits
having topological length $|p|$. With all these approximations,
\eqref{long diag orb 1} reduces to the integral
\begin{equation}
  \Big\langle \left[ \tr \big(\mathcal{G}_{\epsilon}VL \big) \right]^2 \Big\rangle_{k}^{\textrm{diag}} \approx  \frac{\kappa}{2\pi^2} \int_0^\infty \frac{e^{\alpha |p|} d|p|}{|p|} \cdot |p| \frac{\tr (VL)^2}{2B} \cdot e^{-\alpha |p|} \cdot e^{-2\epsilon |p|}.
\end{equation}
Hence
\begin{equation} \label{diag long orbits final}
  \mathcal{F}_V^{\textrm{diag}} \approx \kappa \frac{\tr (VL)^2}{(\tr
    L)^2}.
\end{equation}
This formula, obtained from the long diagonal orbits, coincides with
the prediction \eqref{F Gaussian prediction} of the Gaussian Random
Wave Models \eqref{gaussian model} and \eqref{orthogonal jpdf}. It
predicts asymptotic quantum ergodicity for any increasing sequence of
quantum graphs and a universal rate of convergence $B^{-1}$, as in
\cite{EFKAMM}.

\section{Generating Functions} \label{Generating Functions}

\subsection{Definition and Principles} \label{Definition and
  Principles}

The Green matrices introduced in the subsection \ref{Green Matrices
  and Trace Formulae} can be obtained as the derivatives of certain
determinants. It is convenient first to introduce a Grassmann algebra
$\Lambda$, which can be decomposed as the direct sum of its commuting
sub-algebra $\Lambda_B$, called bosonic, and a set $\Lambda_F$ of
elements anticommuting with each other, called fermionic. Then, the
amplitude space $\mathcal{A}$ can be graded to get $\mathcal{A} \oplus
\mathcal{A}$, and the Grassmann envelope $(\mathcal{A} \oplus
\mathcal{A})(\Lambda)$ defined as in \cite{Berezin} can be built. This
set reads
\begin{equation} \label{Grassmann envelope} (\mathcal{A} \oplus
  \mathcal{A})(\Lambda) \equiv \left\lbrace V = \left(
      \begin{array}{c} V_B \\ V_F \end{array} \right) ; V_{B/F} =
    \sum_{\beta = 1}^{2B} V_{B/F}^\beta |e_\beta\rangle ,
    V_{B/F}^\beta \in \Lambda_{B/F} \right\rbrace,
\end{equation}
where the elements $|e_\beta \rangle$ refer to the elements in \eqref{e
  beta def} of the natural basis of $\mathcal{A}$.  The elements of
$(\mathcal{A} \oplus \mathcal{A})(\Lambda)$ are called supervectors.
The set of endomorphisms on \eqref{Grassmann envelope}, once written
in the natural basis of $\mathcal{A}$, form a set of supermatrices
written $L(\mathcal{A}|\mathcal{A})$. For $q \geq 2$ an integer, let
us introduce complex numbers $j_1, \ldots, j_{q-1}$ and $j_0$,
respectively referred to as retarded and advanced sources, and let us
also consider $q$ directed bonds $\alpha_1, \ldots \alpha_{q-1}$ and
$\alpha_0$.  The corresponding retarded and advanced source
supermatrices are defined by
\begin{eqnarray} \label{Jr def} J_r(\vect{j_r}) & \equiv & 1 + E_B
  \otimes \vect{j_r} \vect{E^{(r)}}, \\ \label{Ja def} J_a(j_a) &
  \equiv & 1 + E_B \otimes j_a E^{(a)},
\end{eqnarray}
where $E_B$ is the projector onto the bosonic sector of $(\mathcal{A}
\oplus \mathcal{A})(\Lambda)$,
\begin{equation} \label{j and E definitions} \vect{j} \equiv \left(
    \begin{array}{c} j_a \\ \vect{j_r} \end{array}\right) \equiv
  \left( \begin{array}{c} j_0 \\ j_1 \\ \vdots \\j_{q-1} \end{array}
  \right) , \quad \vect{E} \equiv \left( \begin{array}{c} E^{(a)} \\
      \vect{E^{(r)}} \end{array} \right) \equiv \left(
    \begin{array}{c} E^{\alpha_0, \alpha_0} \\ E^{\alpha_1, \alpha_1}
      \\ \vdots \\ E^{\alpha_{q-1}, \alpha_{q-1}} \end{array} \right),
\end{equation}
and, for any two directed bonds $\alpha,\alpha' \in \mathbb{N}_{2B}$,
$E^{\alpha, \alpha'}$ stands for the $2B \times 2B$ matrix whose
components are $(E^{\alpha, \alpha'})_{\beta\beta'} \equiv
\delta_{\alpha,\beta} \delta_{\alpha',\beta'}$ in the natural basis of
$\mathcal{A}$. The number $q-1$ of retarded sources corresponds to the
number of matrices $E^{\alpha_j, \alpha_j}$ contained in
$\vect{E^{(r)}}$, so that the product in \eqref{Jr def} makes sense.
In \eqref{Jr def}, \eqref{Ja def} and in what follows, some unit
matrices or supermatrices are not explicitly written in order to keep
the notation as simple as possible. For example, the symbols $1$ in
\eqref{Jr def} and \eqref{Ja def} must be read $\mathds{1}_{BF}
\otimes \mathds{1}_\mathcal{A}$, where $\mathds{1}_{BF} $ is the unit
supermatrix in Bose-Fermi space and $ \mathds{1}_\mathcal{A}$ is the
$2B \times 2B$ unit matrix in amplitude space $\mathcal{A}$.

Let $q \geq 2$ and let $[\vect{\alpha}] \equiv [\alpha_0, \alpha_1,
\ldots, \alpha_{q-1}]$ be a list of $q$ directed bonds. The
corresponding generating function is defined by
\begin{equation} \label{xi def} \xi_{[\vect{\alpha}]}(\vect{j}) \equiv
  \left\langle \sdet^{-1}\Big(1 - J_r(\vect{j_r}) \cdot U_\epsilon(k)
    \Big)\Big(1 - J_a(j_a) \cdot U_\epsilon^\dagger(k) \Big)
  \right\rangle_k,
\end{equation}
where $J_r(\vect{j_r})$ and $J_a(j_a)$ are defined from $\vect{j}
\equiv (j_a,\vect{j_r})^T = (j_0, j_1, \ldots, j_{q-1})^T$ and from
the directed bonds in $[\vect{\alpha}]$ as in \eqref{Jr def} and
\eqref{Ja def}. Notice that this function is well defined in a
neighborhood of the origin, and that it also reads
\begin{equation} \label{xi and G} \xi_{[\vect{\alpha}]}(\vect{j}) =
  \left\langle \textrm{det}^{-1} \Big( 1 - \vect{j_r}\vect{E^{(r)}}
    \big(G(k) - 1\big)\Big) \Big( 1 - j_a E^{(a)} \big(G^\dagger(k) -
    1\big)\Big) \right\rangle_k
\end{equation}
in terms of Green matrices.

It is convenient at this point to give a general rule governing
derivatives of determinants of the form \eqref{xi and G}. An important
quantity is the $\rho$ factor
\begin{equation} \label{rho factor def app} \rho_\alpha(\sigma) \equiv
  \alpha^{\textrm{number of cycles in $\sigma$}}
\end{equation}
defined for any $\alpha \in \mathbb{R}$ and any permutation $\sigma
\in S_s$ of $s \in \mathbb{N}$ elements. This factor can be seen as a
generalization of the signature $(-1)^\sigma$ of $\sigma \in S_s$
since the identity $(-1)^\sigma = (-1)^s \rho_{-1}(\sigma)$ holds.
Now, if $\vect{A} = (A^{(1)},\ldots,A^{(s)})^T$ is a vector containing
$s \in \mathbb{N}$ square matrices $A^{(i)}$ of size $n \in
\mathbb{N}$ and if $\vect{j} \in \mathbb{C}^s$, we have the equality
\begin{equation} \label{der of det} \left. \frac{\partial^s}{\partial
      j_1 \ldots \partial j_s} \det \left( 1 - \vect{j} \vect{A}
    \right)^{-\alpha} \right|_{\vect{j} = 0} = \sum_{\sigma \in S_s}
  \rho_\alpha(\sigma) \prod_{i = 1}^s \sum_{x_i = 1}^n
  A^{(i)}_{x_i,x_{\sigma(i)}}.
\end{equation}
This result can be proved by induction over $s$.  The right-hand side
has a natural diagrammatic representation where each $i \in
\mathbb{N}_s$ is a point and where an arrow is drawn from $i$ to $j$
whenever $\sigma(i) = j$. The sum in \eqref{der of det} is then the
sum over all such diagrams in which each point $i \in \mathbb{N}_s$
has exactly one outgoing and one incoming arrow. The value of each
diagram is a product of traces of the type $\tr \left(
  A^{(i)}A^{\sigma(i)} \cdots A^{\sigma^p(i)} \right)$, with $p$ being
the smallest number in $\mathbb{N}_0$ such that $\sigma^{p+1}(i) = i$,
weighted by its $\rho$ factor, which can be deduced from the number of
connected sub-diagrams.

Let $q \geq 2$ and let $[\vect{\alpha}] \equiv [\alpha_0, \alpha_1,
\ldots, \alpha_{q-1}]$ be a list of $q$ directed bonds.  The rule
\eqref{der of det} can be applied to the expression \eqref{xi and G}
for the generating function, and, making use of \eqref{statistics and
  Green matrices}, one easily gets
\begin{equation} \label{C generated by xi} C_{[\vect{\alpha}]} =
  \lim_{\epsilon\to 0} \frac{(2\epsilon)^{q-1}}{2B(q-1)!}\delta
  \xi_{[\vect{\alpha}]},
\end{equation}
where $C_{[\vect{\alpha}]}$ is the autocorrelation function defined in
\eqref{C alpha def}, and
\begin{equation} \label{delta q def} \delta \xi_{[\vect{\alpha}]}
  \equiv \left. \left( \prod_{s = 0}^{q-1} \frac{\partial}{\partial j_s}
  \right) \xi_{[\vect{\alpha}]}\right|_{\vect{j}=0} .
\end{equation}
The denominator $(q-1)!$ in \eqref{C generated by xi} comes from the
number of diagrams arising when the rule \eqref{der of det} is applied
to the $q-1$ retarded derivatives on
$\xi_{[\vect{\alpha}]}(\vect{j})$. By \eqref{statistics and Green
  matrices}, these diagrams all yield the same contribution.

It is not difficult to check that the generating functions have the
following property. For all $j_a$ and $\vect{j_r} = (j_1, \ldots,
j_{q-1})$ in a sufficiently small neighborhood of the origin,
\begin{equation} \label{fundamental property}
  \xi_{[\vect{\alpha}]}(j_a,\vect{0}) =
  \xi_{[\vect{\alpha}]}(0,\vect{j_r}) = 1.
\end{equation}

For $\sigma \in S_q$ and $[\vect{\alpha}] \equiv [\alpha_0, \alpha_1,
\ldots, \alpha_{q-1}]$, one can introduce a function
$\xi^\sigma_{[\vect{\alpha}]}(\vect{j})$ by the formula \eqref{xi def}
using the matrices $E^{\alpha_j, \alpha_{\sigma(j)}}$ in place of
$E^{\alpha_j,\alpha_j}$ in the source supermatrices, and \eqref{C
  generated by xi} then serves as a definition for
$C_{[\vect{\alpha}]}^\sigma$. The function
$\xi^\sigma_{[\vect{\alpha}]}(\vect{j})$ also satisfies the property
\eqref{fundamental property}, and the identities \eqref{statistics and
  Green matrices} and \eqref{der of det} ensure that
$C_{[\vect{\alpha}]}^\sigma = C_{[\vect{\alpha}]}$ for any $\sigma \in
S_q$. In what follows, the arbitrary choice for $\sigma \in S_q$ in
$\xi^\sigma_{[\vect{\alpha}]}(\vect{j})$ will be  called the choice of
convention. These different but equivalent expressions must not be
confused with the equivalent sums over orientated walks which are the
object of Subsection \ref{Green Matrices and Trace Formulae}. Any
convention $\sigma \in S_q$ for the generating function involves
$(q-1)!$ equivalent sums over orientated walks. However, in the case
$q = 2$, the permutations $\sigma = \textrm{id}$ and $\sigma = (0 \
1)$, which are referred to as parallel and crossed conventions in the
sequel, do correspond to the sums \eqref{autocorrelation G 1} and
\eqref{autocorrelation G 2} respectively.

We started this chapter with the
convention to choose $\sigma \in S_q$ to be the identity
and we will use this convention in most of the following
calculations. This convention is not only singled out by simplicity; it results 
in a generating function \eqref{xi def} that is explicitly
gauge invariant while in other choices the gauge invariance is only
restored in the limit \eqref{C generated by xi}.
It also reduces the complexity of some
calculations because  the  matrices $J_r(\vect{j}_r)$
and $J_a(j_a)$ for the source terms
\eqref{Jr def} are diagonal matrices. While each convention yields a 
different but exactly equivalent expression
approximation schemes may break the exact identity.
This is not worrying as long as the difference is in sub-leading order.
For time-reversal invariant graphs the generating function \eqref{xi def}
is usually not explicitly invariant when one replaces any directed bond
by its reversed partner for $q\ge 3$. The invariance is only revealed once the
derivative in \eqref{C generated by xi} is taken (the limit 
$\epsilon \rightarrow 0$ is not required).  

\subsection{Diagonal Approximation} \label{Diagonal Approximation}

Before working further on the generating function
\eqref{xi def} with the supersymmetry method we
introduce in this subsection similar generating functions
and develop a corresponding trace formula.
The diagonal approximation to this new type of generating functions 
turns out to behave very differently from
the oriented walk representations previously discussed in the 
subsections~\ref{Green Matrices and Trace Formulae} and \ref{Long Diagonal Orbits}. 

The definition \eqref{xi def} of the generating functions, and
the fundamental formula \eqref{C generated by xi} can easily
be generalized to any correlation function \eqref{statistics} with
$p=q$. Moreover, these correlation functions
can also be written in terms of logarithmic derivatives with some
analogy to \eqref{C generated by xi}. 
We will focus on the case $p=q=2$
for which the general correlation function 
can be written as
\begin{equation} 
  \left\langle a_{\alpha_1}^\ast {a_{\alpha_1'}}^\ast a_{\alpha_2}
    a_{\alpha_2'}
  \right\rangle
  = \lim_{\epsilon \to 0} \frac{\epsilon}{B}\ \delta 
  \Xi_{[\alpha_1,\alpha_1';\alpha_2,\alpha_2']},
  \label{2nd order C with new xi}
\end{equation}
where
\begin{eqnarray} 
  \Xi_{[\alpha_1,\alpha_1';\alpha_2,\alpha_2']}(j_a,j_r) \equiv& \Big\langle \log \det \Big( 1
  - \tilde{J}_r(j_r)U_\epsilon(k) \Big) \log \det \Big( 1 -
  \tilde{J}_a(j_a)U_\epsilon^\dagger(k) \Big) \Big\rangle_k \nonumber\\
  =&
  \Big\langle \log \det \Big( 1
  - \tilde{J}_r(j_r)U_\epsilon(k) \Big) \log \det \Big( 1 -
  \tilde{J}_a(j_a)^TU_\epsilon(k) \Big)^\ast \Big\rangle_k,
  \label{new generating function def}
\end{eqnarray}
the source terms are given by
\begin{equation}
  \label{Xi source}
  \tilde{J}_r(j_r)=1 + j_r E^{\alpha_1 ,\alpha_2} \qquad \text{and}
  \qquad
  \tilde{J}_a(j_a)=1 + j_a E^{\alpha_1' ,\alpha_2'} ,
\end{equation}
and  $\delta \Xi
  \equiv 
  \left. \frac{\partial^2}{ \partial
    j_r \partial j_a}
  \Xi(j_a,j_r) \right|_{j_a=j_r=0}$.
The intensity correlation matrix can be obtained 
in two different ways
\begin{eqnarray}
  \label{parallel convention}
  C_{\alpha \alpha'}=&
  \lim_{\epsilon \to 0} \frac{\epsilon}{B}
  \delta
    \Xi_{[\alpha,\alpha';\alpha,\alpha']}
     \\
  \label{crossed convention}
  =&\lim_{\epsilon \to 0} \frac{\epsilon}{B}
  \delta
    \Xi_{[\alpha,\alpha';\alpha',\alpha]}
\end{eqnarray}
referred as parallel and crossed conventions, respectively.
In the orthogonal class 
one has a third representation 
\begin{equation}
  \label{time reversed crossed convention}
  C_{\alpha \alpha'}= \lim_{\epsilon \to 0} \frac{\epsilon}{B}
  \delta
  \Xi_{[\alpha,\hat{\alpha};\alpha',\hat{\alpha}']}
\end{equation} 
called time-reversed crossed convention.

The formula $\log \det = \tr \log$ enables us
to write the new generating function \eqref{new generating function
  def} in terms of generalized periodic orbits on 
the graph. Indeed,
expanding the logarithms and performing the spectral average
yields the trace formula
\begin{equation} \label{Xi trace formula}
  \Xi_{[\alpha_1,\alpha_1';\alpha_2,\alpha_2']}(j_a,j_r) = 
  \sum_{p\in P_{\alpha_1 \alpha_2}}\sum_{p' \in P_{\alpha_2' \alpha_1'}} 
  \sum_{\rho,\rho'=0}^\infty
  \frac{1}{\rho \rho'}
  A_{r,p}(j_r)^\rho \left({A_{a,p'}(j_a)^\ast}\right)^{\rho'}
  \delta_{\rho l_p, \rho' l_{p'}}
\end{equation}
where the retarded and advanced modified 
stability amplitudes $A_{r,p}(j_r)$ and 
$A_{a,p}(j_a)$ of the generalized periodic orbit 
$p=\overline{\beta_1\beta_2\dots\beta_{|p|}}$ are defined by
\begin{equation}
  A_{r,p}(j_r) \equiv \prod_{i = 1}^{|p|} \big[ \tilde{J}_r(j_r) 
  S_\epsilon \big]_{\beta_{i+1} \beta_i} \quad \textrm{and} 
  \quad A_{a, p}(j_a) \equiv \prod_{i = 1}^{|p|} 
  \big[ \tilde{J}_a(j_a)^T S_\epsilon \big]_{\beta_{i+1} \beta_i}\ .
\end{equation}
The periodic orbits $p$ and $p'$ in \eqref{Xi trace formula} are all primitive but can be of a slightly more general type than the primitive periodic orbits considered
in Subsection~\ref{Long Diagonal Orbits}.
Indeed, the retarded source term in \eqref{Xi source} introduces the possibility to jump
$\alpha_2 \curvearrowright \alpha_1$, and similarly the advanced source term introduces the possibility to jump $\alpha_1' \curvearrowright \alpha_2'$. The set 
$P_{\alpha_1 \alpha_2}$ in \eqref{Xi trace formula} then contains all the primitive periodic orbits
that are compatible with the topology of the graph with an additional bridge
$\alpha_2 \curvearrowright \alpha_1$ at the center of these directed bonds.
Note that the two sets $P_{\alpha_1 \alpha_2}$ and
$P_{\alpha_2' \alpha_1'}$ of 
generalized periodic orbits need not be identical.
In the parallel convention for
$C_{\alpha \alpha'}$, the source terms are diagonal matrices,
and $P_{\alpha \alpha}\equiv P$  
reduces to the set of standard primitive periodic orbits, which 
only respect the topology of the graph. 
The length of a generalized periodic orbit is just the sum
of all bond lengths along the periodic orbit, where every jump
$\alpha_2 \curvearrowright \alpha_1$ contributes 
$\frac{1}{2}(L_{\alpha_1}+L_{\alpha_2})$. 
In
the trace formula \eqref{Xi trace formula},
only pairs of primitive orbits contribute such that a repetition 
of one orbit has the same length as a 
repetition of the other. 

The diagonal
approximation to the trace formula \eqref{Xi trace formula}
reduces the sum over pairs of primitive orbits
to either equal orbits $p'=p$, or time-reversed orbits $p'=\hat{p}$.
In both cases, the factor $\delta_{\rho l_p, \rho' l_{p'}}$ enforces $\rho=\rho'$.
Note that $p$ has to be a periodic orbit in
the intersection $p \in P_{\alpha_1 \alpha_2}\cap P_{\alpha_2' \alpha_1'}$
to contribute to the diagonal approximation. 

The remaining sum over periodic orbits can be resummed 
\begin{align} 
  \Xi_{[\alpha_1,\alpha_1';\alpha_2,\alpha_2']}^{\textrm{diag}}(j_a,j_r) = &
   \ \Xi_{[\alpha_1,\alpha_1';\alpha_2,\alpha_2']}^{\textrm{diag},D}(j_a,j_r)& + & \ 
  \Xi_{[\alpha_1,\alpha_1';\alpha_2,\alpha_2']}^{\textrm{diag},C}(j_a,j_r)& + \ 
 \mathcal{C} \nonumber \\
  \label{new xi diag approx}
  = &
  - \log \det\left(1-M^D(j_a,j_r)\right)& -&  \log \det\left(1-M^C(j_a,j_r)\right)& + \ 
  \mathcal{C}
\end{align}
where $M^D(j_a,j_r)$ and $M^C(j_a,j_r)$ are modifications 
of the classical map \eqref{classical map}. They describe
diffuson  and cooperon propagations, which originate from
pairs of periodic orbits with $p'=p$ and $p'=\hat{p}$,
and are given by
\begin{align} 
  M^D(j_a,j_r)_{\beta_1 \beta_2} \equiv 
  &
  \sum_{\beta'} J_r(j_r)_{\beta_1 \beta'} J_a(j_a)_{\beta' \beta_1} 
  |S_{\epsilon,\ \beta' \beta_2}|^2 
  \nonumber 
  \\
  =& \left(1+j_r \delta_{\beta_1 \alpha_1}\delta_{\alpha_1 \alpha_2} +
    j_a  \delta_{\beta_1 \alpha_2'}\delta_{\alpha_1' \alpha_2'}  
  \right)|S_{\epsilon,\ \beta_1 \beta_2}|^2
  +
  \nonumber\\
  \label{modified diffusion classical map}
  &
  j_rj_a \delta_{\beta_1 \alpha_1} 
  \delta_{\alpha_1 \alpha_2'} \delta_{\alpha_1' \alpha_2}|S_{\epsilon,\ \alpha_2 \beta_2}|^2
  \\
  M^C(j_a,j_r)_{\beta_1 \beta_2} \equiv &
  \sum_{\beta'} J_r(j_r)_{\beta_1 \beta'} J_a(j_a)^\mathcal{T}_{\beta' \beta_1} 
  S_{\epsilon,\ \beta' \beta_2} \left(S^{\mathcal{T}}_{\epsilon,\ \beta' \beta_2}\right)^\ast
  \nonumber \\
  =&
  \left(1+j_r \delta_{\beta_1 \alpha_1}\delta_{\alpha_1 \alpha_2} +
   j_a  \delta_{\beta_1 \hat{\alpha}_1'}\delta_{\alpha_1' \alpha_2'}  
  \right)S_{\epsilon,\ \beta_1 \beta_2} \left(
    S_{\epsilon,\ \hat{\beta}_2 \hat{\beta}_1}
  \right)^\ast
  +\nonumber\\
  \label{modified cooperon classical map}
  &
  j_rj_a \delta_{\beta_1 \alpha_1} 
  \delta_{\alpha_1 \hat{\alpha}_1'} \delta_{\alpha_2 \hat{\alpha}_2'}
  S_{\epsilon,\ \alpha_2 \beta_2} \left( S_{\epsilon,\ \hat{\beta}_2 \hat{\alpha}_2} \right)^\ast .
\end{align}
The term
$\mathcal{C}$ in \eqref{new xi diag approx} contains corrections 
for repetitions and
self-retracing orbits ($p=\hat{p}$) which can be shown 
not to contribute to our final result and will be omitted henceforth.

Recall that the classical map is defined by $M_{\beta_1 \beta_2}=|S_{\beta_1 \beta_2}|^2$, 
so that $M^D(0,0)=M_{\epsilon} \equiv e^{-2\epsilon} M$.
For time-reversal invariant systems,
the products of scattering matrices in 
\eqref{modified cooperon classical map} 
reduces to $S_{\epsilon,\ \beta_1 \beta_2} (
  S_{\epsilon,\ \hat{\beta}_2 \hat{\beta}_1}
)^\ast= |S_{\epsilon,\  \beta_1 \beta_2}|^2=M_{\epsilon, \beta_1
  \beta_2}$, so that $\lim_{\epsilon \rightarrow 0} M^C(0,0)=M$. 
If time reversal symmetry is broken
$M^C(0,0)$ does not reduce to $M$ and it does not
describe a Markof process on the graph.
We will see that the cooperon term only contributes to the diagonal approximation
formula if time-reversal symmetry holds.

The derivatives with respect to $j_r$ and $j_a$ can now be taken and
yield
\begin{align}
  \delta \Xi_{[\alpha_1,\alpha_1';\alpha_2,\alpha_2']}^{\textrm{diag},D} =&
  \left[ 
    \tr \left( \frac{1}{1-M_\epsilon} \frac{\partial^2 M^D}{\partial j_+ \partial
        j_-}
    \right) + 
    \tr \left( 
      \frac{1}{1-M_\epsilon} \frac{\partial M^D}{\partial j_+}\frac{1}{1-M_\epsilon} 
      \frac{\partial M^D}{\partial j_-}
    \right)
  \right]_{j_a=j_r=0}\nonumber \\
  =&
  \delta_{\alpha_1 \alpha_2'} \delta_{\alpha_1'
    \alpha_2} 
  \left(\frac{M_\epsilon}{1-M_\epsilon}\right)_{\alpha_2 \alpha_1} 
  + \nonumber \\
  \label{derivatives new xi last formula}
  &\delta_{\alpha_1 \alpha_2} \delta_{\alpha_1' \alpha_2'}
  \left(\frac{M_\epsilon}{1-M_\epsilon}\right)_{\alpha_1 \alpha_1'} 
  \left(\frac{M_\epsilon}{1-M_\epsilon}\right)_{\alpha_1' \alpha_1} 
\end{align}
for the diffuson generating function, and 
\begin{align}
  \delta\Xi_{[\alpha_1,\alpha_1';\alpha_2,\alpha_2']}^{\textrm{diag},C} =&
  \delta_{\alpha_1 \hat{\alpha}_1'} \delta_{\alpha_2
    \hat{\alpha}_2'} 
  \left(\frac{M^C(0,0)}{1-M^C(0,0)}\right)_{\alpha_2 \alpha_1} 
  +\nonumber \\
  \label{derivatives new xi last formula cooperon}
  &\delta_{\alpha_1 \alpha_2} \delta_{\alpha_1' \alpha_2'}
  \left(\frac{M^C(0,0)}{1-M^C(0,0)}\right)_{\alpha_1 \hat{\alpha}_1'} 
  \left(\frac{M^C(0,0)}{1-M^C(0,0)}\right)_{\alpha_1' \hat{\alpha}_1} 
\end{align}
for the cooperon generating function. For broken time-reversal
invariance the classical cooperon map  $M^C(0,0)$ has 
no unit eigenvalues in the limit $\epsilon \rightarrow 0$, and hence, 
\eqref{derivatives new xi last formula cooperon}
identically vanishes in that limit. By contrast, in time-reversal invariant systems
$M^C(0,0)=M_\epsilon$ and the cooperon generating function does contribute.

Finally, only terms in \eqref{derivatives new xi last formula}
and \eqref{derivatives new xi last formula cooperon}
that are singular as $\epsilon \rightarrow 0$ contribute to the
correlation function 
\eqref{2nd order C with new xi}.
In order to isolate these terms,
one makes use of
the decomposition \eqref{classical paths decomposition} of classical
orbits as the sum of a uniform component $|1\rangle \langle 1|$ and a
massive part $R$.
This yields
\begin{align}
  \frac{\epsilon}{B}
  \delta \Xi_{[\alpha_1,\alpha_1';\alpha_2,\alpha_2']}^{\textrm{diag}}
  =&
  \frac{\kappa(1-2\epsilon)}{16 B^3 \epsilon}
  \delta_{\alpha_1 \alpha_2}\delta_{\alpha_1' \alpha_2'} 
  + \frac{1}{4B^2}
  \left(
    \delta_{\alpha_1 \alpha_2'}\delta_{\alpha_1' \alpha_2} + (\kappa-1) 
    \delta_{\alpha_1 \hat{\alpha}_1'}\delta_{\alpha_2 \hat{\alpha}_2'}
   \right)
  +\nonumber\\
  \label{Xi diagonal}
  &+
  \frac{1}{4B^2}\delta_{\alpha_1 \alpha_2}\delta_{\alpha_1' \alpha_2'}
  \left( 
    \left[R_{\alpha_1 \alpha_1'} + R_{\alpha_1' \alpha_1}\right]
    +(\kappa-1)
    \left[ R_{\alpha_1 \hat{\alpha}_1'} + 
      R_{\hat{\alpha}_1' \alpha_1}
    \right]
  \right)+ \mathcal{O}(\epsilon)
\end{align} 
After dropping terms that are $\mathcal{O}(\epsilon)$, \eqref{Xi diagonal} 
may be expected to provide an approximation to
the generating function \eqref{2nd order C with new xi}.
However, its first term diverges like $\epsilon^{-1}$
as $\epsilon \rightarrow 0$. At first sight, this seems
to make periodic-orbit analysis using the trace formula
\eqref{Xi trace formula} for the generating function
much less useful than the
previous trace formulae from Section~\ref{Green Matrices and Trace Formulae}
which behave nicely in the diagonal approximation.
On the other hand, the same divergence also occurs
in the analysis of spectral correlations, which become singular
in the diagonal approximation
at small energy differences. Indeed, one may
obtain the 
corresponding trace formula for the spectral two point correlation function
on a graph $R_2(s)$
by replacing the source terms $\tilde{J}_a(j_a)$ and $\tilde{J}_r(j_r)$ 
appropriately. In this context,  
 a supersymmetry method developed in \cite{GA2, GA1}, which will
be adapted to our purposes in what follows,
cures the divergence. 
One may also try to add off-diagonal terms in the trace formula
in a systematic way but we will not pursue this here.

Note that the Kronecker symbols in the expression \eqref{Xi diagonal} 
force the correlation function $\langle a_{\alpha_1}^\ast a_{\alpha_1'}^\ast
a_{\alpha_2} a_{\alpha_2'} \rangle$ to vanish for all combinations
that are not invariant under all local gauge transformations allowed
by the unitary or orthogonal symmetry class. The non-vanishing
combinations are then equivalent to the three different conventions 
\eqref{parallel convention}, \eqref{crossed convention}
and \eqref{time reversed crossed convention}
for expressing the intensity correlation matrix. These three
conventions lead however to different formulae.

For any of the three conventions \eqref{parallel convention}, \eqref{crossed convention}
and \eqref{time reversed crossed convention}, if one replaces $\epsilon \mapsto \frac{\kappa}{4B}$,
the first line in
\eqref{Xi diagonal} reproduces the prediction of the
Gaussian Random Wave Model up to
corrections that are $\mathcal{O}(B^{-3})$. The second line in 
\eqref{Xi diagonal} then gives a correction in terms of 
system dependent massive modes. This massive correction
turns out to be different for the three conventions
of expressing $C_{\alpha \alpha'}$. 

In the orthogonal class
only the parallel convention \eqref{parallel convention} 
provides an approximated intensity correlation matrix 
that respects the identity $C_{\alpha \alpha'}=C_{\alpha
  \hat{\alpha}'}$
satisfied by the exact intensity correlation matrix.
By contrast,
if either the crossed convention
\eqref{crossed convention} or the time-reversed crossed
convention \eqref{time reversed crossed convention}
is used, the massive terms in the approximated 
intensity correlation matrix explicitly
violate this symmetry.
The origin of this discrepancy is
that, with the parallel convention
\eqref{parallel convention}, each of
the two logarithms in the generating function
$\Xi_{[\alpha,\alpha';\alpha,\alpha']}(j_a,j_r)$ is
invariant under time inversion, while 
in the crossed and time-reversed crossed
conventions  the symmetry is only restored after taking the derivatives and
performing the limit $\epsilon \rightarrow 0$. 

The observations above concerning time-reversal symmetry makes the parallel
convention \eqref{parallel convention}  a privileged choice when it comes to the diagonal approximation.
This convention yields
\begin{equation}
  \label{intensity correlation matrix diagonal}
  C_{\alpha \alpha'}^{\mathrm{diag}, \parallel}
  =\frac{1}{4B^2}\left(
    \frac{\kappa(1-2 \epsilon)}{4B \epsilon}
  +\delta_{\alpha \alpha'} + (\kappa-1) 
      \delta_{\alpha \hat{\alpha}'}
  +R_{\alpha \alpha '}+ R_{\alpha' \alpha}
  +(\kappa-1)\left(R_{\alpha \hat{\alpha} '}+ R_{\hat{\alpha}' \alpha} \right)
  \right).
\end{equation}
The three first terms are universal, and are
equal to the prediction of the Gaussian Random Wave Model
if $\epsilon$ is chosen finite and set equal to $\frac{\kappa}{4B}$ (which we
cannot justify at this stage). The remaining three terms
involve the matrix $R$ and they describe massive
corrections to the universal result. In fact,
these massive contributions may dominate the correlation
functions and, as a consequence,
the rate of convergence for
quantum ergodicity, or they may destroy quantum ergodicity altogether.
This point will be discussed further in
Section~\ref{Criteria and Rates of Universality}. 

\subsection{Nonlinear Supersymmetric $\sigma$ Model} \label{Nonlinear
  Supersymmetric sigma Model}

The generating functions in \eqref{xi def} depend strongly on whether
time-reversal symmetry is broken or conserved. Time inversion acts on
supervectors $\psi$ in the Grassmann envelope $(\mathcal{X} \oplus
\mathcal{X})(\Lambda)$, defined from $\mathcal{X} = \mathcal{A}
\otimes \mathbb{C}^n$ for some $n \in \mathbb{N}_0$ as in
\eqref{Grassmann envelope}, and on supermatrices $A \in
L(\mathcal{X}|\mathcal{X})$ as
\begin{equation} \label{action of time inversion} \mathcal{T} \psi =
  \sigma_1^d \psi^\ast \quad \textrm{and} \quad A^{\mathcal{T}} =
  \sigma_1^d A^T \sigma_1^d.
\end{equation}
In \eqref{action of time inversion}, $\sigma_1^d$ is the first Pauli
matrix acting on the direction space $\mathcal{A}_d$, $\psi^\ast$
denotes the vector obtained from $\psi$ by taking the complex
conjugates of each component, and $A^T$ is the transpose of $A$
defined as in \cite{Efetov} by the condition $(A \psi_1)^T \psi_2 =
\psi_1^T A^T \psi_2$ for all $\psi_1,\psi_2$ in $(\mathcal{X} \oplus
\mathcal{X})(\Lambda)$. Here and henceforth, $\psi^T$ stands for
the row vector obtained from the column vector $\psi \in (\mathcal{X}
\oplus \mathcal{X})(\Lambda)$ by usual transposition. One can now
introduce a 2-dimensional $\mathbb{C}$-linear space $TR$, the
time-reversal space, and the mapping
\begin{equation} \label{TR Psi} \psi \ \mapsto \ \Psi \equiv
  \frac{1}{\sqrt{2}} \left( \begin{array}{c} \psi \\ \mathcal{T} \psi
    \end{array} \right)_{TR} = 
  \frac{1}{\sqrt{2}} \left( \begin{array}{c} \psi \\ \sigma_1^d \psi^\ast
    \end{array} \right)_{TR},
\end{equation}
from $(\mathcal{X} \oplus \mathcal{X})(\Lambda)$ to
$(\mathcal{X}\otimes TR \oplus \mathcal{X}\otimes TR)(\Lambda)$ called
time-reversal doubling. We work with the convention $\chi^{\ast\ast} =
-\chi$ for all $\chi \in \Lambda_F$ as in \cite{Efetov}, and hence,
the Hermitian conjugate of $\Psi$ in \eqref{TR Psi} reads
\begin{equation}
  \bar{\Psi} = \left( \psi^{\dagger} \ , \  \psi^T \sigma_1^d \sigma_3^{BF}  \right),
\end{equation}
where $\psi^\dagger = \psi^{\ast T}$ is the Hermitian conjugate of
$\psi$, and $\sigma_3^{BF}$ stands for the third Pauli matrix acting
on the Bose-Fermi space.  Similarly, the time-reversal doubling of a
supermatrix $A \in L(\mathcal{X}|\mathcal{X})$ is defined by
\begin{equation} \label{observables doubling} A \ \mapsto \
  \mathcal{A} \equiv \left( \begin{array}{cc} A & 0 \\ 0 &
      A^{\mathcal{T}} \end{array} \right)_{TR} = \left(
    \begin{array}{cc} A & 0 \\ 0 & \sigma_1^d A^{T} \sigma_1^d
    \end{array} \right)_{TR},
\end{equation} 
and is an element of $ L(\mathcal{X}\otimes TR|\mathcal{X} \otimes
TR)$. In \eqref{TR Psi}, \eqref{observables doubling} and in what
follows, an index $TR$ added to a supermatrix means that this
supermatrix is explicitly written in the $TR$ space, and the same
notational trick is used for any other space. The components in
time-reversal space will be indexed by $t \in \{ \up, \down \}$. The
definitions above call for a notion of generalized transposition $
\mathcal{A}^\tau$ of $ \mathcal{A} \in L(\mathcal{X}\otimes
TR|\mathcal{X} \otimes TR)$, which is defined as in \cite{GA1} by
\begin{equation}
  \mathcal{A}^\tau \equiv \tau \mathcal{A}^T \tau^{-1}, \quad \textrm{where} \quad \tau \equiv \sigma_1^d \left( \begin{array}{cc} 0 & \sigma_{3}^{BF} \\ \mathds{1}_{BF} & 0
    \end{array}\right)_{TR}.
\end{equation} 
This definition implies that the equality $\bar{\Psi}_{1} \mathcal{A}
\Psi_{2} = \bar{\Psi}_{2} \mathcal{A}^\tau \Psi_{1}$ holds for any
couple of supervectors $\Psi_1, \Psi_2 \in (\mathcal{X}\otimes TR
\oplus \mathcal{X}\otimes TR)(\Lambda)$ and for any supermatrix $
\mathcal{A} \in L(\mathcal{X}\otimes TR|\mathcal{X} \otimes TR)$. It
follows that $(\mathcal{A}^{\tau})^{\tau} = \mathcal{A}$ and
$(\mathcal{AB})^\tau = \mathcal{B}^\tau \mathcal{A}^\tau$ for any such
supermatrices. Moreover, using the property $({A^{T}})^T = \sigma_3^{BF} A
\sigma_3^{BF}$ of the transposition in $L(\mathcal{X}|\mathcal{X})$,
it is easy to check that a supermatrix $\mathcal{A}$ obtained from
some $A \in L(\mathcal{X}|\mathcal{X})$ by time-reversal doubling
\eqref{observables doubling} is invariant under generalized
transposition.

Now, the generating functions can be written
\begin{equation} \label{xi with TR} \xi_{[\vect{\alpha}]}(\vect{j}) =
  \sdet^{-1} J_r J_a \left\langle \sdet^{-\frac{1}{2}}\Big(\tJpi -
    \tUe(k) \Big)\Big(\tJmi - \tUed(k) \Big) \right\rangle_k,
\end{equation}
where $\mathcal{J}_{r/a}$ and $\mathcal{U}_\epsilon(k)$ are the
time-reversal doubles of $J_{r/a}$ and $U_\epsilon(k)$. Following the
scheme developed in \cite{GA2} and \cite{GA1}, the generating
functions \eqref{xi with TR} can be represented in terms of a
nonlinear supersymmetric $\sigma$ model.

First, it is convenient to make use of the equality
\begin{equation} \label{sdet of block matrices} \sdet \left(
    \begin{array}{cc} A& B \\ C & D \end{array} \right) = \sdet(AD)
  \sdet (1 - A^{-1} B D^{-1} C)
\end{equation}
that holds for any square supermatrices $A,B,C$ and $D$ 
of the same size, and write the retarded and advanced
superdeterminants in \eqref{xi with TR} as
\begin{equation} \label{Retarded auxiliary doubling}
  \sdet^{-1/2}(\tJpi-\tUe) = \sdet^{-1/2} \left( \begin{array}{cc} 1 &
      \sqtSe T \\ T \sqtSe & \tJpi \end{array}\right),
\end{equation}
and
\begin{equation} \label{Advanced auxiliary doubling}
  \sdet^{-1/2}(\tJmi-\tUed) = \sdet^{-1/2} \left( \begin{array}{cc} 1
      & \sqtSed T^\dagger \\ T^\dagger \sqtSed& \tJmi
    \end{array}\right).
\end{equation}
In these expressions, $\mathcal{S}_\epsilon$ is the time-reversal
double of $S_\epsilon$, and the square root of a matrix is defined by
keeping the same eigenvectors and by taking the square roots of the
eigenvalues fixing the half-line singularity of the logarithm to
$(-\infty,0]$. It is not difficult to check that this definition of
the square root leads to the natural properties $\sqrt{A}\sqrt{A} = A$
and $\sqrt{A^\dagger} = \sqrt{A}^\dagger$. These two properties have
been used in order to obtain \eqref{Retarded auxiliary doubling} and
\eqref{Advanced auxiliary doubling}. Besides, if $\mathcal{A}$ is the
time-reversal double of $A$, then $\sqrt{\mathcal{A}}$ is the
time-reversal double of $\sqrt{A}$.  The 2-dimensional structure
introduced in the right-hand sides of \eqref{Retarded auxiliary
  doubling} and \eqref{Advanced auxiliary doubling} is referred to as
the auxiliary space $X$, and the components with respect to the basis
of $X$ used in these two matrix expressions will be indexed by $x \in
\{1,2\}$. The formulae \eqref{xi with TR}, \eqref{Retarded auxiliary
  doubling} and \eqref{Advanced auxiliary doubling} now enable us to
express the generating functions as the Gaussian superintegrals
\begin{equation} \label{xi first integral formula}
  \xi_{[\vect{\alpha}]}(\vect{j}) = \sdet^{-1} J_r J_a \cdot \int
  d\psi \left\langle e^{-S[\Psi]} \right\rangle_k,
\end{equation}
where $d\psi = d\psi_r d\psi_a$, $d\psi_r$ and $d\psi_a$ being two
Berezin measures \cite{Berezin} on $(\mathcal{A} \oplus
\mathcal{A})(\Lambda)$, $\Psi$ is the time-reversal double of $\psi$,
and
\begin{eqnarray} \label{first action} S[\Psi] & \equiv & \left(
    \bar{\Psi}_{r1} \ \bar{\Psi}_{r2} \right) \left(
    \begin{array}{cc} 1 & \sqtSe  T \\  T \sqtSe & \tJpi \end{array} \right)
  \left( \begin{array}{c} \Psi_{r1} \\ \Psi_{r2} \end{array} \right)
  \nonumber\\
  &  & + \left( \bar{\Psi}_{a1} \ \bar{\Psi}_{a2} \right) \left(
    \begin{array}{cc} 1 & \sqtSed T^\dagger  \\  T^\dagger \sqtSed & \tJmi
    \end{array} \right) \left( \begin{array}{c} \Psi_{a1} \\ \Psi_{a2}
    \end{array} \right).
\end{eqnarray}
The indices $r$ and $a$ of $\psi$ and $\Psi$ refer to the retarded and
advanced components of these supervectors used to write \eqref{Retarded
  auxiliary doubling} and \eqref{Advanced auxiliary doubling} as
Gaussian superintegrals respectively.

In the quadratic form \eqref{first action}, the off-diagonal couplings
depend on $k$ through the variables $(kL_1, \ldots, kL_B)$ in the
propagation matrix $T$. Using the fact that $T$ and $\sqtSe$ are
invariant under generalized transposition, these off-diagonal terms
can be written
\begin{equation}
  S_{\textrm{cf}}[\Psi]  =  2\bar{\Psi}_{r1}  \sqtSe T \Psi_{r2} + 2\bar{\Psi}_{a2} 
  T^\dagger \sqtSed   \Psi_{a1}.
\end{equation}
Since the bond lengths are assumed incommensurate the invariant
measure of the automorphism $k \mapsto (kL_1, \ldots, kL_B)$
$(\textrm{mod } 2\pi)$ on the $B$-torus is merely the product of $B$
Haar measures on the circle \cite{BarraGaspard}. Hence,
\begin{equation} \label{Barra Gaspard transform} \left\langle
    e^{-S_{\textrm{cf}}[\Psi]} \right\rangle_k = \prod_{b = 1}^B
  \int_0^{2\pi} \frac{d \varphi_b}{2\pi}
  e^{-S_{\textrm{cf}}^b[\Psi_b;\varphi_b]},
\end{equation}
where
\begin{equation} \label{S before color-flavor}
  S_{\textrm{cf}}^b[\Psi_b;\varphi_b] \equiv 2 \sum_{d = \pm} \left[
    \left( \bar{\Psi}_{r1}\sqtSe \right)_{bd}e^{i\varphi_b} \Psi_{r2;
      bd} + \bar{\Psi}_{a2; bd} e^{-i\varphi_b} \left( \sqtSed
      \Psi_{a1} \right)_{bd} \right].
\end{equation}
Then, the color-flavor transformation \cite{Zirn} can be applied
separately to each integral in the right-hand side of \eqref{Barra
  Gaspard transform}. This procedure introduces supermatrix variables
$Z_b$ and $\tZ_b$, which lie in $L(\mathcal{A}_d\otimes
TR|\mathcal{A}_d \otimes TR)$, and yields
\begin{equation} \label{Scf after cf} \left\langle
    e^{-S_{\textrm{cf}}[\Psi]} \right\rangle_k = \prod_{b = 1}^B \int
  d(Z_b, \tZ_b) \sdet (1 - Z_b \tZ_b)
  e^{-S_{\textrm{cf}}^b[\Psi_b;Z_b,\tZ_b]},
\end{equation}
with
\begin{eqnarray} \label{S after color-flavor}
  S_{\textrm{cf}}^b[\Psi_b;Z_b,\tZ_b] & = & 2 \sum_{d = \pm} \left( \bar{\Psi}_{r1}\sqtSe \right)_{bd} Z_{b;dd'} \left(   \sqtSed   \Psi_{a1} \right)_{bd'} \nonumber\\
  && + 2 \sum_{d = \pm} \bar{\Psi}_{a2; bd} \tZ_{b,dd'} \Psi_{r2; bd}
\end{eqnarray}
Basically, the retarded and advanced components of $\Psi$, which are
uncoupled in \eqref{S before color-flavor} become coupled in \eqref{S
  after color-flavor}, and conversely, the components in auxiliary
space, which are mixed in \eqref{S before color-flavor}, are
diagonalized by the color-flavor transformation. The reason for
resorting to this transformation is to get an action with
saddle-points, which is not the case in \eqref{S before color-flavor}.
The integration in \eqref{Scf after cf} must be performed over the set
of supermatrices $(Z_b , \tZ_b)$ satisfying the conditions
\begin{equation} \label{color-flavor requirements} \tZ_{BB} =
  Z_{BB}^\dagger, \quad \tZ_{FF} = -Z_{FF}^\dagger,
\end{equation}
and such that the eigenvalues of the positive hermitian matrix
$Z_{BB}^\dagger Z_{BB}$ are less than unity. The measure $d(Z_b,
\tZ_b)$ is then the Berezin measure over this set.

In order to simplify the notation, one can introduce the new
supermatrix fields
\begin{equation} \label{fields Z and tZ def} Z = \bigoplus_{b = 1}^B
  Z_b \quad \textrm{and} \quad \tilde{Z} = \bigoplus_{b = 1}^B \tZ_b,
\end{equation}
which belong to $L(\mathcal{A}\otimes TR|\mathcal{A}\otimes TR)$.
These supermatrices still satisfy the color-flavor requirements
\eqref{color-flavor requirements}. From \eqref{S after color-flavor}
and the diagonal terms of \eqref{first action}, one gets the new
quadratic form
\begin{eqnarray} \label{action after average} S[\Psi;Z,\tZ] & = &
  \left( \bar{\Psi}_{r1} \ \bar{\Psi}_{a1} \right) \left(
    \begin{array}{cc} 1 & \sqtSe Z \sqtSed \\ \sqtSed Z^\tau \sqtSe &
      1 \end{array} \right) \left( \begin{array}{c} \Psi_{r1} \\
      \Psi_{a1} \end{array} \right)
  \nonumber\\
  && + \left( \bar{\Psi}_{r2} \ \bar{\Psi}_{a2} \right) \left(
    \begin{array}{cc} \tJpi & \tZ^\tau \\ \tZ & \tJmi \end{array} \right)
  \left( \begin{array}{c} \Psi_{r2}
      \\ \Psi_{a2} \end{array} \right).
\end{eqnarray}
The integral over $\psi$ in \eqref{xi first integral formula} remains
Gaussian after the color-flavor transformation, and, from the explicit
formula \eqref{action after average}, the generating functions become
\begin{eqnarray}
  \xi_{[\vect{\alpha}]}(\vect{j})
  & = & \sdet^{-1} J_r J_a \int d(Z,\tZ) \ \sdet(1-Z\tZ)  \sdet^{-1/2}\left(
    \begin{array}{cc} \tJpi & \tZ^\tau \\ \tZ & \tJmi \end{array} \right)\nonumber\\
  && \quad \sdet^{-1/2}\left( \begin{array}{cc} 1 & \sqtSe Z \sqtSed \\
      \sqtSed Z^\tau \sqtSe & 1 \end{array} \right).
\end{eqnarray}
The first superdeterminant in the integrand comes from the $B$
superdeterminant factors introduced in \eqref{Scf after cf}.  Making
use of the rule \eqref{sdet of block matrices} once again and
resorting to the well-known formula $\sdet = \exp \str \log$ enables
us to write
\begin{equation} \label{exact xi superintegral}
  \xi_{[\vect{\alpha}]}(\vect{j}) = \int d(Z,\tZ) \ e^{-S[Z,\tZ]},
\end{equation}
where the function $S[Z,\tZ]$, called the action, or the exact action in order
to distinguish between $S[Z,\tZ]$ and its subsequent approximations,
is defined by
\begin{eqnarray} \label{Exact action} S[Z,\tZ] & = & -\str\log
  (1-Z\tZ) + \frac{1}{2}\str\log(1-Z \tSed
  Z^\tau \tSe) \nonumber\\
  && + \frac{1}{2} \str\log (1-\tJp \tZ^\tau \tJm \tZ) .
\end{eqnarray}
Notice that, if the sources $\vect{j_r}$ and $j_a$ are set to zero,
the resulting source-free action $S_0[Z,\tZ]$ is precisely the one
obtained in \cite{GA2} and \cite{GA1} for the generating function of
the spectral two-point correlation function.

The different conventions $\sigma \in S_q$ for the generating
functions discussed at the end of Subsection~\ref{Definition and
  Principles} can also be written in terms of the nonlinear
supersymmetric $\sigma$ model \eqref{exact xi superintegral}. Indeed,
in order to get $\xi_{[\vect{\alpha}]}^\sigma(\vect{j})$, it suffices
to replace $J_r$ and $J_a$ with $J_r^\sigma$ and $J_a^\sigma$ in the
exact action, where these two new source supermatrices are defined as
in \eqref{Jr def} and \eqref{Ja def} using the matrices $E^{\alpha_j,
  \alpha_{\sigma(j)}}$ instead of $E^{\alpha_j, \alpha_j}$ for all $0
\leq j \leq q-1$.

\section{Mean Field Theory} \label{Mean Field Theory}

\subsection{The Zero Mode}

The first step of our approximation scheme consists of restricting the
superintegral \eqref{exact xi superintegral} to the subset of
supermatrices $(Z_0,\tZ_0)$ around which the first variations
\begin{equation}
  \lim_{\eta \to 0} \frac{S_0 [Z_0 + \eta W, \tZ_0] - S_0 [Z_0, \tZ_0]}{\eta} \quad \textrm{and} \quad \lim_{\eta \to 0} \frac{S_0 [Z_0, \tZ_0 + \eta W] - S_0 [Z_0, \tZ_0]}{\eta}
\end{equation}
of the exact source-free action $S_0[Z,\tZ]$ vanish as $\epsilon\to 0$
for all supermatrices $W$ in $L(TR \otimes \mathcal{A} | TR \otimes
\mathcal{A})$. This subset of mean field configurations, called the
zero mode, was identified in \cite{GA2} and \cite{GA1} and consists of
the supermatrices satisfying
\begin{eqnarray}
  & Z_0 = \mathds{1}_\mathcal{A} \otimes Y \quad \textrm{and} \quad \tZ_0 = \mathds{1}_\mathcal{A} \otimes \tY, & \nonumber\\
  &\textrm{with} \quad Y,\tY \in L(TR|TR) \quad \textrm{such that} \quad  \tY = Y^\tau. &
\end{eqnarray}
Moreover, $Y$ and $\tY$ must be diagonal in $TR$ space if
time-reversal symmetry is broken. Of course, the color-flavor
relations \eqref{color-flavor requirements} must still be satisfied,
that is, the identities $\tY_\textrm{BB} = Y_\textrm{BB}^\dagger$ and
$\tY_\textrm{FF} = -Y_\textrm{FF}^\dagger$ are fulfilled, and the
eigenvalues of $Y_\textrm{BB}^\dagger Y_\textrm{BB}$ must have moduli
smaller than one. The supermatrices $(Y, \tY)$ satisfying these
relations parametrize a manifold, the so-called Efetov $\sigma$ model
space. Efetov's $\sigma$ model space with unitary symmetry has 4
commuting and 4 anticommuting parameters, whereas 8 commuting and 8
anticommuting parameters are involved in the orthogonal symmetry
class.

Let us introduce a 2-dimensional $\mathbb{C}$-linear space $RA$,
called retarded-advanced space, and let us consider the supermatrices
in $L(RA \otimes TR|RA \otimes TR)$
\begin{equation} \label{R def} R \equiv \left( \begin{array}{cc} 1 & Y
      \\ \tilde{Y} & 1 \end{array} \right)_{RA} \quad \textrm{and}
  \quad R^{-1} = \left( \begin{array}{cc} \frac{1}{1 - Y \tY} & -Y
      \frac{1}{1 - \tY Y} \\ - \tY \frac{1}{1 - Y \tY} & \frac{1}{1 -
        \tY Y} \end{array} \right)_{RA},
\end{equation}
Then, one can set \cite{Zirn}
\begin{equation} \label{Q matrix def} Q \equiv R \sigma_3^{RA} R^{-1},
\end{equation}
where $\sigma_3^{RA}$ stands for the third Pauli matrix in
retarded-advanced space. By construction, these matrices satisfy $Q^2
= Q$. Moreover, if for a supermatrix $A$ having a retarded-advanced
structure $\bar{A}$ denotes the supermatrix
\begin{equation}
  \bar{A} \equiv \mathcal{K} A^\dagger \mathcal{K}, \quad \textrm{where} \quad \mathcal{K} \equiv \left( \begin{array}{cc} \sigma_3^{RA} & 0 \\ 0 & 1 \end{array} \right)_{BF},
\end{equation}
the Efetov $\sigma$ model space is characterized by the constraints
$\bar{Q} = Q$, $Q^\tau = \sigma_3^{RA} Q \sigma_3^{RA}$, and $Q$
diagonal in $TR$ space for the unitary symmetry class.  Efetov's polar
coordinates \cite{Efetov} then involve writing
\begin{equation} \label{Q in Efetov coordinates} Q = UQ_{0}\bar{U},
\end{equation}
with
\begin{equation} \label{Q0 def} Q_{0} \equiv \left( \begin{array}{cc}
      \cos \hat{\theta} & i\sin\hat{\theta} \\ -i\sin\hat{\theta} &
      -\cos \hat{\theta} \end{array} \right)_{RA}, \quad \hat{\theta}
  \equiv \left( \begin{array}{cc} i \theta_{B} & 0 \\ 0 & \theta_{F}
    \end{array} \right)_{BF}.
\end{equation}
The equations $Q_0^2 = Q_0$ and $\bar{Q}_{0} = Q_0$ are automatically
fulfilled for any real symmetric matrices $\theta_B$ and $\theta_F$
acting on the $TR$ space if
\begin{equation} \label{U u v} U \equiv U_1 U_2 \equiv \left(
    \begin{array}{cc} u & 0 \\ 0 & v \end{array} \right)_{RA} \equiv
  \left( \begin{array}{cc} u_1 & 0 \\ 0 & v_1 \end{array} \right)_{RA}
  \left( \begin{array}{cc} u_2 & 0 \\ 0 & v_2 \end{array} \right)_{RA}
\end{equation}
are required to satisfy $\bar{U}_1 U_1 = 1$ and $\bar{U}_2 U_2 = 1$,
that is $\bar{u}_i \equiv u_i^\dagger = u_i^{-1}$ and $\bar{v}_i
\equiv \sigma_3^{BF} v_i^\dagger \sigma_3^{BF} = v_i^{-1}$, for $i \in
\{ 1,2 \}$. The purpose of $U_1$ is to diagonalize $Q$ in Bose-Fermi
space, and hence, this supermatrix contains all the anticommuting
parameters. One can for example choose
\begin{equation} \label{u1 supermatrix} u_1 \equiv \left(
    \begin{array}{cc} 1 - 2 \eta^\dagger \eta + 6 (\eta^\dagger \eta
      )^2 & -2 (1 - 2 \eta^\dagger \eta )\eta^\dagger \\ 2 \eta (1 - 2
      \eta^\dagger \eta) & 1 - 2 \eta \eta^\dagger + 6 (\eta
      \eta^\dagger)^2 \end{array} \right)_{BF}, \quad \eta \equiv
  \left( \begin{array}{cc} \eta_1^\ast & \eta_2 \\ \eta_2^\ast &
      \eta_1 \end{array} \right)_{TR} ,
\end{equation}
with $\eta_i, \eta_i^\ast \in \Lambda_F$, $i \in \{ 1,2 \}$, and
define $v_1$ by substituting $i\kappa_1$ for $\eta_1$ and $i\kappa_2$
for $\eta_2$. For the Efetov space with unitary symmetry, one sets
$\eta_2, \kappa_2 \to 0$, in which case $\eta \eta^\dagger \eta$ and
$\eta^\dagger \eta \eta^\dagger$ vanish, and similarly for $\kappa$
and $\kappa^\dagger$. Requiring $Q_0$ to carry the additional symmetry
$Q_0^\tau = \sigma_3^{RA} Q_0 \sigma_3^{RA}$ amounts to writing the
matrix angles $\theta_B$ and $\theta_F$ in \eqref{Q0 def} as
\begin{equation} \label{orth, thetaB thetaF} \theta_B = \left(
    \begin{array}{cc} \theta_1 & \theta_2 \\ \theta_2 & \theta_1
    \end{array} \right)_{TR}, \qquad \theta_F = \left(
    \begin{array}{cc} \theta &0 \\ 0 & \theta \end{array}
  \right)_{TR},
\end{equation}
with $\theta_1, \theta_2 > 0$ and $\theta \in [0,2\pi]$, and set
$\theta_2 \to 0$ in the unitary symmetry case. Together with the
property $U_1^\tau = \bar{U}_1$ which follows from the definitions of
$u_1$ and $v_1$ above, and from imposing $U_2^\tau = \bar{U}_2$ on
$U_2$, this symmetry implies that the required equality $Q^\tau =
\sigma_3^{RA} Q \sigma_3^{RA}$ indeed holds. There are still 2 and 5
remaining commuting parameters that have to be included in $U_2$ in
order to span the full Efetov space for unitary and orthogonal
symmetries respectively. It is not difficult to check that, for any
matrix $V$ in SU(2), and for any $\xi, \chi \in [0,2\pi]$,
\begin{equation} \label{u_2 supermatrix} u_2 \equiv \left(
    \begin{array}{cc} e^{i\xi \sigma_3^{TR}} & 0 \\ 0 & V \end{array}
  \right)_{BF} \quad \textrm{and} \quad v_2 \equiv \left(
    \begin{array}{cc} e^{i\chi \sigma_3^{TR}} & 0 \\ 0 & 1 \end{array}
  \right)_{BF}
\end{equation}
lead to a supermatrix $U_2$ in \eqref{U u v} with the required
symmetries $\bar{U}_2 = U_2^\tau = U_2^{-1}$. This finishes the
description of the Efetov space with orthogonal symmetry in terms of
polar coordinates. In the unitary symmetry case, one can take $\chi =
0$ and $V = e^{i\phi \sigma_3^{TR}}$.

\subsection{The Mean Field Autocorrelation Functions} \label{The Mean
  Field Autocorrelation Functions}

The restriction of the superintegral \eqref{exact xi superintegral}
onto the zero mode defines mean field generating functions
$\xi_{[\vect{\alpha}]}^{\textrm{MF}}(\vect{j})$ which, using \eqref{R
  def} and \eqref{Q matrix def}, and after some algebra, can be put
on the form
\begin{equation} \label{mean field generating function def}
  \xi_{[\vect{\alpha}]}^{\textrm{MF}}(\vect{j}) \equiv \int dQ \
  e^{-S_0^{\textrm{MF}}[Q]}P_{[\vect{\alpha}]}(\vect{j}),
\end{equation}
where
\begin{equation}
  S_0^{\textrm{MF}}[Q] = \frac{B \epsilon}{2} \str\  \hat{Q} + \mathcal{O}(\epsilon)
\end{equation}
is the source-free action $S_0$ at the configuration $Q$ of the zero
mode, $\hat{Q}$ denotes the supermatrix $\sigma_3^{RA}Q - \mathds{1}$,
$\hat{Q}^B$ stands for its Bose-Bose block, and
$P_{[\vect{\alpha}]}(\vect{j})$ is the supersymmetry breaking factor
\begin{equation}
  P_{[\vect{\alpha}]}(\vect{j}) = \det \left[ \mathds{1} - \frac{1}{2} \left( \begin{array}{cc} \vect{j_r}\cdot \vect{\tEp} & 0 \\ 0 & j_a \tEm \end{array} \right)_{RA}\hat{Q}^B \right]^{-\frac{1}{2}}.
\end{equation}
In this last expression $\tEm$ is the time-reversal double of
$E^{(a)}$ in \eqref{j and E definitions}, and similarly $ \vect{\tEp}$
is the vector containing the time-reversal doubles of the $q-1$
matrices entering the vector $\vect{E^{(r)}}$, and in \eqref{mean field
  generating function def}, $dQ$ is the measure $d(Z, \tZ)$ in
\eqref{exact xi superintegral} induced on the zero mode manifold.
Notice that the scattering matrix $S$ does not enter the mean field
generating function. It can indeed be seen in \eqref{Exact action}
that, after being commuted with $Z_0$ and $\tZ_0$, $S$ meets its
adjoint in the mean field action and thus disappears by unitarity.

The formula \eqref{C generated by xi} applied to the mean field
generating functions instead of the exact ones generates mean field
autocorrelation functions $C_{[\vect{\alpha}]}^\mf$, and commuting the
derivatives with respect to the sources with the superintegral in
\eqref{mean field generating function def} yields
\begin{equation} \label{autocorrelation mean field def}
  C_{[\vect{\alpha}]}^\mf = \lim_{\epsilon\to 0}
  \frac{(2\epsilon)^{q-1}}{2B(q-1)!} \int dQ \ e^{-S_0^{\textrm{MF}}}
  \delta P_{[\vect{\alpha}]},
\end{equation}
where $\delta P_{[\vect{\alpha}]}$ denotes the derivatives
\begin{equation} \label{delta q P def} \delta P_{[\vect{\alpha}]}
  \equiv \left( \prod_{s = 0}^{q-1} \frac{\partial}{\partial j_s}
  \right) P_{[\vect{\alpha}]}(\vect{0}).
\end{equation}
These derivatives can easily be calculated by means of the general
rule \eqref{der of det}. For any integer $q \geq 2$, one gets
\begin{equation} \label{delta q P beta factorized} \delta
  P_{[\vect{\alpha}]} = \frac{1}{2^q} \sum_{\sigma \in S_q}
  \rho_{\frac{1}{2}}(\sigma) \sum_{\vect{t} \in \{ \up,\down \}^q}
  F_{[\vect{\alpha}]}(\vect{t},\sigma) \pi(\vect{t},\sigma),
\end{equation}
where, for $\alpha \in \mathbb{R}$ and $\sigma \in S_q$,
$\rho_\alpha(\sigma)$ denotes the $\rho$ factor defined in \eqref{rho
  factor def app}, and for any vector $\vect{t} \in \{ \up,\down
\}^q$,
\begin{equation} \label{pi factor def} \pi(\vect{t},\sigma) \equiv
  \left\{ \begin{array}{cc} \hat{Q}^B_{aa;t_0,t_0} \prod_{j=1}^{q-1}
      \hat{Q}^B_{rr;t_j,t_{\sigma(j)}} & \textrm{if } \sigma(0)=0 \\
      \hat{Q}^B_{ar;t_0,t_{\sigma(0)}}\hat{Q}^B_{ra;t_i,t_0}
      \prod_{\begin{subarray}{c} j=1\\ j \neq i \end{subarray}}^{q-1}
      \hat{Q}^B_{rr;t_j,t_{\sigma(j)}} & \textrm{if } \sigma(i)=0 , i
      \in \mathbb{N}_{q-1}\end{array} \right.
\end{equation}
involves a product of $q$ components of the $Q$ matrix and
\begin{equation} \label{F factor def}
  F_{[\vect{\alpha}]}(\vect{t},\sigma) \equiv
  \sum_{\vect{\gamma}\in(\mathbb{N}_{2B})^{q}} \prod_{j=0}^{q-1}
  \Big[\mathcal{E}^{\alpha_j, \alpha_j} \Big]_{\begin{subarray}{c}
      \gamma_j, \gamma_{\sigma(j)} \\ t_j, t_j \end{subarray} } =
  \prod_{j=0}^{q-1} \delta_{ t_j[\alpha_j] ,
    t_{\sigma(j)}[\alpha_{\sigma(j)}]},
\end{equation}
is the $[\vect{\alpha}]$-dependent factor. In \eqref{F factor def},
the new notations $\up[\beta] = \beta$ and $\down[\beta] =
\hat{\beta}$ for a directed bond $\beta \in \mathbb{N}_{2B}$ have been
introduced. These results enable us to rewrite the mean field
autocorrelation functions \eqref{autocorrelation mean field def} as
\begin{equation} \label{mf C final formula} C_{[\vect{\alpha}]}^\mf =
  \sum_{\sigma \in S_q} \rho_{\frac{1}{2}}(\sigma) \sum_{\vect{t} \in
    \{ \up, \down \}^q } F_{[\vect{\alpha}]}(\vect{t}, \sigma)
  I_\pi(\vect{t}, \sigma),
\end{equation}
where, for any $\vect{t} = (t_0, \ldots, t_{q-1})$ in $\{ \up, \down
\}^q$, and for any $\sigma \in S_q$,
\begin{equation} \label{integral I pi def} I_\pi(\vect{t}, \sigma)
  \equiv \lim_{\epsilon \to 0} \frac{\epsilon^{q-1}}{4B(q-1)!} \int dQ
  \ e^{-S_0^\mf} \pi(\vect{t}, \sigma).
\end{equation}

The superintegrals $I_\pi(\vect{t}, \sigma)$ are mean field
superintegrals in the $\epsilon \to 0$ regime, and their values depend
on the symmetry class. In the unitary symmetry class, the measure $dQ$
reads \cite{Efetov}
\begin{equation} \label{dQ unitary} dQ = \frac{1}{2^6 \pi^2}
  \frac{d\lambda_1 d\lambda}{(\lambda_1 - \lambda)^2} d\eta_1
  d\eta_1^\ast d\kappa_1^\ast d\kappa_1 d\phi d\xi
\end{equation}
in terms of Efetov's polar coordinates, where $\lambda_1 \equiv \cosh
\theta_1$ and $\lambda \equiv \cos \theta$, and the mean field
source-free action $S_0^\mf$ is
\begin{equation} \label{S0 unitary} S_0^\mf = 2B\epsilon \big(
  \lambda_1 - \lambda \big).
\end{equation}
It can be checked, and it is stated in \cite{Mirlin}, that, in the
unitary mean field superintegral \eqref{integral I pi def}, the lowest
order term in $\epsilon$ is obtained by only retaining in
$\pi(\vect{t}, \sigma)$ its terms of highest order in $\lambda_1$ and
by replacing the expression $(\lambda_1 - \lambda)$ in \eqref{dQ
  unitary} and \eqref{S0 unitary} with $\lambda_1$. Therefore, in the
expressions
\begin{eqnarray} \label{unitary QhatB}
  \hat{Q}_{rr}^B & = & u_{1BB} u_{2B} \cosh \theta_B \bar{u}_{2B} \bar{u}_{1BB} + u_{1BF} u_{2F} \cos \theta_F \bar{u}_{2F} \bar{u}_{1FB} -1 \nonumber\\
  \hat{Q}_{aa}^B  & = & v_{1BB} v_{2B} \cosh \theta_B \bar{v}_{2B} \bar{v}_{1BB} + v_{1BF} v_{2F} \cos \theta_F \bar{v}_{2F} \bar{v}_{1FB} -1\nonumber\\
  \hat{Q}_{ra}^B & = & - u_{1BB} u_{2B} \sinh \theta_B \bar{v}_{2B} \bar{v}_{1BB} + u_{1BF} u_{2F} i \sin \theta_F \bar{v}_{2F} \bar{v}_{1FB} \nonumber\\
  \hat{Q}_{ar}^B & = & -v_{1BB} v_{2B} \sinh \theta_B \bar{u}_{2B}
  \bar{u}_{1BB} + v_{1BF} v_{2F} i \sin \theta_F \bar{u}_{2F}
  \bar{u}_{1FB}
\end{eqnarray}
for the components of $\hat{Q}^B$, which follow from \eqref{Q in Efetov
  coordinates}, \eqref{Q0 def} and \eqref{U u v}, only the first
terms of the right-hand sides contribute in the limit 
\eqref{integral I pi def}.
Moreover, for the same reason one
can replace $\sinh \theta_1$ with $\cosh \theta_1$. These remarks,
together with the formulae \eqref{u1 supermatrix}, \eqref{orth, thetaB
  thetaF} and \eqref{u_2 supermatrix}, lead to
\begin{equation} \label{unitary QB simplified} \hat{Q}^B \sim U_{2B}
  \cdot x \cdot U_{2B}^\dagger,
\end{equation}
with
\begin{equation} \label{unitary U2B} x \equiv \lambda_1 \cdot \left(
    \begin{array}{cc} |\eta|^2 & - |\eta| |\kappa| \\ - |\eta|
      |\kappa| & |\kappa|^2 \end{array} \right)_{RA}, \quad U_{2B}
  \equiv \left( \begin{array}{cc} e^{i\xi \sigma_3^{TR}} & 0 \\ 0 &
      e^{i\chi \sigma_3^{TR}} \end{array} \right)_{RA},
\end{equation}
and
\begin{equation}
  |\eta| \equiv 1 - 2 \eta_1^\ast \eta_1, \quad|\kappa| \equiv 1 + 2 \kappa_1^\ast \kappa_1
\end{equation}
In \eqref{unitary QB simplified} and henceforth, for $a$ and $b$ two
functions of the Efetov polar coordinates, the equivalence $a \sim b$
means that $b$ can be substituted for $a$ in the integrand of the mean
field integral without modifying the result.  The equivalence
\eqref{unitary QB simplified} implies that $\pi(\vect{t},\sigma)$ in
\eqref{pi factor def} satisfies
\begin{eqnarray} \label{unitary pi simplified} \pi(\vect{t}, \sigma) &
  \sim & \prod_{j=0}^{q-1} \delta_{t_j, t_{\sigma(j)}} \left\{
    \begin{array}{ll} x_{aa} x_{rr}^{q-1} & \textrm{if } \sigma(0) = 0
      \\ x_{ar} x_{ra} x_{rr}^{q-2} & \textrm{if } \sigma(i)=0 , i \in
      \mathbb{N}_{q-1}\end{array} \right. \nonumber\\ \label{unitary
    pi after second equivalence}
  & \sim & |\eta|^{2(q-1)}|\kappa|^2 \cdot \lambda_1^q \cdot  \prod_{j=0}^{q-1} \delta_{t_j, t_{\sigma(j)}}  \nonumber\\
  & \sim & 2^4 (q-1) \cdot \kappa_1 \kappa_1^\ast \eta_1^\ast \eta_1
  \cdot \lambda_1^q \cdot \prod_{j=0}^{q-1} \delta_{t_j,
    t_{\sigma(j)}}
\end{eqnarray}
The last equivalence expresses the fact that only the term containing
all the anticommuting parameters can contribute to the superintegral
\eqref{integral I pi def}. Combining \eqref{integral I pi def},
\eqref{dQ unitary} and \eqref{S0 unitary} with $(\lambda_1 - \lambda)
\to \lambda_1$, and \eqref{unitary pi simplified} together, one
arrives at
\begin{equation} \label{unitary I pi result} I_\pi(\vect{t}, \sigma) =
  \lim_{\epsilon \to 0} \frac{\epsilon^{q-1}}{2B(q-2)!} \int_0^\infty
  \ e^{-2B\epsilon \lambda_1}\lambda_1^{q-2} d\lambda_1 \cdot
  \prod_{j=0}^{q-1} \delta_{t_j, t_{\sigma(j)}} = \frac{1}{(2B)^{q}}
  \cdot \prod_{j=0}^{q-1} \delta_{t_j, t_{\sigma(j)}} .
\end{equation}
In the orthogonal symmetry class, a similar calculation leads to
$I_\pi(\vect{t}, \sigma) = (2B)^{-q}$. This result can also be
inferred from \eqref{mf C final formula}, \eqref{unitary I pi result},
and from the expectation that the mean field intensity 
correlation matrix should satisfy $C_{\alpha\alpha'}^\mf = 
C_{\hat{\alpha}\alpha'}^\mf$ and should not depend on the symmetry class 
if $\alpha$ and $\alpha'$ are supported on two different bonds.
 
By \eqref{mf C final formula} and by the results found above for
$I_\pi(\vect{t}, \sigma)$, the mean field autocorrelation functions
become
\begin{equation} \label{mean field C} C_{[\vect{\alpha}]}^\mf =
  \frac{1}{(2B)^q} \sum_{\begin{subarray}{c} \sigma \in S_q \\
      \vect{t} \in \{ \up,\down \}^q \end{subarray}}
  \rho_{\frac{1}{2}}(\sigma) \prod_{j=0}^{q-1} \left\lbrace
    \begin{array}{ll} \delta_{t_j,t_{\sigma(j)}} \delta_{\alpha_j,
        \alpha_{\sigma(j)}} & (U) \\
      \delta_{t_j[\alpha_j],t_{\sigma(j)}[\alpha_{\sigma(j)}]} & (O)
    \end{array} \right.
\end{equation}
In order to get some explicit formulae out of \eqref{mean field C},
one can for example apply the rule \eqref{der of det} once more, and
notice that
\begin{equation} \label{C beta q integral form}
  C_{[\vect{\alpha}]}^\mf = \frac{1}{(2B)^q} \left. \left(
      \prod_{k=0}^{q-1} \frac{\partial}{\partial j_k} \right)\det
    \left[ 1 - \sum_{k=0}^{q-1} j_k N(\alpha_k) \right]^{-\frac{1}{2}}
  \right|_{\vect{j} = \vect{0}}, \quad
\end{equation}
where, for $\beta \in \mathbb{N}_{2B}$ a directed bond, $N(\beta)$ is
the matrix acting on $\mathcal{A} \otimes TR$ defined by
\begin{equation} \label{N beta def}
  N(\beta) \equiv \left\{ \begin{array}{cl} E^{\beta,\beta}\otimes \mathds{1}_{TR} & (U) \\
      \left(\begin{array}{cc} E^{\beta,\beta} & E^{\beta,\beta} \\
          E^{\hat{\beta},\hat{\beta}} & E^{\hat{\beta},\hat{\beta}}
        \end{array} \right)_{TR} & (O) \end{array} \right.
\end{equation}
Let us first consider the unitary symmetry class $(U)$, and let us
characterize the list $[\vect{\alpha}]$ of directed bonds by another
list $\vect{\beta} = (\beta_1, \ldots, \beta_n)$ of distinct directed
bonds and a vector of integers $\vect{q} = (q_1, \ldots, q_n)$ such
that $\beta_j$ occurs exactly $q_j$ times in $[\vect{\alpha}]$. With
the notation $C^\mf_{\vect{\beta}}(\vect{q}) =
C_{[\vect{\alpha}]}^\mf$, an explicit calculation of the determinant
in \eqref{C beta q integral form} shows that
\begin{equation} \label{unitary mf correlations final result}
  C_{\vect{\beta}}^\mf (\vect{q}) = \frac{1}{(2B)^q} \prod_{k=1}^n
  \left. \frac{\partial^{q_k}}{\partial j_k^{q_k}} \frac{1}{1 - j_k}
  \right|_{j_k = 0} = \prod_{j = 1}^n \frac{q_j!}{(2B)^{q_j}}.
\end{equation}
Let us now consider the orthogonal symmetry class $(O)$, and let us
characterize the list $[\vect{\alpha}]$ by the list $\vect{\beta}$ of
distinct directed bonds and the vector $\vect{q}$ of integers
\begin{equation} \label{mf gen autocorr orth}
  \left\{ \begin{array}{ll} \vect{\beta} = ( \vect{\beta}_1, \ldots, \vect{\beta}_m, \beta_{m+1}, \ldots, \beta_n) \\
      \vect{q} = (\vect{q}_1, \ldots , \vect{q}_m, q_{m+1}, \ldots, q_{n}) \end{array} \right.,  \quad \left\{ \begin{array}{l} \vect{\beta}_j\equiv(\beta_j, \hat{\beta}_j) \\
      \vect{q}_j \equiv (q_j, \hat{q}_j) \end{array} \right., \ j \in
  \mathbb{N}_m
\end{equation}
such that the components of $\vect{q}$ indicate the number of
occurrences of the corresponding elements in $\vect{\beta}$. A first
inspection of the formula \eqref{C beta q integral form} shows that
$C^\mf_{\vect{\beta}}(\vect{q})$ factorizes as
\begin{equation}
  C^\mf_{\vect{\beta}}(\vect{q}) = \prod_{k = 1}^m C_{\beta_k, \hat{\beta}_k}^\mf(q_k, \hat{q}_k) \prod_{k = m+1}^n C^\mf_{\beta_k}(q_k)
\end{equation}
Then, a calculation of the determinant shows that the correlation
functions $C_{\beta_k}^\mf (q_k)$ are given by the unitary formula
\eqref{unitary mf correlations final result}, and
\begin{equation} \label{orthogonal mf two bonds int form} C_{\beta,
    \hat{\beta}}^\mf(q, \hat{q}) = \frac{1}{(2B)^{q+ \hat{q}}} \left.
    \frac{\partial^{q}}{\partial j^{q}}
    \frac{\partial^{\hat{q}}}{\partial \hat{j}^{\hat{q}}}\frac{1}{1 - j -
      \hat{j}} \right|_{j = \hat{j} = 0} = \frac{(q+
    \hat{q})!}{(2B)^{q+\hat{q}}}
\end{equation}
It can be checked that these formulae coincide precisely with the
predictions of the Gaussian Random Waves Models \eqref{gaussian model}
and \eqref{orthogonal jpdf}.

Moreover, it can be checked in \eqref{mean field C} that the
autocorrelation functions $C^{\sigma}_{[\vect{\alpha}]}$ defined at
the end of Subsection \ref{Definition and Principles} give rise to
mean field autocorrelation functions $C^{\sigma,
  \mf}_{[\vect{\alpha}]}$ that do not depend on the particular
convention $\sigma \in S_q$ chosen.

\section{The Gaussian Correction} \label{The Gaussian Correction}

\subsection{Beyond Mean Field Theory}

It is known \cite{BKS2} that not all increasing sequences of quantum
graphs are quantum ergodic, and hence, the mean field theory does not
always yield the main contributions to the autocorrelation functions.
Therefore, it is necessary to estimate the importance of the
supermatrices $Z$ and $\tZ$ lying off the zero mode manifold. For this
purpose, let us write
\begin{equation} \label{Z decomposition} Z \equiv Z_0 + \delta Z,
  \quad \tZ \equiv \tZ_0 + \delta \tZ,
\end{equation}
with $(Z_0, \tZ_0)$ on the zero mode manifold and with $(\delta Z,
\delta \tZ)$ an orthogonal deviation, and let us expand the exact
action $S[Z, \tZ]$ in \eqref{Exact action} up to second order in
$\delta Z$ and $\delta \tZ$ around $(Z_0, \tZ_0)$. The truncated
action $\tilde{S}[\delta Z, \delta \tZ, Z_0, \tZ_0]$ obtained in this
way leads to a generating function
\begin{equation} \label{truncated S and xi} \int d(Z_0, \tZ_0) \
  d(\delta Z, \delta \tZ) \ e^{-\tilde{S}[\delta Z, \delta \tZ ; Z_0,
    \tZ_0]}.
\end{equation}
Suppose for the moment that $Z$ and $\tZ$ in \eqref{truncated S and
  xi} are also required to satisfy $\tZ = Z^\tau$. It follows that
the partial traces over $\mathcal{A}$ of the supermatrices $\delta Z$
and $\delta \tZ$ in \eqref{Z decomposition} must vanish. This property
implies that the truncated action $\tilde{S}$ has no linear terms in
$\delta Z$ and $\delta \tZ$. A direct but tedious calculation shows
that, if the sources in the truncated action are set to zero, one gets
\begin{equation}
  \tilde{S}_0 \left[ \delta Z, \delta \tZ ; Z_0, \tZ_0 \right] = S_0^{\mf} \left[ Z_0, \tZ_0 \right] + S^{(2)}_0 \left[W ; \tilde{W}  \right], 
\end{equation} 
where $S_0^{\mf} [ Z_0, \tZ_0 ]$ is the mean field source-free action,
\begin{equation} \label{from Z0 and delta Z to W} W \equiv \big( 1 -
  Z_0 \tZ_0 \big)^{-\frac{1}{2}} \delta Z \big( 1 - \tZ_0 Z_0
  \big)^{-\frac{1}{2}} , \quad \tilde{W} \equiv \big( 1 - \tZ_0 Z_0
  \big)^{-\frac{1}{2}} \delta \tilde{Z} \big( 1 - Z_0 \tZ_0
  \big)^{-\frac{1}{2}},
\end{equation}
and $S_0^{(2)}[W, \tilde{W}]$ is the term of the exact source-free
action $S_0[W, \tilde{W}]$ of second order in $W$ and $\tilde{W}$
around the origin. If $e^{- \tilde{S}_0}$ was to be integrated as in
\eqref{truncated S and xi}, the changes of variables \eqref{from Z0 and
  delta Z to W}, which both have unit Jacobian, would factorize the
superintegrals over the zero mode $(Z_0, \tZ_0)$ and over the
orthogonal deviation $(W, \tilde{W})$. This factorization occurs
because the domain of the superintegral over $(W, \tilde{W})$ is
independent of $(Z_0, \tZ_0)$. Indeed, it can be readily seen that the
equality $\tilde{Z} = Z^\tau$ merely becomes $\tilde{W} = W^\tau$.
Moreover, if all the Grassmann generators of $\Lambda$ are sent to
zero, which is what really matters for the domain of a superintegral,
the color-flavor requirements \eqref{color-flavor requirements} on
$(\delta Z, \delta \tZ)$ become $\tilde{W}_{BB} = W^\dagger_{BB}$ and
$\tilde{W}_{FF} = -W^\dagger_{FF}$, and there is no further condition
concerning the eigenvalues of the positive Hermitian matrix
$W^\dagger_{BB} W_{BB}$.  Finally, the condition ensuring that
$(\delta Z, \delta \tZ)$ is orthogonal to the zero mode manifold
forces $W$ and $\tilde{W}$ to have vanishing partial traces over
$\mathcal{A}$.

By analogy with the situation described above, where the sources are
set to zero and where the supermatrix variables are constrained to
satisfy $\tilde{Z} = Z^\tau$, one defines the truncated generating
functions
\begin{equation}
  \tilde{\xi}_{[\vect{\alpha}]}(\vect{j}) \equiv \xi^\mf_{[\vect{\alpha}]}(\vect{j}) \cdot \xi^{G}_{[\vect{\alpha}]}(\vect{j}),
\end{equation}
where the Gaussian generating function is defined by
\begin{equation} \label{Gaussian generating function def}
  \xi^{G}_{[\vect{\alpha}]}(\vect{j}) \equiv \int d^{G}(W,\tilde{W}) \
  e^{-S^{(2)} [W,\tilde{W}] }
\end{equation}
and $S^{(2)} [W,\tilde{W}]$ is the term of the exact action $S[W,
\tilde{W}]$ in \eqref{Exact action} of second order in $W$ and
$\tilde{W}$ around the origin, namely
\begin{equation} \label{2nd order action} S^{(2)}[W, \tilde{W}] = \str
  \left( W \tilde{W} - \frac{1}{2} \tJp \tilde{W}^\tau \tJm \tilde{W}
    - \frac{1}{2} W \tSed W^\tau \tSe \right).
\end{equation}
The integration in \eqref{Gaussian generating function def} is over
all supermatrices $W$ and $\tilde{W}$ in $L(TR \otimes \mathcal{A}| TR
\otimes \mathcal{A})$ that are diagonal in $\mathcal{A}_b$, define a
configuration $(W, \tilde{W})$ orthogonal to the zero mode manifold,
and satisfy the color-flavor conditions $\tilde{W}_{BB} =
W^\dagger_{BB}$ and $\tilde{W}_{FF} = -W^\dagger_{FF}$. The measure
$d^{G}(W,\tilde{W})$ is then the product of the flat Berezin measures
over the independent components of $W$ and $\tilde{W}$.

In \eqref{fundamental property}, it is claimed that if either the
advanced or the retarded sources are sent to zero, the exact
generating function becomes identically equal to one in a neighborhood
of the origin. It can be checked that the same property holds
separately for the mean field and the Gaussian generating functions.
Therefore, if the formula \eqref{C generated by xi} is used to define
truncated autocorrelation functions $\tilde{C}_{[\vect{\alpha}]}$ from
$\tilde{\xi}_{[\vect{\alpha}]}$, one gets
\begin{equation} \label{truncated autocorrelations def}
  \tilde{C}_{[\vect{\alpha}]} = C_{[\vect{\alpha}]}^\mf +
  C_{[\vect{\alpha}]}^{G},
\end{equation}
where $C_{[\vect{\alpha}]}^\mf $ are the mean field autocorrelation
functions found in the previous section, and
\begin{equation}
  C_{[\vect{\alpha}]}^{G} =  \lim_{\epsilon\to 0} \frac{(2\epsilon)^{q-1}}{2B(q-1)!}\delta \xi_{[\vect{\alpha}]}^{G}, \quad \delta \xi_{[\vect{\alpha}]}^{G} \equiv \left( \prod_{s = 0}^{q-1} \frac{\partial}{\partial j_s}\right) \xi^{G}_{[\vect{\alpha}]}(\vect{0}),
\end{equation}
are the Gaussian autocorrelation functions.

In fact, in order to calculate the Gaussian generating functions
\eqref{Gaussian generating function def}, one can first calculate the
second order generating function
\begin{equation} \label{Second order generating function def}
  \xi^{(2)}_{[\vect{\alpha}]}(\vect{j}) \equiv \int
  d^{(2)}(Z,\tilde{Z}) \ e^{-S^{(2)} [Z,\tilde{Z}] },
\end{equation}
defined from \eqref{Gaussian generating function def} by relaxing the
constraint that $(Z, \tilde{Z})$ must be orthogonal to the zero mode,
and then divide by the second order mean field generating function
\begin{equation} \label{Second order mean field generating function
    def} \xi^{\mf (2)}_{[\vect{\alpha}]}(\vect{j}) \equiv \int d^{\mf
    (2)}(Y,\tilde{Y}) \ e^{-S^{\textrm{MF} (2)} [Y,\tilde{Y}] },
\end{equation}
which contains the zero mode contribution to \eqref{Second order
  generating function def}. With the notations $Z_0 =
\mathds{1}_\mathcal{A} \otimes Y$ and $\tZ_0 = \mathds{1}_\mathcal{A}
\otimes \tY$ for the supermatrix variables in the zero mode, $d^{\mf
  (2)}(Y,\tilde{Y})$ is the measure induced by $d^{(2)}(Z,\tilde{Z})$
on the zero mode manifold. Similarly, the mean field second order
action $S^{\textrm{MF}(2)}$ is obtained by restricting $S^{(2)}$, that
is
\begin{equation} \label{second order mf action} S^{\mf(2)}[Y,
  \tilde{Y}] = \frac{1}{2} \str \left( (2 - e^{-2\epsilon})
    \mathds{1}_{\mathcal{A}} \otimes Y\tY - \tJp Y \tJm \tY \right).
\end{equation}
The generating functions \eqref{Second order generating function def}
and \eqref{Second order mean field generating function def} are
identically equal to one in a neighborhood of the origin if either the
advanced or the retarded sources are set to zero. Therefore, the
Gaussian autocorrelation functions can be written
\begin{equation} \label{Gaussian C from xi2 and xi2 mf}
  C_{[\vect{\alpha}]}^{G} = \lim_{\epsilon\to 0}
  \frac{(2\epsilon)^{q-1}}{2B(q-1)!} \left( \delta
    \xi_{[\vect{\alpha}]}^{(2)} - \delta \xi_{[\vect{\alpha}]}^{\mf
      (2)} \right)
\end{equation}
with the obvious definitions for $\delta \xi_{[\vect{\alpha}]}^{(2)}$
and $\delta \xi_{[\vect{\alpha}]}^{\mf (2)}$.

\subsection{Diagonal Modes in Direction Space} \label{Diagonal Modes
  in Direction Space}

Let us first only consider the subset of supermatrix variables $(Z,
\tZ)$ that are diagonal in the whole amplitude space $\mathcal{A} =
\mathcal{A}_b \otimes \mathcal{A}_d$. If $d^{(2)}_{(dd)}(Z,\tilde{Z})$
denotes the measure induced from $d^{(2)}(Z,\tilde{Z})$ on this
subset, the goal is to calculate the $(dd)$ second order generating
functions
\begin{equation} \label{Second order (dd) generating function def}
  \xi^{(2)}_{[\vect{\alpha}], (dd)}(\vect{j}) \equiv \int
  d^{(2)}_{(dd)}(Z,\tilde{Z}) \ e^{-S^{(2)} [Z,\tilde{Z}] }.
\end{equation}
The diagonal modes $Z$ and $\tZ$ are parametrized in time-reversal
space as follows,
\begin{equation} \label{TR space structure}
  Z = \left( \begin{array}{cc} Z_{1} & Z_{2}^\dagger \\
      Z_{3}^\mathcal{T} \sigma_{3}^{\textrm{BF}} & Z_{4}^{\dagger
        \mathcal{T}} \end{array} \right) \quad \textrm{and} \quad \tZ
  = \left( \begin{array}{cc} \tZ_{1} & \sigma_{3}^{\textrm{BF}}
      \tZ_{3}^\mathcal{T} \\ \tZ_{2}^\dagger & \tZ_{4}^{\dagger
        \mathcal{T}} \end{array} \right),
\end{equation}
and their generalized transposes read
\begin{equation} \label{TR space gen transp} Z^\tau = \left(
    \begin{array}{cc} Z_{4}^\dagger & \sigma_{3}^{\textrm{BF}}
      Z_{2}^{\dagger \mathcal{T}} \\ Z_{3} & Z_{1}^\mathcal{T}
    \end{array} \right) \quad \textrm{and} \quad
  \tZ^\tau = \left( \begin{array}{cc} \tZ_{4}^\dagger & \tZ_{3} \\
      \tZ_{2}^{\dagger \mathcal{T}} \sigma_{3}^{\textrm{BF}} &
      \tZ_{1}^\mathcal{T} \end{array} \right).
\end{equation}
The modes $Z_2^\dagger$, $\tZ_2^\dagger$, $Z_3$ and $\tZ_3$  are only 
considered if time-reversal invariance is conserved, so 
that $Z$, $\tilde{Z}$ and their generalized transposes all 
become diagonal in time-reversal space. When these formulae are
substituted into $S^{(2)}[Z, \tilde{Z}]$ given by \eqref{2nd order
  action}, the diagonal modes in time-reversal space, which are
indexed by 1 and 4, are coupled together, and do not mix with the
off-diagonal ones indexed by 2 and 3. After some algebra, one finds
\begin{equation}
  S_{(dd)}^{(2)} = S^{(2)D}_{(dd)} + (\kappa - 1) S^{(2)C}_{(dd)}
\end{equation}
with the diffusion action $S^{(2)D}_{(dd)}$ and the cooperon action
$S^{(2)C}_{(dd)}$ defined by
\begin{eqnarray} \label{S2 diff} S^{(2)D}_{(dd)} & = & \str\left(
    Z_{1}\tZ_{1} + Z_{4}^\dagger \tZ_{4}^\dagger - J_r \tZ_{4}^\dagger
    J_a \tZ_{1} - Z_{1} \Sed Z_{4}^\dagger \Se \right) \\ \label{S2
    coop} S^{(2)C}_{(dd)} & = & \str\left( Z_{2}^\dagger
    \tZ_{2}^\dagger + Z_{3}\tZ_{3} - J_r \tZ_{3} J_{a}^{\mathcal{T}}
    \tZ_{2}^\dagger - Z_{2}^\dagger \Sedtau Z_{3} \Se \right)
\end{eqnarray}
Hence, the generating functions $\xi_{[\vect{\alpha}],(dd)}^{(2)}$
factorize as
\begin{equation} \label{xi 2 facorization}
  \xi_{[\vect{\alpha}],(dd)}^{(2)} = \xi_{ [\vect{\alpha}]
    ,(dd)}^{(2)D} \cdot \left( \xi_{ [\vect{\alpha}],(dd)}^{(2)C}
  \right)^{\kappa - 1}
\end{equation}  
where, for $\circ \in \{ D,C \}$,
\begin{equation} \label{xi 2 dd circ superintegral}
  \xi_{[\vect{\alpha}],(dd)}^{(2)\circ}(\vect{j}) = \int
  d^{(2)\circ}_{(dd)} (Z,\tZ)\ e^{-S^{(2)\circ}_{(dd)}[Z, \tZ]}.
\end{equation}
In \eqref{xi 2 dd circ superintegral}, the diffusion measure is the
product of the Berezin measures of the independent components in the
supermatrix variables $Z_1$, $Z_4^\dagger$, $\tZ_1$ and
$\tZ_4^\dagger$, and the cooperon measure is similarly formed with the
independent components in $Z_2^\dagger$, $Z_3$, $\tZ_2^\dagger$ and
$\tZ_3$.

Notice that the cooperon generating functions \eqref{xi 2 dd circ
  superintegral}, which only exist if $S^\mathcal{T} = S$, can be
obtained from the corresponding diffusion generating functions by
replacing $J_a$ with $J_a^\mathcal{T}$. One can thus focus on the
diffusion generating functions and infer the cooperon result from this
remark.

In order to perform the diffusion and cooperon integrals in \eqref{xi 2
  dd circ superintegral}, the supertraces in \eqref{S2 diff} and
\eqref{S2 coop} must be explicitly expanded in Bose-Fermi space.
Notice that the color-flavor conditions on the Bose-Bose and
Fermi-Fermi components of $Z$ and $\tZ$ in \eqref{TR space structure}
become $\tZ_{jBB} = Z_{jBB}^\dagger = Z_{jBB}^\ast$ and $\tZ_{jFF} = -
Z_{jFF}^\dagger = Z_{jFF}^\ast$ for all $j \in \mathbb{N}_4$. Let us
define for each directed bond $\beta \in \mathbb{N}_{2B}$
\begin{equation} \label{z commuting}
  \begin{array}{ccc}
    \tilde{z}_{1 \bar{0} \beta} =  (Z_{1BB}^\ast \, , \, Z_{1FF}^\ast)_\beta &, & z_{1 \bar{0} \beta} = \left(\begin{array}{c} Z_{1BB} \\ Z_{1FF} \end{array}\right)_\beta, \\
    z_{4 \bar{0} \beta}^{\dagger} =  (Z_{4BB}^\ast \, , \,  -Z_{4FF}^\ast)_\beta &, & \tilde{z}_{4 \bar{0} \beta}^\dagger = \left(\begin{array}{c} Z_{4BB} \\  - Z_{4FF} \end{array}\right)_\beta.
  \end{array}
\end{equation}
The vectors $\tilde{z}_{1 \bar{0} \beta}$ and $z_{1 \bar{0} \beta}$
contain the commuting parameters of $\tZ_{1\beta}$ and $Z_{1\beta}$
respectively, and the vectors $z_{4 \bar{0} \beta}^{\dagger}$ and
$\tilde{z}_{4 \bar{0} \beta}^\dagger$ contain those of
$Z_{4\beta}^\dagger$ and $\tZ_{4\beta}^\dagger$. Similarly, the
anticommuting variables of the diffusion action are arranged in the
vectors
\begin{equation} \label{z anticommuting}
  \begin{array}{ccc}
    \tilde{z}_{1 \bar{1} \beta} =  (\tZ_{1BF} \, , \, \tZ_{1FB})_\beta &, & z_{1 \bar{1} \beta} = \left(\begin{array}{c} Z_{1BF} \\ Z_{1FB} \end{array}\right)_\beta, \\
    z_{4 \bar{1} \beta}^{\dagger} =  (Z_{4BF}^\ast \, , \,  Z_{4FB}^\ast)_\beta &, & \tilde{z}_{4 \bar{1} \beta}^\dagger = \left(\begin{array}{c} \tZ_{4BF}^\ast \\  \tZ_{4FB}^\ast \end{array}\right)_\beta.
  \end{array}
\end{equation}
Collecting the $2B$ row-vectors $\tilde{z}_{1 \bar{0}\beta}$ (resp.
$z_{4 \bar{0}\beta}^\dagger$) together, one can write a larger row
vector $\tilde{z}_{1 \bar{0}}$ (resp. $z_{4 \bar{0}}^\dagger$). The
column-vectors $z_{1\bar{0}}$ and $\tilde{z}_{4 \bar{0}}^\dagger$ are
formed similarly from $z_{1\bar{0} \beta}$ and $\tilde{z}_{4 \bar{0}
  \beta}^\dagger$, and one proceeds in the same way with the
anticommuting variables in \eqref{z anticommuting}.  Let us also
introduce a $2B \times 2B$ matrix $s$ defined from the Bose-Bose
blocks of the source supermatrices $J_{a}$ and $J_r$ by
\begin{equation} \label{s diffusion} s(\vect{j})_{\beta\beta'} \equiv
  J_{a}(j_a)_{BB,\beta'\beta}J_r(\vect{j_r})_{BB,\beta\beta'}.
\end{equation}
A direct expansion of \eqref{S2 diff} in Bose-Fermi space then leads
to
\begin{equation} \label{S2 diff comm} S^{(2)D}_{(dd)\bar{0}} = (
  \tilde{z}_{1 \bar{0}} \, , \, z_{4 \bar{0}}^\dagger) \left(
    \begin{array}{cc} \mathds{1}_{\mathcal{A}}\otimes
      \mathds{1}_{2\times 2} & -\left( \begin{smallmatrix}
          s(j_a,\vect{j_r}) & \\ & s(0,\vect{0}) \end{smallmatrix}
      \right) \\ -M_\epsilon \otimes \mathds{1}_{2\times 2} &
      \mathds{1}_{\mathcal{A}}\otimes \mathds{1}_{2\times 2}
    \end{array} \right) \left( \begin{array}{c} z_{1 \bar{0}} \\
      \tilde{z}_{4 \bar{0}}^\dagger \end{array} \right)
\end{equation}
for the part of $S^{(2)D}_{(dd)}$ involving the commuting variables,
and
\begin{equation} \label{S2 diff anticomm} S^{(2)D}_{(dd)\bar{1}} = (
  \tilde{z}_{1 \bar{1}} \, , \, z_{4 \bar{1}}^\dagger) \left(
    \begin{array}{cc} \mathds{1}_{\mathcal{A}}\otimes \left(
        \begin{smallmatrix} 0 & 1 \\ -1 & 0 \end{smallmatrix} \right)
      & \left( \begin{smallmatrix} s(j_a,\vect{0}) & \\ &
          s(0,\vect{j_r}) \end{smallmatrix} \right) \\ M_\epsilon
      \otimes \mathds{1}_{2\times 2} & \mathds{1}_{\mathcal{A}}
      \otimes \left( \begin{smallmatrix} 0 & 1 \\ -1 & 0
        \end{smallmatrix} \right) \end{array} \right) \left(
    \begin{array}{c} z_{1 \bar{1}} \\ \tilde{z}_{4 \bar{1}}^\dagger
    \end{array} \right)
\end{equation}
for the part involving the anticommuting variables. Notice that these
formulae depend on the scattering matrix $S_\epsilon = e^{-\epsilon}
S$ only through the classical map $M_\epsilon = e^{-2\epsilon}M$ it
generates. It is straightforward to calculate the diffusion
superintegral \eqref{xi 2 dd circ superintegral} from these quadratic
forms, and the cooperon generating functions are found by substituting
the matrix $J_a^\mathcal{T}$ for the matrix $J_a$ in the diffusion
results. If $\circ \in \{ D,C \}$, one gets
\begin{equation} \label{xi 2 diff}
  \xi_{[\vect{\alpha}],(dd)}^{(2)\circ} (j_a,\vect{j_r}) = \frac{\det
    \Big( \mathds{1}_{\mathcal{A}} - s^\circ(j_a,\vect{0})M_\epsilon
    \Big) \det \Big( \mathds{1}_{\mathcal{A}} -
    s^\circ(0,\vect{j_r})M_\epsilon \Big)}{\det \Big(
    \mathds{1}_{\mathcal{A}} - M_\epsilon \Big) \det \Big(
    \mathds{1}_{\mathcal{A}} -s^\circ(j_a,\vect{j_r}) M_\epsilon
    \Big)},
\end{equation}
where $s^D \equiv s$ in \eqref{s diffusion}, and $s^C$ is obtained
from $s$ by replacing $J_a$ with $J_a^\mathcal{T}$.

In order to unveil the mean field contribution to \eqref{xi 2 diff},
one restricts the superintegral in \eqref{Second order (dd) generating
  function def} to the zero mode $Z_0 = \mathds{1}_\mathcal{A}
\otimes Y$, $\tZ_0 = \mathds{1}_\mathcal{A} \otimes \tY$. The
supermatrices $Y$ and $\tY$ are required to satisfy $\tY = Y^\tau$,
and can thus be parametrized by
\begin{equation} \label{mean field Y parametrization in TR} Y = \left(
    \begin{array}{cc} \Yd & \Yc \\ \tYctau \sigma_3^{BF} & \tYdtau
    \end{array} \right) \quad \textrm{and} \quad \tY = \left(
    \begin{array}{cc} \tYd & \sigma_3^{BF}\Yctau \\ \tYc & \Ydtau
    \end{array} \right)
\end{equation}
in time-reversal space. Then, the second order mean field action
\eqref{second order mf action} splits into a diffusion part,
containing the supermatrices $\Yd$ and $\tYd$, and a cooperon part,
involving $\Yc$ and $\tilde{Y}_C$. One gets
\begin{eqnarray} \label{S2 mean field diffusion}
  S^{\mf(2)D} & = & \str \left( \left(2 - e^{-2\epsilon} \right) \Yd \tYd \otimes \mathds{1}_{\mathcal{A}}  - J_r \Yd J_a \tYd \right) \\
  S^{\mf(2)C} & = & \str \left( \left(2 - e^{-2\epsilon} \right) \Yc
    \tYc \otimes \mathds{1}_{\mathcal{A}} - J_r \Yc J_a^{\mathcal{T}}
    \tYc \right).
\end{eqnarray}
It follows that the mean field contribution to \eqref{xi 2 diff}
factorizes into a diffusion and a cooperon factor, as in \eqref{xi 2
  facorization}.  The expressions \eqref{S2 mean field diffusion} can
be developed in Bose-Fermi space, and the resulting quadratic forms
have inverse superdeterminant
\begin{equation} \label{mean field second order xi 2 deg}
  \xi^{\mf(2)\circ}_{[\vect{\alpha}]}(j_a,\vect{j_r}) = \frac{\Big(
    1-e^{-2\epsilon} - \sigma^\circ(j_a,\vect{0}) \Big) \Big(
    1-e^{-2\epsilon} - \sigma^\circ(0,\vect{j_r}) \Big) }{ \Big(
    1-e^{-2\epsilon} \Big) \Big( 1-e^{-2\epsilon} -
    \sigma^\circ(j_a,\vect{j_r}) \Big)}, \quad \circ \in \{ D,C \},
\end{equation}
where
\begin{equation}
  \sigma^\circ(j_a,\vect{j_r}) \equiv \frac{1}{2B}\sum_{\beta, \beta' = 1}^{2B} s^\circ(j_a,\vect{j_r})_{\beta\beta'} - 1.
\end{equation}
Notice that no index $(dd)$ has been added in the left-hand side of
\eqref{mean field second order xi 2 deg}. The reason is that the zero
mode supermatrices $Z_0$ and $\tZ_0$ are always diagonal in the
direction space $\mathcal{A}_d$, and hence \eqref{mean field second
  order xi 2 deg} is also the mean field contribution to the full
second order generating functions \eqref{Second order generating
  function def}.

The diffusion and cooperon generating functions in \eqref{xi 2 diff}
and \eqref{mean field second order xi 2 deg} become identically one in
a neighborhood of the origin if either the advanced or the retarded
sources are set to zero. Hence, the $(dd)$ Gaussian autocorrelation
functions defined as in \eqref{Gaussian C from xi2 and xi2 mf} from
the $(dd)$ second order generating functions \eqref{xi 2 facorization}
split
\begin{equation} \label{dd Gaussian C splitting} C^{G}_{[\alpha],
    (dd)} = C^{G, D}_{[\alpha], (dd)} + (\kappa - 1) C^{G,
    C}_{[\alpha], (dd)},
\end{equation}
where the diffusion and cooperon autocorrelation functions in the
right-hand side are defined by the formula \eqref{Gaussian C from xi2
  and xi2 mf} applied to the diffusion and cooperon versions of
\eqref{xi 2 diff}, that is
\begin{equation} \label{Gaussian C diff and coop from xi2 and xi2 mf}
  C_{[\vect{\alpha}], (dd)}^{G \circ} \equiv \lim_{\epsilon\to 0}
  \frac{(2\epsilon)^{q-1}}{2B(q-1)!} \left( \delta
    \xi_{[\vect{\alpha}],(dd)}^{(2)\circ} - \delta
    \xi_{[\vect{\alpha}]}^{\mf (2)\circ} \right), \quad \circ \in \{
  D,C \}.
\end{equation}
 
The next step towards the calculation of \eqref{Gaussian C diff and
  coop from xi2 and xi2 mf} is to calculate the derivatives of the
diffusion and cooperon $(dd)$ second order generating functions
\eqref{xi 2 diff}. Performing explicitly the unique advanced
derivative, one easily finds that for all integers $q\geq 2$, and for
all sets of $q$ directed bonds $[\vect{\alpha}]$,
\begin{equation} \label{advanced derivative done} \delta
  \xi_{[\vect{\alpha}], (dd)}^{(2)\circ} = \prod_{s = 1}^{q-1} \left.
    \frac{\partial}{\partial j_s} \tr \left[ \frac{\partial
        s^\circ}{\partial j_a}(0, \vect{j_r}) M_\epsilon \frac{1}{1 -
        s^\circ (0, \vect{j_r}) M_\epsilon} \right]
  \right|_{\vect{j_r} = \vect{0}},
\end{equation}  
where $\vect{j_r} \equiv (j_1, \ldots, j_{q-1})^T$ are the $q-1$
retarded sources.

Let us first consider the situation $q = 2$ and $[\vect{\alpha}] =
[\alpha, \alpha']$. Then, the unique retarded derivative in
\eqref{advanced derivative done} can be performed, and one gets
\begin{equation} \label{delta xi2 circ with s}
  \delta\xi^{(2)\circ}_{[\alpha,\alpha'],(dd)} = \tr \left[
    \frac{M_{\epsilon}}{1-M_{\epsilon}} s^\circ_{r}
    \frac{M_{\epsilon}}{1-M_{\epsilon}} s^\circ_{a} +
    \frac{M_{\epsilon}}{1-M_{\epsilon}} s^\circ_{ra} \right],
\end{equation}
where $s^\circ_{r}$, $s^\circ_{a}$ and $s^\circ_{ra}$ respectively
denote the derivatives of $s^\circ$ with respect to $j_r$, $j_a$, and
$j_{r}$ and $j_a$, all evaluated at $j_r = j_a = 0$. If these
derivatives are calculated using the parallel convention for $J_a$ and
$J_r$, the diffusion $(dd)$ derivatives take the form
\begin{equation} \label{delta 1 1 xi D}
  \delta\xi^{(2)D}_{[\alpha,\alpha'],(dd)} = \delta_{\alpha,\alpha'}
  \left( \frac{M_\epsilon}{1 - M_\epsilon} \right)_{\alpha\alpha} +
  \left( \frac{M_\epsilon}{1 - M_\epsilon} \right)_{\alpha\alpha'}
  \left( \frac{M_\epsilon}{1 - M_\epsilon} \right)_{\alpha'\alpha},
\end{equation}
and the cooperon $(dd)$ derivatives is given by the same formula with
$\hat{\alpha}$ in place of $\alpha$. Notice that this expression
agrees with the derivatives 
 \eqref{derivatives new xi last formula} 
and \eqref{derivatives new xi last formula cooperon}
of the diagonal approximation to the trace formula
for the generating function in Section~\ref{Diagonal
  Approximation}. If the derivatives of $s^\circ$ found with the
crossed convention are plugged into \eqref{delta xi2 circ with s}, one
gets the diffusion $(dd)$ derivatives
\begin{equation} \label{delta 1 1 xi D crossed} \delta\xi^{\times
    (2)D}_{[\alpha,\alpha'],(dd)} \equiv \left( \frac{M_\epsilon}{1 -
      M_\epsilon} \right)_{\alpha\alpha'} +
  \delta_{\alpha,\alpha'}\left( \frac{M_\epsilon}{1 - M_\epsilon}
  \right)_{\alpha\alpha} \left( \frac{M_\epsilon}{1 - M_\epsilon}
  \right)_{\alpha\alpha}
\end{equation}
and the corresponding expression with $\hat{\alpha}$ in place of
$\alpha$ for the cooperon $(dd)$ derivatives. This is again the result
obtained in \eqref{derivatives new xi last formula} 
and \eqref{derivatives new xi last formula cooperon}
for the
derivatives of the diagonal approximation to
the generating function. The derivatives of
the second order mean field generating function \eqref{mean field
  second order xi 2 deg} have to be removed from the previous
formulae. They read
\begin{equation} \label{delta 1 1 xi 2 mf} \delta
  \xi^{\mf(2)\circ}_{[\alpha,\alpha']} = \left( \frac{1}{1 -
      e^{-2\epsilon}} \right)^2 \sigma_r \sigma_a + \left( \frac{1}{1
      - e^{-2\epsilon}} \right) \sigma_{ra},
\end{equation}
where the indices $r$ and $a$ denote the derivatives taken on
$\sigma$, which are all evaluated at the origin. These derivatives can
be calculated according to the parallel or the crossed conventions.
The results obtained are given by the formulae \eqref{delta 1 1 xi D}
and \eqref{delta 1 1 xi D crossed} by systematically replacing the sum
of classical walks $M_\epsilon(1 - M_\epsilon)^{-1}$ with the uniform
component of $(1 - M_\epsilon)^{-1}$, which is defined as in
\eqref{classical paths decomposition} and reads $(1 - e^{-2
  \epsilon})^{-1} |1 \rangle \langle 1 |$. This draws a parallel
between the zero mode and the uniform component. Finally, when the
formula \eqref{Gaussian C diff and coop from xi2 and xi2 mf} is
applied, the terms in \eqref{delta 1 1 xi D} and \eqref{delta 1 1 xi D
  crossed} that are too singular in $\epsilon$ are exactly
compensated by the mean field derivatives, and one is left with the
finite result
\begin{equation} \label{delta 11 xi diff first} C^{G,D}_{\alpha,
    \alpha',(dd)} = \frac{R_{\alpha \alpha'} +
    R_{\alpha'\alpha}}{(2B)^2} - \frac{2}{(2B)^3} \quad \textrm{and}
  \quad C^{\times G,D}_{\alpha \alpha',(dd)} = \delta_{\alpha,
    \alpha'} C^{G,D}_{\alpha \alpha',(dd)}
\end{equation}
for the parallel and crossed diffusion $(dd)$ Gaussian 
intensity correlation matrix. In these formulae, the matrix $R$ denotes
the massive component defined by the decomposition \eqref{classical
  paths decomposition}. The cooperon contributions to the
Gaussian intensity correlation
matrix read
\begin{equation} \label{delta 11 xi coop first} C^{G,C}_{\alpha
    \alpha',(dd)} = \frac{R_{\hat{\alpha} \alpha'} + R_{\alpha'
      \hat{\alpha}}}{(2B)^2} - \frac{2}{(2B)^3} \quad \textrm{and}
  \quad C^{\times G,C}_{\alpha \alpha',(dd)} = \delta_{\alpha,
    \alpha'} C^{G,C}_{\alpha \alpha',(dd)}
\end{equation}

The surprising second terms of the formulae in \eqref{delta 11 xi diff
  first} and \eqref{delta 11 xi coop first} originate from the fact
that the zero mode contribution to \eqref{delta 1 1 xi D} and
\eqref{delta 1 1 xi D crossed} is the uniform component of $(1 -
M_\epsilon)^{-1}$ and not $M_\epsilon(1 - M_\epsilon)^{-1}$. This
discrepancy is due to the additional symmetry $\tY = Y^\tau$ of the
zero mode. Notice however that these second terms are of higher order
in $(2B)^{-1}$, and are thus of minor importance when the large graph
limit is considered.

Using the same strategy as above, it is not difficult to calculate
more retarded derivatives in \eqref{advanced derivative done} and to
remove the mean field contributions. If $[\vect{\alpha}]$ is a list of
$q \geq 2$ directed bonds, and if the convention $\sigma = \textrm{id}
\in S_q$ for the generating function \eqref{xi 2 diff} is chosen, one
gets the formula
\begin{equation} \label{C beta q dd D final}
  C_{[\vect{\alpha}],(dd)}^{G,D} = \frac{1}{(q-1)(2B)^{q}}
  \sum_{\begin{subarray}{c} k,l = 0 \\ k \neq l
    \end{subarray}}^{q-1}R_{\alpha_k \alpha_l} -\frac{q}{(2B)^{q+1}}
\end{equation}
for the diffusion $(dd)$ Gaussian autocorrelation function of degree
$q$, and the same formula with $\hat{\alpha}_0$ in place of $\alpha_0$
for the cooperon $(dd)$ Gaussian autocorrelation function of degree
$q$.

\subsection{Off-Diagonal Modes in Direction Space} \label{Off-Diagonal
  Modes in Direction Space}

Let us now investigate the full second order generating functions
taking into account the modes $Z$ and $\tZ$ that are off-diagonal in
direction space. The parametrizations \eqref{TR space structure} of
$Z$ and $\tZ$ in time-reversal space can be kept, and hence the
formulae \eqref{S2 diff} and \eqref{S2 coop} also hold in the presence
of off-diagonal modes. This implies in particular that the second
order generating functions factorize into diffusion and cooperon
generating functions as in \eqref{xi 2 facorization}, and that the
cooperon formulae, which are considered only if time-reversal invariance is
conserved, can be found from their diffusion counterparts by replacing
$J_a(j_a)$ with $J_a(j_a)^\mathcal{T}$. One can thus temporarily
concentrate on the diffusion modes only.

One can distinguish between diagonal and off-diagonal modes and
introduce the notations
\begin{equation} \label{Z and tZ doubling} Z^{\textrm{diag}}_\beta
  \equiv Z_{\beta\beta} \quad \textrm{and} \quad
  Z^{\textrm{off}}_\beta \equiv Z_{\beta\hat{\beta}},
\end{equation}
and similarly for $\tZ$. The quadratic action couples diagonal modes
with themselves, which is precisely the part treated in the previous
subsection, off-diagonal modes with themselves, and diagonal modes
with off-diagonal modes.

The integration scheme used here is similar to the one that leads to
the explicit formula \eqref{xi 2 diff} for the $(dd)$ Gaussian
generating functions in terms of four determinants. Let us first focus
on the commuting components $Z_{jss}$, $s \in \{B,F \}$, of the
fields.  The row and column vectors defined in \eqref{z commuting},
whose purpose is to write the diagonal action $S^{(2)D}_{(dd)}$ as a
quadratic form, are adapted to the situation where the fields $Z$ and
$\tZ$ are as in \eqref{Z and tZ doubling}. Let us define
\begin{equation} \label{z commuting off} \tilde{w}_\beta =
  \left(\tilde{z}_{1 \bar{0}}^\textrm{diag}, \, z_{4
      \bar{0}}^{\dagger\textrm{diag}}, \, \tilde{z}_{1
      \bar{0}}^\textrm{off}, \, z_{4
      \bar{0}}^{\dagger\textrm{off}}\right)_\beta
\end{equation}
where $\tilde{z}_{1 \bar{0}}^\textrm{diag}$ and $z_{4
  \bar{0}}^{\dagger\textrm{diag}}$ are formed with the diagonal modes
of $\tilde{z}_{1 \bar{0}}$ and $z_{4 \bar{0}}^{\dagger}$ defined in
\eqref{z commuting}, and $\tilde{z}_{1 \bar{0}}^\textrm{off}$ and
$z_{4 \bar{0}}^{\dagger\textrm{off}}$ are formed with the off-diagonal
ones. We proceed in the same way with the column vectors and introduce
\begin{equation}
  w_\beta = \left( z_{1 \bar{0}}^{\textrm{diag}\ T} , \, \tilde{z}_{4 \bar{0} \beta}^{\dagger\textrm{diag} \ T}, \,  z_{1 \bar{0}}^{\textrm{off} \ T}, \, \tilde{z}_{4 \bar{0} \beta}^{\dagger\textrm{off} \ T} \right)_\beta^T.
\end{equation}
Then, a careful inspection of the diffusion second order action
\eqref{S2 diff} and some algebra show that the part of this action
involving the commuting variables is the quadratic form
\begin{equation} \label{quadratic commuting action} S^{(2)D}_{\bar{0}}
  = \sum_{\beta, \beta' = 1}^{2B} \tilde{w}_\beta
  \mathcal{B}_{\beta\beta'} w_{\beta'},
\end{equation}
defined by the $16B \times 16B$ matrix
\begin{equation} \label{S2 diff comm off quad form final} \mathcal{B}
  = \left( \begin{array}{cc|cc} \mathds{1}_{\mathcal{A}} \otimes
      \mathds{1}_{2\times 2} & - \left( \begin{smallmatrix}
          s(\vect{j}) \\ & \mathds{1}_{\mathcal{A}}\end{smallmatrix}
      \right) &
      0 & - \left( \begin{smallmatrix} a(\vect{j}) \\ & 0 \end{smallmatrix} \right) \\
      - M_{\epsilon} \otimes\mathds{1}_{2\times 2} &
      \mathds{1}_{\mathcal{A}} \otimes \mathds{1}_{2\times 2} & -
      P_\epsilon \cdot \mathds{1}_{2 \times 2} & 0 \\ \hline 0 & -
      \left( \begin{smallmatrix} b(\vect{j}) \\ & 0 \end{smallmatrix}
      \right) &
      \mathds{1}_{\mathcal{A}} \otimes \mathds{1}_{2\times 2} & - \left( \begin{smallmatrix} c(\vect{j})  \\ &  \mathds{1}_{\mathcal{A}} \end{smallmatrix} \right) \\
      - Q_\epsilon \cdot \mathds{1}_{2 \times 2} & 0 & -
      K_{\epsilon}\otimes\mathds{1}_{2\times 2} &
      \mathds{1}_{\mathcal{A}} \otimes \mathds{1}_{2\times 2}
    \end{array}\right).
\end{equation}
Notice that the block coupling the diagonal modes together is
precisely \eqref{S2 diff comm}. In the quadratic form \eqref{S2 diff
  comm off quad form final}, the $2B \times 2B$ matrices
$P_\epsilon$, $Q_\epsilon$ and $K_\epsilon$ are defined by
\begin{equation}
  P_{\epsilon \beta \beta'} \equiv S_{\epsilon \beta \beta'} S_{\epsilon \beta \hat{\beta}'}^\ast, \quad Q_{\epsilon \beta \beta'} \equiv S_{\epsilon \beta \beta'} S_{\epsilon \hat{\beta} \beta'}^\ast \quad \textrm{and} \quad K_{\epsilon \beta \beta'} \equiv  \delta_{\hat{\beta},\beta'} S_{\epsilon \beta\hat{\beta}} S^\ast_{\epsilon \hat{\beta} \beta}.
\end{equation}
In fact, $P_\epsilon$ and $Q_\epsilon$ both vanish since we only
consider simple graphs. The square of $K \equiv \lim_{\epsilon \to 0}
K_\epsilon$ is the diagonal matrix
\begin{equation} \label{K square} \left( K^2 \right)_{\beta\beta'} =
  \delta_{\beta,\beta'} M_{\beta\hat{\beta}}M_{\hat{\beta}\beta},
\end{equation}
which only depends on $S$ through the classical map $M$. It can be
deduced from \eqref{K square} that the spectrum of $K$ is real and
is contained in $(-1,1)$ if the graph is ergodic. In \eqref{S2 diff comm off
  quad form final}, $s(\vect{j})$ is the matrix defined in \eqref{s
  diffusion}, $c(\vect{j})$ is another matrix satisfying $c(\vect{0})
= 1$, and $a(\vect{j})$ and $b(\vect{j})$ are given by
\begin{equation} \label{function a def} a(j_a,
  \vect{j_r})_{\beta\beta'} \equiv \delta_{\beta\beta'} j_a E^{(a)}_{
    \hat{\beta} \beta} + \delta_{\beta \hat{\beta}'}
  \left(\vect{j_r}\vect{E^{(r)}} \right)_{ \beta\hat{\beta}}
\end{equation}
and
\begin{equation} \label{function b def} b(j_a,
  \vect{j_r})_{\beta\beta'} \equiv \delta_{\beta\beta'} j_a
  E^{(a)}_{\beta \hat{\beta} } + \delta_{\beta \hat{\beta}'}\left(
    \vect{j_r} \vect{E^{(r)}} \right)_{ \beta\hat{\beta}}.
\end{equation}
It can be checked that $a(\vect{j})$ and $b(\vect{j})$ both vanish if
the convention $\sigma = \textrm{id}$ for the generating functions is
used.

The determinant of $\mathcal{B}$ in \eqref{S2 diff comm off quad form
  final} can be calculated using the rule \eqref{sdet of block
  matrices} adapted to conventional determinants. The result is the
product
\begin{equation} \label{det of B} \det \mathcal{B} = \Delta_{FF-FF}
  \cdot \Delta_{BB-BB}(\vect{j})
\end{equation}
where $\Delta_{FF-FF} \equiv \Delta_{BB-BB}(\vect{0})$,
\begin{equation}
  \Delta_{BB-BB}(\vect{j}) = \det \left( \mathds{1}_{\mathcal{A}} - s(\vect{j}) M_\epsilon \right)  \det \left( \mathds{1}_{\mathcal{A}} - c(\vect{j}) K_\epsilon \right)  \det \left( \mathds{1}_{\mathcal{A}} -  N_\epsilon (\vect{j}) \right),
\end{equation}
and
\begin{equation}
  N_\epsilon (\vect{j}) \equiv M_\epsilon  \frac{1}{1 - s(\vect{j})M_\epsilon} a(\vect{j}) K_\epsilon \frac{1}{1 - c(\vect{j}) K_\epsilon} b(\vect{j}).
\end{equation}
The first factor in \eqref{det of B} comes from the couplings between
$FF$ and $FF$ components of the variables in the vectors
$\tilde{w}_\beta$ and $w_\beta$.  The second factor comes from the
couplings between $BB$ and $BB$ components. The fact that the
contribution of the $FF-FF$ couplings can be found from the
contribution of the $BB-BB$ couplings by setting all the sources to
zero can actually already be observed on the formula \eqref{S2 diff}
for the diffusion second order action $S^{(2) D}$. A further look at
this formula enables one to deduce the contributions of the couplings
between the anticommuting variables. It can be seen that the matrix
mixing the $BF$ components of the row vectors $\tilde{w}_\beta$ and
the $FB$ components of column vectors $w_\beta$ has determinant
$\Delta_{BB-BB}(j_a, \vect{0})$, and similarly, the matrix mixing the
$FB$ components of the row vectors $\tilde{w}_\beta$ and the $BF$
components of the column vectors $w_\beta$ has determinant
$\Delta_{BB-BB}(0, \vect{j_r})$. Hence, the diffusion second order
generating function reads
\begin{equation} \label{xi 2 diff factorization} \xi^{(2)
    D}_{[\vect{\alpha}]} = \xi^{(2) D}_{[\vect{\alpha}], (dd)} \cdot
  \xi^{(2) D}_{[\vect{\alpha}], (oo)} \cdot \xi^{(2)
    D}_{[\vect{\alpha}], (do)},
\end{equation}
where the first factor in the right-hand side is the diffusion $(dd)$
second order generating function \eqref{xi 2 diff},
\begin{equation} \label{xi 2 oo} \xi^{(2)
    D}_{[\vect{\alpha}],(oo)}(j_a, \vect{j_r}) \equiv \frac{\det \Big(
    \mathds{1}_{\mathcal{A}} - c (j_a,\vect{0})K_{\epsilon} \Big) \det
    \Big( \mathds{1}_{\mathcal{A}} - c(0,\vect{j_r})K_{\epsilon} \Big)
  }{\det \Big( \mathds{1}_{\mathcal{A}} - K_{\epsilon} \Big) \det
    \Big( \mathds{1}_{\mathcal{A}} - c(j_a,\vect{j_r}) K_{\epsilon}
    \Big)},
\end{equation}
and
\begin{equation} \label{xi 2 do} \xi^{(2) D}_{
    [\vect{\alpha}],(do)}(j_a,\vect{j_r}) \equiv \frac{ \det \Big(
    \mathds{1}_{\mathcal{A}} - N_\epsilon (j_a, \vect{0}) \Big) \det
    \Big( \mathds{1}_{\mathcal{A}} - N_\epsilon (0, \vect{j_r}) \Big)
  } {\det \Big( \mathds{1}_{\mathcal{A}} - N_\epsilon (0, \vect{0})
    \Big) \det \Big( \mathds{1}_{\mathcal{A}} - N_\epsilon (j_a,
    \vect{j_r}) \Big)}.
\end{equation}

These functions all have the property that they become identically one
in a neighborhood of the origin if either the advanced or the retarded
derivatives are set to zero. Hence, their product \eqref{xi 2 diff
  factorization}, and the cooperon analogs, share the same property.
It follows that the derivatives of these functions satisfy
\begin{equation} \label{Gaussian C splitting} \delta
  \xi^{(2)}_{[\alpha]} = \sum_{x \in \{ dd, oo, do \} } \delta
  \xi^{(2) D}_{[\alpha], (x)} + (\kappa - 1) \sum_{x \in \{ dd, oo, do
    \} } \delta \xi^{(2) C}_{[\alpha], (x)}.
\end{equation}
Moreover, since $K$ has no eigenvalue unity, the $(do)$ generating
functions at $\epsilon = 0$ are analytic in a neighborhood of the
origin, and hence, their derivatives cannot contribute to the Gaussian
autocorrelation functions \eqref{Gaussian C from xi2 and xi2 mf}.

If the convention $\sigma = \textrm{id}$ is used for the generating
functions, then the functions $a$ and $b$ in \eqref{function a def}
and \eqref{function b def} vanish. In this case, the $(do)$ generating
function \eqref{xi 2 do} is equal to one, its derivatives vanish, and
only the $(dd)$ derivatives remain in \eqref{Gaussian C splitting}.
Therefore, from \eqref{Gaussian C splitting}, \eqref{C beta q dd D
  final} and this remark, the Gaussian autocorrelation functions
\eqref{Gaussian C from xi2 and xi2 mf} of degree $q$ read
\begin{equation} \label{Final Gaussian C} C_{[\vect{\alpha}]}^{G} =
  C_{[\vect{\alpha}],(dd)}^{G} = \frac{1}{(q-1)(2B)^{q}}
  \sum_{\begin{subarray}{c} k,l = 0 \\ k \neq l
    \end{subarray}}^{q-1}R_{\alpha_k \alpha_l} -\frac{q}{(2B)^{q+1}} +
  (\kappa - 1) \Big\{ \alpha_0 \to \hat{\alpha}_0\Big\}.
\end{equation}
Here, the last term $\{ \alpha_0 \to \hat{\alpha}_0 \}$ stands for the
right-hand side with $\alpha_0$ replaced with $\hat{\alpha}_0$.  In
summary, the off-diagonal modes do not bring any additional
contribution to the Gaussian autocorrelation functions if the
convention $\sigma = \textrm{id}$ is used.

Let us now consider the situation where $q = 2$, $[\vect{\alpha}] =
[\alpha,\alpha']$, and the generating function is defined with the
crossed convention. In this case, the two derivatives on the $(do)$
generating function \eqref{xi 2 do} give
\begin{equation} \label{derivatives on xi do}
  \delta\xi^{(2)}_{[\alpha,\alpha'],(do)} = \tr \left[
    \frac{1}{1-N_\epsilon} N_{\epsilon,0} \frac{1}{1-N_\epsilon}
    N_{\epsilon,1} + \frac{1}{1-N_\epsilon} N_{\epsilon,01} \right],
\end{equation}
where $N_{\epsilon}$ denotes the value of the function
$N_\epsilon(j_0,j_1)$ at the origin, and $N_{\epsilon, 0}$,
$N_{\epsilon, 1}$ and $N_{\epsilon, 01}$ stand for its derivatives at
the origin. Since $a$ and $b$ are zero at the origin, $N_{\epsilon}$,
$N_{\epsilon, 0}$ and $N_{\epsilon, 1}$ vanish. Therefore, only the
second term in the trace of \eqref{derivatives on xi do} contributes,
and a short calculation shows that the diffusion $(do)$ Gaussian
approximation to the intensity correlation matrix in the crossed
convention reads
\begin{equation}
  C_{\alpha \alpha',(do)}^{\times G, D} =  \lim_{\epsilon \to 0}\frac{\epsilon}{B}  \delta_{\alpha',\hat{\alpha}}  \left( \frac{M_\epsilon}{1 - M_\epsilon} \right)_{\alpha\hat{\alpha}} \left[ \left( \frac{K}{1 - K} \right)_{\alpha\alpha} + \left( \frac{K}{1 - K} \right)_{\hat{\alpha}\hat{\alpha}} \right] 
\end{equation}
The cooperon result turns out to be the same. With the decomposition
\eqref{classical paths decomposition} of the classical walks, and
using the fact that the diagonal elements of $K^n$ vanish if the
integer $n$ is odd, one gets
\begin{equation} \label{C do crossed}
  C_{\alpha \alpha',(do)}^{\times G} = \kappa
  \frac{\delta_{\alpha,\hat{\alpha}'}}{(2B)^2} \left[
    R^K_{\alpha\alpha} +R^K_{\hat{\alpha}\hat{\alpha}} \right],
\end{equation}
where
\begin{equation}
  R^K_{\alpha\alpha'} \equiv \left(  \frac{K^2}{1 - K^2} \right)_{\alpha\alpha'} = \delta_{\alpha,\alpha'} \frac{M_{\alpha\hat{\alpha}}M_{\hat{\alpha}\alpha}}{1 - M_{\alpha\hat{\alpha}}M_{\hat{\alpha}\alpha}}.
\end{equation}
The matrix $R^K$, called the back-scattering matrix, is formed with
all the oriented walks followed with the classical map $M$ which
involve only back-scatterings $\beta \to \hat{\beta}$ and no
transmission. Together with \eqref{delta 11 xi diff first} and
\eqref{delta 11 xi coop first}, \eqref{C do crossed} yields the
Gaussian contribution 
\begin{eqnarray} \label{Gaussian C crossed}
  C_{\alpha\alpha'}^{\times G} & = & \frac{ \delta_{\alpha,\alpha'}}{(2B)^2} \Big[ 2R_{\alpha\alpha} + (\kappa - 1) \big( R_{\hat{\alpha}\alpha} + R_{\alpha\hat{\alpha}} \big)\Big]- \delta_{\alpha,\alpha'}\frac{2 \kappa}{(2B)^3} \nonumber\\
  && + \kappa \frac{\delta_{\alpha,\hat{\alpha}'}}{(2B)^2} \Big[
  R^K_{\alpha\alpha} +R^K_{\hat{\alpha}\hat{\alpha}} \Big]
\end{eqnarray}
to the intensity correlation matrix in the crossed convention.

For $q = 2$ and $\alpha = \alpha'$, the parallel and crossed results
\eqref{Final Gaussian C} and \eqref{Gaussian C crossed} coincide.
Notice that in this case, the parallel and crossed sums over oriented
walks represented in Figure~\ref{G matrices and paths} are also the
same.

\section{Criteria and Rates of Universality} \label{Criteria and Rates
  of Universality}

\subsection{Full Universality and Criterion for Ergodicity}

The calculation scheme summarized in \eqref{truncated autocorrelations
  def} leads to truncated autocorrelation functions, which are sums
of the mean field contributions obtained in Section~\ref{Mean Field
  Theory} and the Gaussian contributions obtained in Section~\ref{The
  Gaussian Correction}. On one hand, the mean field results coincide with the universal Gaussian Random Wave Models introduced in Subsection~\ref{Gaussian Random Waves Models}.
On the other hand, the Gaussian quantities in \eqref{Final Gaussian C} depend on the quantum graph, but do so only through its classical map $M$, and more 
precisely through its matrix $R$, which is defined in (35) and represents the 
massive component of the sum of classical paths 
$\frac{M_\epsilon}{1 - M_\epsilon}$.
The
importance of the truncated autocorrelation functions is twofold.
Firstly, their Gaussian contributions can
be compared with their universal mean field parts in the limit of
large graphs. These comparisons lead to conditions on the increasing
sequence of quantum graphs $\{(G_l,S_l)\}_{l\in\mathbb{N}}$ to
asymptotically follow the predictions of the universal Gaussian Random Wave Models, in which
case we say that full universality is met. Secondly, for a class of
increasing sequences larger than this universal class, the truncated
quantities approximate the exact autocorrelation functions \eqref{C alpha def}.

Let $(G,S)$ be an ergodic simple quantum graph, and let
$[\vect{\alpha}]$ be a list of $q$ directed bonds for some integer $q
\geq 2$.  The mean field autocorrelation functions found in
Subsection~\ref{The Mean Field Autocorrelation Functions} and the
Gaussian autocorrelation functions in \eqref{Final Gaussian C} lead to
the truncated autocorrelation functions
\begin{equation} \label{Final Truncated C} \tilde{C}_{[\vect{\alpha}]}
  = \frac{c \left(\kappa, [\vect{\alpha}] \right)}{(2B)^q} +
  \frac{1}{(q-1)(2B)^{q}} \sum_{\begin{subarray}{c} k,l = 0 \\ k \neq
      l \end{subarray}}^{q-1} \left[ R_{\alpha_k \alpha_l} + (\kappa -
    1) \Big\{ \alpha_0 \to \hat{\alpha}_0\Big\} \right] -\frac{\kappa
    q}{(2B)^{q+1}} .
\end{equation} 
The parameter $\kappa$ is equal to one or two depending whether time-reversal 
symmetry is broken or conserved, and $c \left(\kappa, [\vect{\alpha}] \right)$ is the combinatorial
factor defined in the following way.
Suppose that each directed bond
$\beta$ appears exactly $q_\beta$ times in the list $[\vect{\alpha}]$,
then $c \left(1, [\vect{\alpha}] \right) \equiv \prod_{\beta = 1}^{2B}
q_\beta !$. Suppose now that $p_b$ denotes the number of directed
bonds in the list $[\vect{\alpha}]$ supported on bond $b$, then $c
\left(2, [\vect{\alpha}] \right) \equiv \prod_{b = 1}^{B} p_b !$.
The intensity correlation matrix and the moments implied by
\eqref{Final Truncated C} read
\begin{equation}
  \label{final correlation matrix}
  \tilde{C}_{\alpha \alpha'}=\frac{1}{4B^2}
  \left(1+\delta_{\alpha \alpha'} +R_{\alpha \alpha'} + R_{\alpha'\alpha} \right)
  + \frac{\kappa-1}{4B^2}\left( \delta_{\alpha \hat{\alpha}'} +
    R_{\alpha \hat{\alpha}'} + R_{\hat{\alpha}'\alpha} \right)
\end{equation}
and
\begin{equation}
  \label{final moments}
  \tilde{M}_{q, \alpha}=\frac{q!}{(2B)^q}
  \left(1+ \frac{1}{(q-1)!} R_{\alpha \alpha} + \frac{\kappa-1}{q!}(
    R_{\alpha \hat{\alpha}} +R_{\hat{\alpha} \alpha} + (q-2)R_{\alpha \alpha} )  \right) .
\end{equation} 
In these two expressions, the last term of \eqref{Final Truncated C}, which is of higher order in the inverse number of bonds $B^{-1}$, has been neglected since we are ultimately interested in the large graph limit $B \to \infty$.

\begin{figure}
  \begin{center}
    \includegraphics[width=0.49\textwidth]{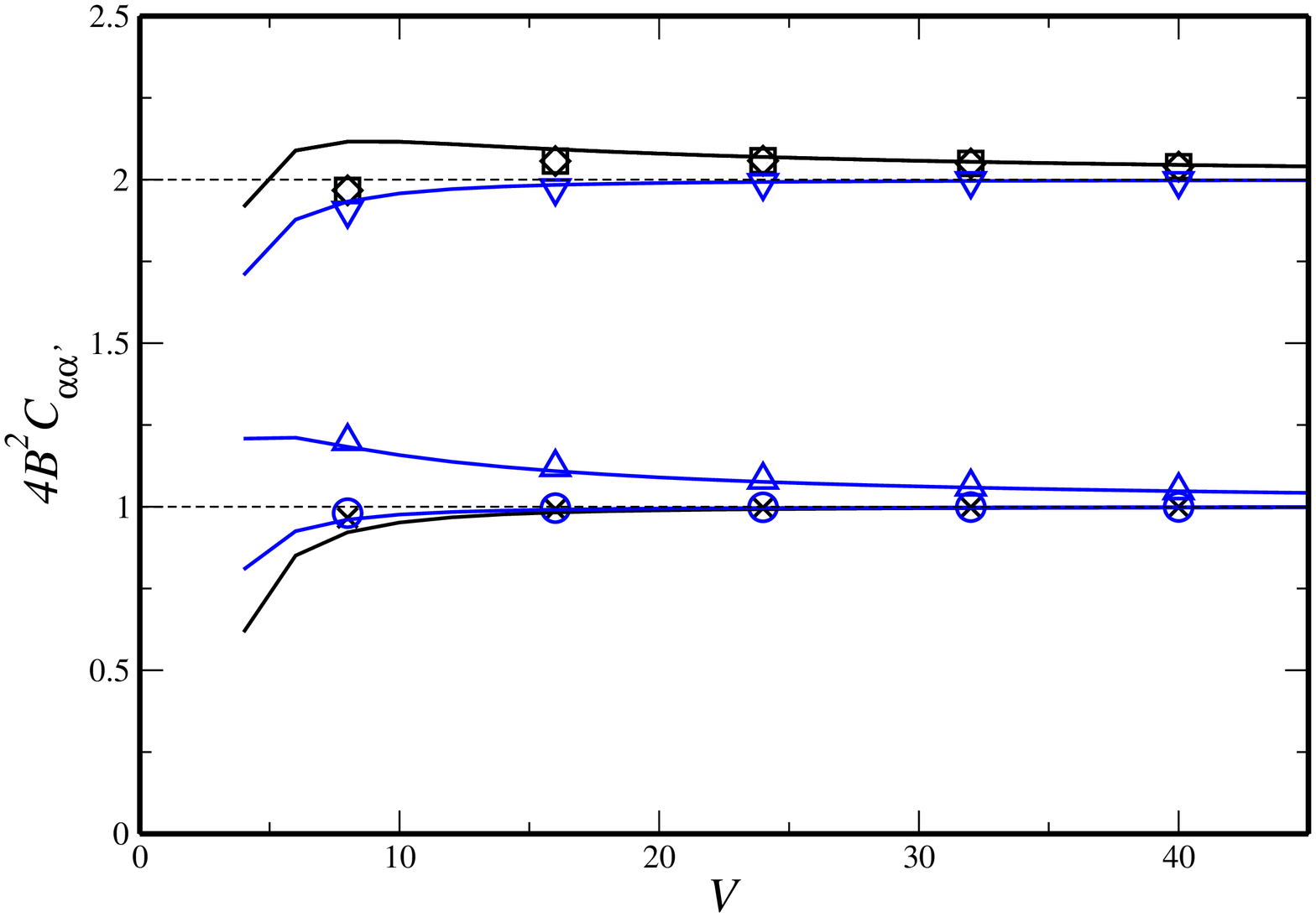}
    \includegraphics[width=0.49\textwidth]{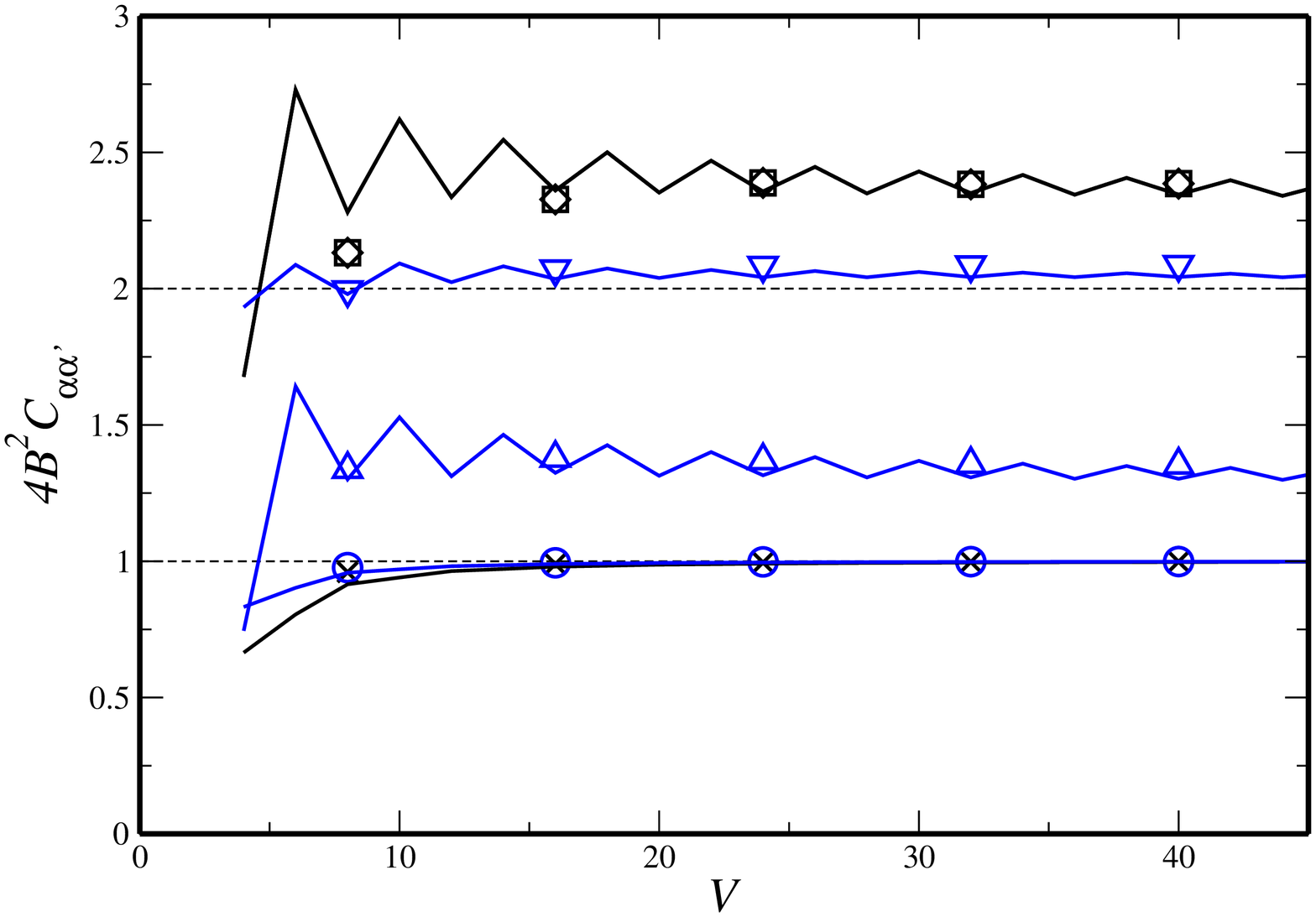}\\
    \includegraphics[width=0.49\textwidth]{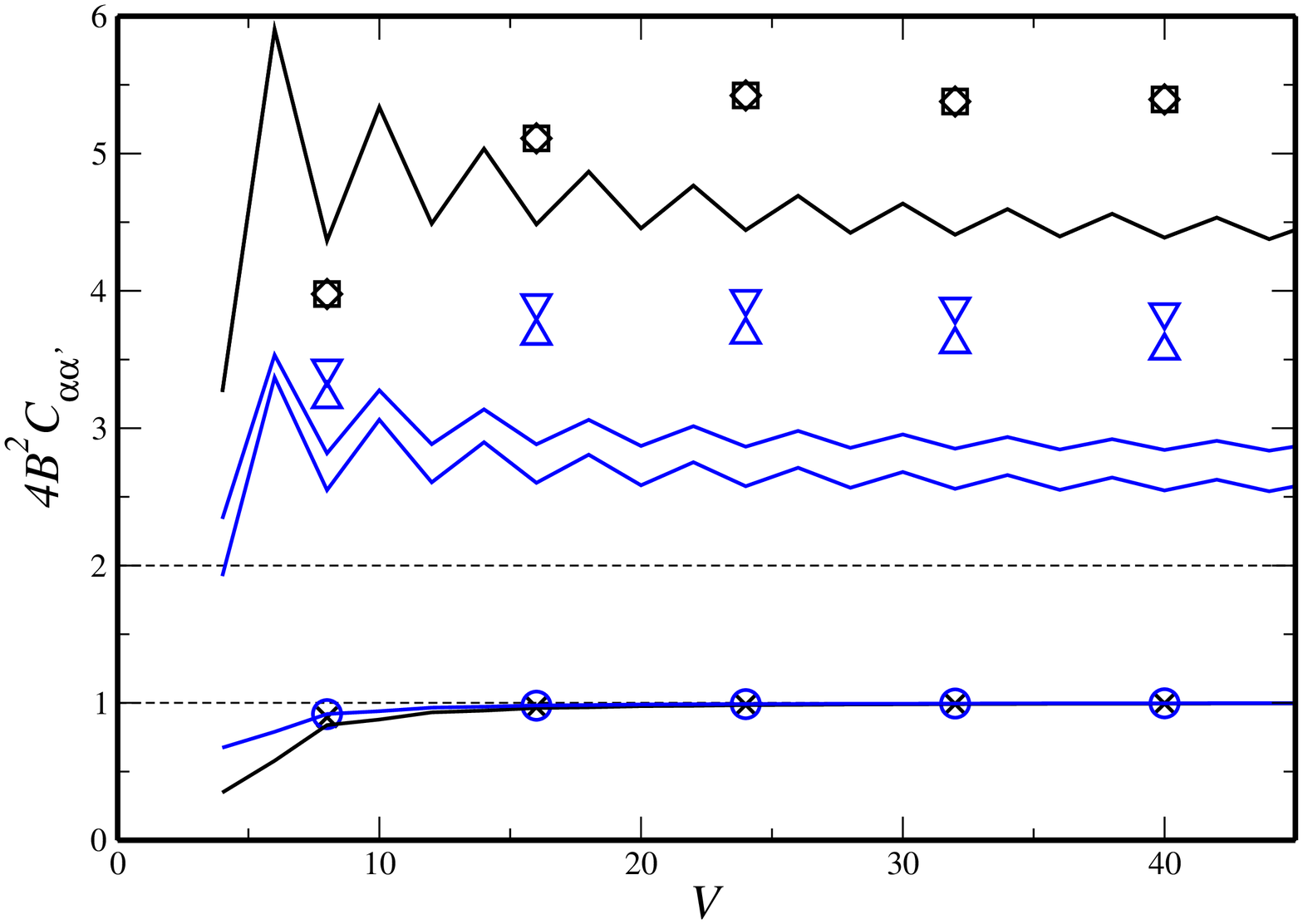}
    \includegraphics[width=0.49\textwidth]{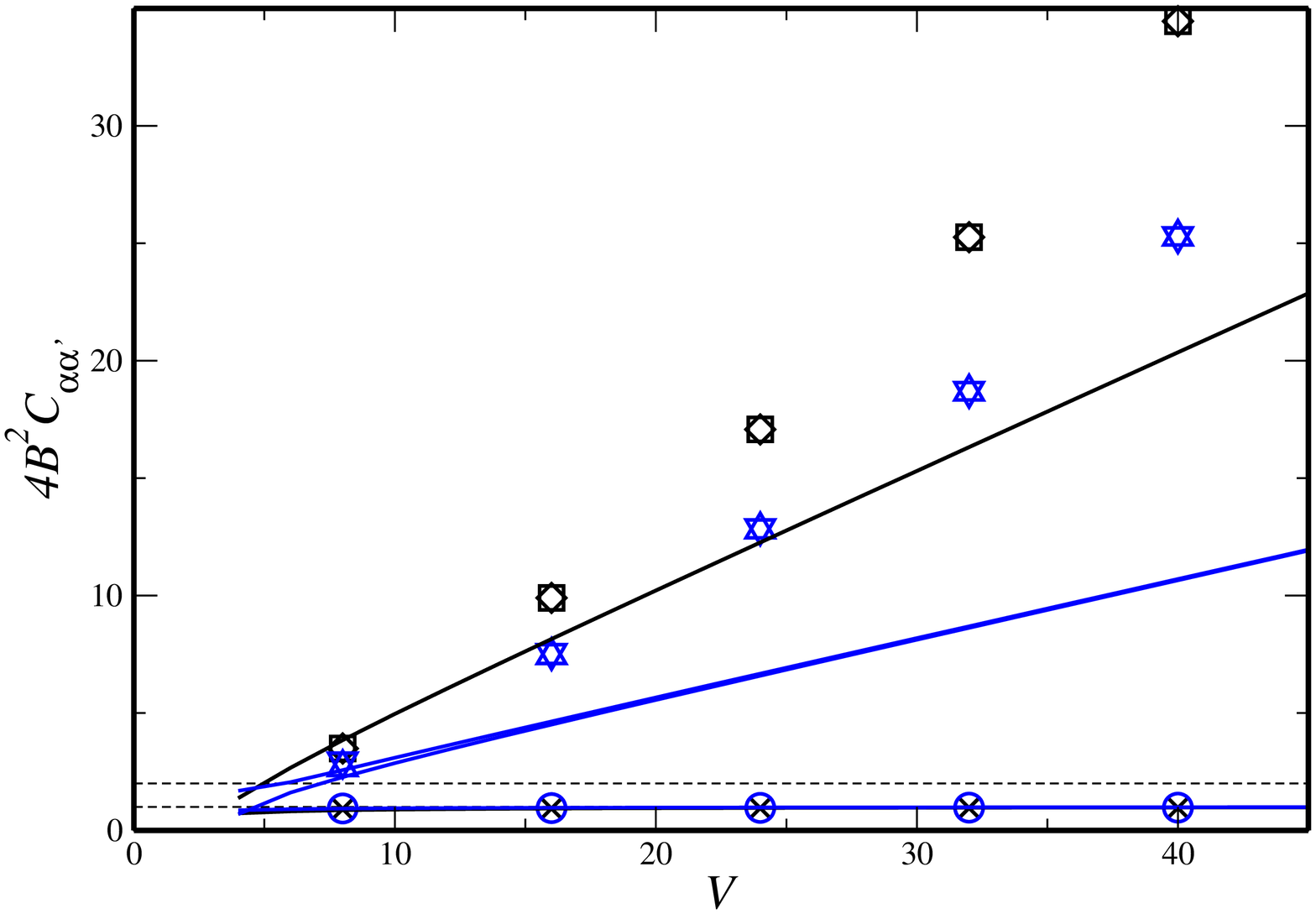}
    \caption{ 
      Rescaled matrix elements of the intensity 
      correlation matrix for complete graphs with $V$ vertices.\newline
      The four panels correspond to four different choices of the
      scattering matrix. The upper left panel is a complete DFT graph. For
      the upper right and lower left panel the scattering matrix that 
      corresponds
      to one vertex has been chosen by
      the following unitary transformation of a DFT matrix
      $\sigma^i(\lambda)= e^{-i\pi/4}\frac{1-\lambda + e^{i\pi/4}(1+\lambda) 
        \sigma^{\mathrm{DFT}}}{1+\lambda + e^{i\pi/4}(1-\lambda) 
        \sigma^{\mathrm{DFT}}}$ with the values $\lambda=0.5$ for the
      upper right panel and $\lambda=0.25$ in the lower left panel.
      For the lower right panel Neumann scattering matrices have 
      been chosen.\newline
      An additional magnetic field was applied to break time-reversal symmetry.
      Black symbols and lines correspond to results for the orthogonal
      class and blue symbols and lines to results for the unitary class
      where the symbols correspond to numerically obtained 
      intensity correlations and full lines to the
      corresponding prediction \eqref{final correlation matrix}.
      The squares (orthogonal) and downwards pointing triangles (unitary class)
      give the average rescaled
      diagonal element $2B \sum_{\alpha=1}^{2B} C_{\alpha \alpha}$
      (the Gaussian Random Wave Model predicts the value 2 
      indicated by the upper dashed line).  
      The diamonds (orthogonal) and upwards pointing triangles (unitary class)
      give the average rescaled
      time-reversed diagonal element 
      $2B \sum_{\alpha=1}^{2B} C_{\alpha \hat{\alpha}}$
      (the Gaussian Random Wave Model predicts the values 1 in the unitary and 2 
      in the orthogonal class).
      Note that the diamonds and squares always lie on top of each other.
      Eventually, crosses (orthogonal) and circles (unitary class)
      give the average rescaled
      element $\frac{B}{B-4} \sum_{\alpha',\alpha=1: \alpha\neq \alpha', 
        \alpha \neq \hat{\alpha}'}^{2B} C_{\alpha \alpha'}$
      (the Gaussian Random Wave Model predicts the value 1). The corresponding
      predictions from \eqref{final correlation matrix} are given
      by the full lines.  
    } 
    \label{numerics}
  \end{center}
\end{figure}

For graphs in the orthogonal class, the exact autocorrelation functions $C_{[\vect{\alpha}]}$ in \eqref{C alpha def} do reflect the symmetry $|a_{\alpha}|^2 =|a_{\hat{\alpha}}|^2$ of the wave function intensities.
The
intensity correlation matrix \eqref{final correlation matrix} 
indeed satisfies $ \tilde{C}_{\alpha \alpha'} =  
\tilde{C}_{\hat{\alpha} \alpha'}$ if $\kappa = 2$, but the truncated 
autocorrelation functions \eqref{final moments} of degree $q\ge 3$ do
not obey such a symmetry in general.
A comparison of 
\eqref{final correlation matrix}
with a numerically obtained
intensity correlation matrix for complete quantum graphs with
various choices of scattering matrices reveals that
\eqref{final correlation matrix} captures the asymptotics 
$B \rightarrow \infty$ very well at least as long as the intensity matrix is
not dominated by the massive contributions (see Figure~\ref{numerics}). 
A numerical comparison of higher moments to \eqref{final moments}
(not shown) reveals that their massive contributions are not as well 
approximated by our theory.
We should however emphasize that the numerical
evaluation of higher moments is not very stable. The statements we
put forward below are all consistent with the numerically
obtained data we have.

Let $\{(G_l,S_l)\}_{l\in\mathbb{N}}$ be a sequence of increasing
ergodic simple graphs that are either all in the unitary or in the
orthogonal class.  The formula \eqref{Final Truncated C} suggests that
the mean field term is the dominant one if and only if the sequence of
matrices $\{R_l \}_{l \in \mathbb{N}}$ converges to zero as $l \to
\infty$. Consequently, we conjecture that the Gaussian Random Wave
Models \eqref{gaussian model} or \eqref{orthogonal jpdf} are
asymptotically met in the sequence $\{(G_l,S_l)\}_{l\in\mathbb{N}}$ if
and only if the norm of the matrices $R_l$ decay as $l \to \infty$.
Note that $R_l \rightarrow 0$ introduces a small parameter
in which a systematic expansion may be performed. We strongly believe
that our approach may be extended to a rigorous proof that  $R_l \rightarrow 0$
implies convergence to the Gaussian Random Wave Model.

Notice that, by the definition \eqref{classical paths decomposition}
of the matrix $R_\epsilon$, the decay of the sequence 
$R_{\epsilon, l}$ is equivalent to
\begin{equation}
  \label{A}
  \lim_{l\to\infty} \frac{M_{\epsilon,l}}{1 - M_{\epsilon,l}} = \mathcal{O}(\epsilon). 
\end{equation}
This is also equivalent to the decay of 
$R_l \equiv \lim_{\epsilon \to 0} R_{\epsilon,l}$. Since all the 
components of the matrix $M_{\epsilon,l} \equiv e^{-2\epsilon} M_l$ are 
non-negative, \eqref{A} implies that
\begin{equation}
  \lim_{l\to\infty} M_l = 0. 
  \label{B}
\end{equation}
A necessary condition for this property to occur, and hence for full 
universality to be met, is that the valencies of the vertices all tend to 
infinity. This condition has already been derived at the end of 
Subsection~\ref{Gaussian Random Waves Models},
where the Random Wave Models are built. Conversely, suppose that \eqref{B}
is fulfilled, then the equality \eqref{A} also holds, and thus $R_l$ decays. 
Therefore, the necessary and sufficient condition for full universality 
that is conjectured above is actually equivalent to \eqref{B}.

The expression \eqref{final correlation matrix},
together with the formula \eqref{fluctuations def}, generate a
truncated expression $\tilde{\mathcal{F}}_V$ for the fluctuations of
an observable $V$. The asymptotic quantum ergodicity problem described
in Subsection~\ref{Asymptotic Quantum Ergodicity} can be addressed in
terms of these truncated fluctuations. Moreover, in the situations
where $\tilde{\mathcal{F}}_V$ decays, $\tilde{\mathcal{F}}_V$ is
expected to approximate the exact fluctuations $\mathcal{F}_V$.  A
direct calculation shows that, for an observable $V$ with $\bar{V} =
0$,
\begin{equation} \label{truncated fluctuations result}
  \tilde{\mathcal{F}}_V = \kappa \frac{\tr (VL)^2}{(\tr L)^2} + 2
  \kappa \frac{\sum_{\beta,\beta'}\big[VL\cdot R \cdot
    VL\big]_{\beta\beta'}}{(\tr L)^2}
\end{equation}
This formula motivates the following criterion for asymptotic quantum
ergodicity to be met in an increasing sequence of quantum graphs. We
conjecture that an increasing sequence
$\{(G_l,S_l)\}_{l\in\mathbb{N}}$ of ergodic simple graphs is
asymptotically quantum ergodic if and only if
\begin{equation}
  \label{criterion QE}
  \lim_{l \to \infty} \frac{\sum_{\beta,\beta'}\big[V_lL_l \cdot R_l  \cdot V_lL_l\big]_{\beta\beta'}}{(\tr L_l)^2} = 0
\end{equation}
for any acceptable sequence $\{V_l\}_{l\in\mathbb{N}}$ with $\bar{V}_l
= 0$.
We will give a slightly more detailed variant of this conjecture 
(and a discussion of possible obstruction to its validity) below.

Moreover, if the stronger condition $R_l \to 0$ is fulfilled, the
increasing sequence of graphs is fully universal, and the convergence
rate of $\tilde{\mathcal{F}}_V$, and hence of $\mathcal{F}_V$, is then
also universal. As in the case of the validity of the Gaussian 
Random Wave Model we strongly believe that our approach can be extended
to a rigorous proof using $R$ (or an equivalent quantity) as the small
parameter.

Note that the crossed formulae for the massive (Gaussian) contribution 
to the intensity correlation matrix in
\eqref{Gaussian C crossed} differ from the formulae in \eqref{Final
  Gaussian C}, which are used above. 
The crossed
expressions only involve the diagonal components $(\alpha, \alpha)$
and the components $(\alpha,\hat{\alpha})$ of the matrix $R$ --
they also contain a new backscattering term. These do
not obey $C_{\alpha \alpha}= C_{\alpha \hat{\alpha}}$
and, indeed, they do not capture the massive corrections 
in the exact 
correlation matrix as well as the parallel convention
in a numerical test (Figure~\ref{numerics} only presents the results
for the parallel convention).

Let us now consider observables $V$ such that $V_b L_b = \frac{\tr
  L}{2B}$ on half of the bonds, and $V_b L_b = - \frac{\tr L}{2B}$ on
the other half. The set of such observables is actually sufficiently
large to compare the intensities of the wave function on the different
bonds. Moreover, they provide acceptable sequences, according to
\eqref{acceptable seq of obs}. For such observables, \eqref{truncated
  fluctuations result} yields
\begin{equation} \label{truncated crossed fluctuations with special
    obs}
  \tilde{\mathcal{F}}_V \approx 
  \kappa \frac{1}{2B} + \frac{2 \kappa \tr (R+R\sigma_1^d )}{(2B)^2}, 
\end{equation}
where we neglected almost all off-diagonal terms (apart from those obeying
$\alpha'=\hat{\alpha}$) in the double sum
in  \eqref{truncated fluctuations result}. 
An increasing sequence
$\{(G_l,S_l)\}_{l\in\mathbb{N}}$ of simple graphs is then expected to be
asymptotically quantum ergodic 
if and
only if both
\begin{equation} \label{Crossed Criterion for QE} \lim_{l \to \infty}
  \frac{\tr\ R_l}{(2B_l)^2} = \lim_{l \to \infty}
  \frac{1}{(2B_l)^2}\sum_{i = 2}^{2B_l} \frac{1 - m_{l,i}}{m_{l,i}} =
  0
\end{equation} 
and
\begin{equation} \label{Crossed Criterion for QE orthogonal} \lim_{l
    \to \infty} \frac{\tr\ R_l \sigma_1^d}{(2B)^2} = 0
\end{equation}
hold. In \eqref{Crossed Criterion for QE}, the complex numbers 
$m_{l,i}$, $2 \leq i \leq 2B$, are the $2B - 1$ non-zero masses, 
that is the $2B - 1$ eigenvalues of the matrix $1 - M_l$.

\subsection{Quantum Ergodicity and the Classical Spectral Gap}
\label{Asymptotic Quantum Ergodicity and Classical Spectral Gap}

Sufficient conditions for the condition 
\eqref{Crossed Criterion for QE} 
to be fulfilled or violated in an increasing sequence
$\{(G_l,S_l)\}_{l\in\mathbb{N}}$ of ergodic simple graphs can be given
in terms of the sequence $\{ \Delta_{M_l} \}_{l \in \mathbb{N}}$ of
spectral gaps of $1 - M_l$.

Let us first consider the
case that all
non-zero masses stay away from the origin. The sum in
\eqref{Crossed Criterion for QE} behaves like $2B_l$, and hence, after
dividing by $(2B_l)^2$, the large graphs limit vanishes, and
\eqref{Crossed Criterion for QE} (and 
similarly \eqref{Crossed Criterion for QE orthogonal}) holds.

Now let us turn to the case that some masses approach zero
as $B_l \rightarrow \infty$. For sake of simplicity, the index $l$ of 
the quantum graph $(G_l,S_l)$ will be dropped.
Let us order the spectrum $\{ m_i \}_{i \in \mathbb{N}_{2B}}$ of $1 -
M$ such that $|m_i| \leq |m_{i+1}|$ for all $i \in \mathbb{N}_{2B }$,
and let us now suppose that the spectral gap $\Delta_{M} \equiv |m_2|$
approaches the origin with an exponential rate $\alpha > 0$, that is
\begin{equation}
  |m_2| \sim (2B)^{-\alpha}.
  \label{alpha scaling}
\end{equation}
 The matrix $R$ is real since $M$ is real and the
vector $|1\rangle$ is also real. It follows that the massive
contribution of the fluctuations \eqref{truncated crossed fluctuations
  with special obs} can be written
\begin{equation} \label{massive fluctuations splitting pos neg}
  \tilde{\mathcal{F}}_{V}^M \equiv \frac{2 \kappa}{(2B)^2} \sum_{i = 2}^{2B}
  \Re \frac{1 - m_i}{m_i} = \frac{2 \kappa}{(2B)^2} \sum_{i = 2}^{2B} \frac{
    \Re m_i}{|m_i|^2} - \frac{2 \kappa (2B - 1)}{(2B)^2}.
\end{equation}
The second term of the right-hand side behaves like $(2B)^{-1}$, so
that \eqref{Crossed Criterion for QE} is satisfied if and only if the
first term of the right-hand side, denoted by
$\hat{\mathcal{F}}_{V}^M$ in what follows, decays.  With the obvious
inequality $\Re m_i \leq |m_i|$, one gets
\begin{equation}
  \hat{\mathcal{F}}_V^M \leq \frac{2 \kappa}{(2B)^2}  \sum_{i = 2}^{2B} \frac{ 1}{|m_i|} \leq  \frac{2 \kappa}{(2B)^2}   \frac{ 2B - 1}{|m_2|} \sim (2B)^{\alpha - 1}.
\end{equation}
Therefore, if $\alpha < 1$, $\hat{\mathcal{F}}_{V}^M$ decays and
\eqref{Crossed Criterion for QE} is fulfilled.

Since there are $2B-1$ non-zero masses, and since these masses are either real
or appear in complex conjugated pairs, there is at least one mass
$m_{l,i}$ such that
\begin{equation} \label{t i finite} t_i \equiv \sup_{l \in \mathbb{N}}
  \Big| \tan \arg m_{l,i} \Big| < \infty.
\end{equation}
Remember that the non-zero masses are confined in the open disc of
radius 1 centered at 1. It follows that $|m_i| \leq(1 + t_i) \Re m_i$,
and thus
\begin{equation} \label{new lower bound of hat F}
  \hat{\mathcal{F}}_V^M > \frac{2}{(2B)^2} \frac{ \Re m_i }{|m_i|^2}
  \geq \frac{2}{(2B)^2} \frac{1}{(1 + t_i) |m_i|} \sim (2B)^{\alpha -
    2}.
\end{equation}
Therefore, if $\alpha \geq 2$, $\hat{\mathcal{F}}_V^M$ does not 
decay, \eqref{Crossed
  Criterion for QE} is violated, and the increasing sequence of
quantum graphs is not asymptotically quantum ergodic.

\subsection{The Four Possible Regimes}
\label{four regimes}

To summarise our findings and give a more detailed 
account of our conjecture and possible obstructions to its validity 
we have found the following four regimes 
\begin{enumerate}
\item The fully universal 
  regime $R_l \rightarrow 0$. 
  Equivalently,  all matrix elements of the classical map
  converge to zero, or
  all eigenvalues
  apart from the (non-degenerate) eigenvalue one
  of the classical map $M_l$ converge to zero, or all $2B-1$
  non-zero masses converge to one. 
  In this case the
  eigenvalues of the classical map may be used as a small parameter
  for a systematic expansion. Our theory then shows 
  that all autocorrelation functions
  converge to the universal predictions of the Gaussian
  Random Wave Model, and we believe that the scheme used
  here can be extended to a rigorous proof.
  Such graphs are also asymptotically 
  quantum ergodic with
  a universal rate of convergence.\\
  This class
  includes the complete DFT graphs
  (or complete quantum graphs such that nonvanishing
  elements of the classical map are of order $\mathcal{O}(V^{-1})$).
\item The large mass regime, characterized by $\alpha<1$ in \eqref{alpha scaling}
  and 
  $\lim_{l \to \infty}  R_l \neq 0$. Some non-zero masses do not converge to 
  one, but none of them approaches
  zero too fast. The complete Neumann graph is an example
  in this class with $\alpha=\frac{1}{2}$.
  Our theory predicts massive 
  corrections to the predictions of the Gaussian Random Wave Model
  which persist in the asymptotic regime $B\rightarrow \infty$.
  This implies that the Gaussian Random Wave Model is not applicable
  but asymptotic quantum ergodicity still holds. This statement 
  has the status of a 
  conjecture which may be very hard to prove because there is no
  obvious small parameter. 
  As a consequence our result \eqref{Final Truncated C}
  may not estimate the massive contribution 
  accurately.
  Our numerical data (see Figure~\ref{numerics}) indeed show that the 
  intensity correlation 
  matrix for a complete Neumann graph 
  is of a similar
  order of magnitude as predicted by our theory 
  but its massive contribution is
  underestimated. 
  More work needs to be done to capture the massive contributions for 
  higher moments correctly. Our theory may be improved considerably
  by starting from a different exact expression or
  by going beyond the Gaussian approximation in the massive modes. 
  In the orthogonal
  case one should start from an expression 
  that incorporates the symmetries of the wave
  function in all orders. 
\item The crossover regime, characterized
  $1 \le \alpha \le 2$ in \eqref{alpha scaling}. 
  As $\lim R_l \neq 0$ the Gaussian
  Random Wave Model does not hold. In this regime we conjecture
  that the criteria \eqref{criterion QE}
  or \eqref{Crossed Criterion for QE} decide whether a sequence
  of graphs is asymptotically quantum ergodic or not. This conjecture 
  for the crossover regime should be taken with much more care
  than the previous
  conjecture for the large mass regime. It does work
  for Neumann star graphs which have exponent $\alpha=1$ and 
  for which different methods revealed 
  that asymptotic quantum ergodicity 
  does not hold \cite{BBK}. These graphs have indeed 
  a large number of masses with $m \sim 1/B$. 
  This number is of order $\mathcal{O}(B)$
  such that the limit in \eqref{Crossed Criterion for QE}
  gives a constant.
  However, it has also been brought
  to our attention \cite{private} that analogous criteria derived
  in \cite{GA2, GA1} for the validity of Gaussian Random Matrix
  predictions for spectral correlation functions may lead
  to wrong conclusions
  for some borderline cases for which the analogous massive
  contributions are overestimated using the saddle-point approximation
  to the corresponding exact variant of the supersymmetric
  $\sigma$-model. To some extent the prediction for the
  massive correction may be improved as outlined in the large mass regime.  
\item The non-universal small mass regime, $\alpha\ge 2$ 
  in \eqref{alpha scaling}. We conjecture that
  neither the Gaussian Random Wave Model nor asymptotic quantum ergodicity
  hold. In this regime the saddle-point analysis to the exact supersymmetric
  $\sigma$-model may break down completely. While it may not be trivial to
  prove this part of our conjecture rigorously our results give very strong
  evidence in favour of the conjecture.
\end{enumerate}

\section{Discussions} \label{Discussions}

Our main results are the formula \eqref{Final Truncated C} for the
autocorrelation functions $C_{[\vect{\alpha}]}$ defined in \eqref{C
  alpha def}, and the formula \eqref{truncated fluctuations result}
for the fluctuations of
an observable defined in \eqref{fluctuations def}. These formulae
depend on the quantum graph only through the matrix $R$, and this
matrix, which is defined in \eqref{classical paths decomposition},
only involves the underlying classical dynamics $M$. Hence, our
results relate the statistical properties of the quantum energy
eigenfunctions to properties of the classical dynamics on the graph.
Moreover, they also reveal that the system dependency has no chance to
vanish, and hence, a finite graph cannot be entirely described by the
Gaussian Random Wave Model developed in Section~\ref{Gaussian Random
  Waves Models} or even be quantum ergodic. These properties can only
be met asymptotically in increasing sequences of graphs, that is in
sequences of graphs where the number of bonds tends to infinity. In
Section~\ref{Criteria and Rates of Universality}, classical criteria
for such a sequence to be asymptotically described by the Gaussian
Random Wave Model or to be asymptotically quantum ergodic are
formulated (the section concludes with a summary of the criteria
and connected conjectures).
The condition $R \to 0$ for full universality, that is for
the Gaussian model to be satisfied in the large graph limit, is more
restrictive than the criteria for asymptotic quantum ergodicity. This
is understandable since this latter property only depends on the
second moment of the intensities, and the fluctuations
$\tilde{\mathcal{F}}_V$ in \eqref{truncated fluctuations result},
which measure the deviation to ergodicity, can also decay in a
non-universal way.

The general formulae \eqref{Final Truncated C} for the autocorrelation
functions and, in particular, \eqref{truncated fluctuations result}
for the fluctuations, have been obtained by a saddle-point analysis of
the exact field-theoretical expression \eqref{exact xi superintegral}.
A comparison with the two periodic orbits approaches in the
subsections \ref{Long Diagonal Orbits} and \ref{Diagonal
  Approximation} reveals how the field-theoretical scheme exactly
proceeds. The first term of $\tilde{\mathcal{F}}_V$ in \eqref{truncated
  fluctuations result}, which is universal, originates from our exact
calculation on the zero mode manifold, and it coincides with the
result predicted by the long diagonal orbits in Subsection~\ref{Long
  Diagonal Orbits}. This draws a parallel between the zero mode, that
is the uniform component of the classical map $M$, and long diagonal
orbits. This is in fact not surprising since the zero mode is
precisely the one that does not decay, and can thus survive in long
orbits. The second term of $\tilde{\mathcal{F}}_V$ involves the
system-dependent matrix $R$, that is the non-zero masses, and it
coincides with the system-dependent contribution of the diagonal
approximation exposed in Subsection~\ref{Diagonal Approximation}.
Hence, one deduces that our field-theoretical approach discriminates
between the different modes of the classical map $M$. The uniform
component of $M$ is treated in an exact way, which the diagonal
approximation in \ref{Diagonal Approximation} cannot do, while the
massive decaying modes are treated in a perturbative way.

It is also interesting to compare our results with those obtained by
S.~Gnutzmann and A.~Altland in \cite{GA2} and \cite{GA1} concerning
the asymptotic spectral two-point correlation function $R_2(s)$ in a
sequence of increasing quantum graphs. Their theory relates the function
$R_2(s)$ to the sequence of spectral gaps $\Delta$ of the matrices $1
- M$.  If the spectral gaps stay away from zero, the random matrix
two-point correlation function is obtained in the limit of large
graphs. Our condition $R \to 0$ for full universality requires all the
non-zero eigenvalues of $1 - M$ to tend to one, which is obviously
much stronger. Hence, even in situations where the Gaussian Random Wave
Model does not hold, there is a possibility for random matrix theory
to describe $R_2(s)$, but if the Gaussian Wave Model does hold, then
$R_2(s)$ must be universal. Moreover, if the sequence of spectral gaps
vanishes as $\Delta_M \sim B^{-\alpha}$ as the number of bonds $B$
becomes large, Gnutzmann and Altland's theory predicts different
outcomes for $R_2(s)$ depending on the value of the positive number
$\alpha$. If $\alpha < \frac{1}{2}$, a random matrix behavior is
reached, whereas a non-zero system-dependent contribution always
remains if $\alpha \geq 1$. In the intermediate regime $\alpha \in
[\frac{1}{2}, 1)$, the asymptotic two-point function $R_2(s)$ depends
on the proportion of vanishing modes, as explained in \cite{GA1}. In
Subsection~\ref{Asymptotic Quantum Ergodicity and Classical Spectral
  Gap}, we found that $\alpha<1$ implies asymptotic quantum
ergodicity, whereas $\alpha \geq 2$ generally forbids ergodicity.
Therefore, universality for $R_2(s)$ implies asymptotic quantum
ergodicity. However, in the domain $\alpha \in [\frac{1}{2}, 1)$,
quantum ergodicity is always reached, whereas $R_2(s)$ can be
non-universal.

To conclude, let us mention some possible improvements of our method
and some interesting directions for further research. In the main
formula \eqref{Final Truncated C}, the system-dependent terms
correspond to a Gaussian approximation around $Q= \sigma_3^{RA}$ in
the directions that are transverse to the saddle-point manifold. A
true Gaussian approximation should expand the exact action to second
order around every point of the saddle-point manifold. The
correspondence between these two procedures has only been verified on
the submanifold $\tZ = Z^\tau$ with vanishing sources. Moreover and 
more importantly, in
this expansion around the zero mode manifold, the higher order terms
have not been controlled. Estimating these terms remains a major
problem of this field-theoretical method. Note that only
in the fully universal case $R \to 0$ one knows a small parameter
($R$ itself) that one may use to order a systematic expansion.  
Besides, we have also shown
that different  but equivalent conventions in \eqref{statistics and
  Green matrices}  lead to different outcomes by our second order
expansion scheme.  
This implies that  our second order expansion
is not a systematic expansion  in any intrinsic parameter of the quantum graph.
Another question is  whether the formula \eqref{Final Truncated C} is
suitable to describe  other quantum systems if the matrix $M$ is
replaced with the  Perron-Frobenius operator of a chaotic Hamiltonian system. 
The field-theoretical
method used here  is probably difficult to generalize to other systems.
An idea would be to develop a single periodic orbit approach that
reproduces \eqref{Final Truncated C} and transfer it to other types of
systems.

\paragraph{Acknowledgments} This work has been partially supported by
ORSAS and the Swiss Society of Friends of the Weizmann Institute of
Science. The authors are grateful to Yan Fyodorov, Martin Sieber
and Uzy Smilansky for their helpful comments.

\end{document}